\documentclass[11pt]{article} 
\usepackage{a4wide}

\RequirePackage[latin1]{inputenc}     
 
\usepackage[english]{babel} 
\usepackage{amsmath} 
\usepackage{amsfonts} 
\usepackage{amssymb} 
\usepackage{xspace} 
\usepackage{latexsym} 
\usepackage{url} 
\usepackage{xspace} 
\usepackage[all]{xy} 
\usepackage{graphics}
\usepackage[usenames,dvipsnames]{color}              
\RequirePackage[latin1]{inputenc}

\newenvironment{restate-proposition}[2][{}]{\noindent\textbf{Proposition~{#2}}\;\textbf{#1}\  
}{\vskip 1em} 
 
\newenvironment{restate-theorem}[2][{}]{\noindent\textbf{Theorem~{#2}}\;\textbf{#1}\  
}{\vskip 1em} 
 
\newenvironment{restate-corollary}[2][{}]{\noindent\textbf{Corollary~{#2}}\;\textbf{#1}\  
}{\vskip 1em}

\newcommand{\Proofitem}[1]{\medskip \noindent $#1\;$} 
\newcommand{\Proofitemf}[1]{\noindent $#1\;$} 
\newcommand{\Defitem}[1]{\smallskip \noindent $#1\;$} 
 
\newcommand{\Defitemf}[1]{\noindent $#1\;$}

\newcommand{\hbra}{\noindent\hbox to \textwidth{\leaders\hrule height1.8mm depth-1.5mm\hfill}} 
\newcommand{\hket}{\noindent\hbox to \textwidth{\leaders\hrule height0.3mm\hfill}} 
\newcommand{\ratio}{.3}

 \newtheorem{theorem}{Theorem} 
  
 \newtheorem{definition}[theorem]{Definition} 
 \newtheorem{lemma}[theorem]{Lemma} 
  
 \newtheorem{proposition}[theorem]{Proposition} 
 \newtheorem{example}[theorem]{Example} 
  
 \newtheorem{remark}[theorem]{Remark} 
 
 \newtheorem{proviso}[theorem]{Proviso}

\newcommand{\Proof}{\noindent {\sc Proof}. } 
 
 \newcommand{\qed}{\hfill${\Box}$}

\newcommand{\Figbar}{{\center \rule{\hsize}{0.3mm}}}    
 
\newcommand{\cl}[1]{{\cal #1}}          

\newcommand{\cmp}[1]{{\cal C}(#1)}
\newcommand{\evn}[2]{{\cal E}_{{\it vn}}(#1,#2)}   
\newcommand{\cps}[1]{{\cal C}_{{\it cps}}(#1)}
\newcommand{\h}[1]{{\cal C}_{{\it h}}(#1)}      
\newcommand{\cc}[1]{{\cal C}_{{\it cc}}(#1)}      
\newcommand{\rg}[1]{{\cal C}_{\it rg}(#1)}   
\newcommand{\vn}[1]{{\cal C}_{{\it vn}}(#1)}     
\newcommand{\sco}{\mid}     
\newcommand{\rb}[1]{{\cal R}(#1)}             
\newcommand{\ms}[1]{|#1|}                    
\newcommand{\fv}[1]{{\sf fv}(#1)}            
\newcommand{\ftv}[1]{{\sf ftv}(#1)}           
\newcommand{\frv}[1]{{\sf frv}(#1)}            
\newcommand{\zero}{{\bf 0}}                  
 
\newcommand{\Gives}{\vdash}             
\newcommand{\aGives}{\vdash^{{\it vn}}} 
\newcommand{\rGives}{\vdash^{{\it rg}}} 

\newcommand{\arrow}{\rightarrow}        
\newcommand{\trarrow}{\stackrel{*}{\rightarrow}}        
 
\newcommand{\lets}[3]{{\sf let} \ #1 = #2 \ {\sf in} \ #3}    
\newcommand{\pack}[1]{{\sf pack}(#1)}
\newcommand{\unpack}[1]{{\sf unpack}(#1)}
\newcommand{\letall}[2]{{\sf let} \ {\sf all}(#1) \ {\sf in} \ #2}
\newcommand{\dis}[2]{{\sf dis}(#1) \ {\sf in} \ #2}

\newcommand{\Alt}{ \mid\!\!\mid  } 
 
\newcommand{\infer}[2]{\begin{array}{c} #1 \\ \hline #2 \end{array}} 
 
\newcommand{\lb}[2]{#1>#2}
\newcommand{\plb}[2]{#2>#1}
\newcommand{\lab}[1]{{\cal L}(#1)}
\newcommand{\labi}[2]{{\cal L}_{#1}(#2)}
\newcommand{\er}[1]{{\it er}(#1)}
\newcommand{\rer}[1]{{\it rer}(#1)} 
\newcommand{\ren}[1]{{\it ren}(#1)} 
\newcommand{\at}[1]{{\sf at}(#1)}

\newcommand{\union}{\cup}               
\newcommand{\inter}{\cap}               
\newcommand{\minus}{\backslash}         
\newcommand{\comp}{\circ}               
 
\newcommand{\prj}[2]{\pi_{#1}(#2)}     

\newcommand{\rl}[1]{\;#1\;}             
 
\newcommand{\mand}{\mbox{ and }} 
\newcommand{\w}[1]{{\it #1}}    

\newcommand{\xst}[2]{\exists\, #1\;\: #2}

\newcommand{\s}[1]{{\sf #1}}    

\newcommand{\sem}[1]{{\cal I}(#1)} 

\newcommand{\act}[1]{\xrightarrow{#1}} 

\newcommand{\set}[1]{\{#1\}}

 \newcommand{\wact}[1]{\stackrel{#1}{\Rightarrow}} 

\newcommand{\eval}{\Downarrow}

\newcommand{\C}{{\sf C}}

\newcommand{\Cminor}{{\sf Cminor}}
\newcommand{\mips}{{\sf Mips}}
\newcommand{\eighty}{{\sf 8051}}
\newcommand{\cerco}{{\sf CerCo}}
\newcommand{\coq}{{\sf Coq}}
\newcommand{\ocaml}{{\sf OCaml}}

\begin{document} 
 
\title{Certifying and reasoning about \\ cost annotations of functional programs
\footnote{An extended abstract with the same title without proofs and {\em not} accounting for the typing 
of the compilation chain and the memory management of the compiled code 
has appeared in \cite{ARG11}. Also the present version introduces a prototype
implementation available in \cite{YRG11}.
The authors were supported by the {\em Information and Communication
  Technologies (ICT) Programme} as Project FP7-ICT-2009-C-243881 $\cerco$. 
}}



\author{Roberto M. Amadio$^{(1)}$~~~~~~~~~       Yann~R\'egis-Gianas$^{(2)}$ \\
{\small $^{(1)}$ Universit\'e Paris Diderot (UMR-CNRS 7126)}\\
{\small $^{(2)}$ Universit\'e Paris Diderot (UMR-CNRS 7126) and INRIA (Team ${\pi}r^2$)}}

\maketitle

\begin{abstract}
We present a so-called labelling method to insert cost annotations
in a higher-order functional program, to certify their correctness
with respect to a standard, typable compilation chain 
to assembly code including safe memory management, 
and to reason about them in a higher-order Hoare logic.
\end{abstract}

\section{Introduction}
In previous work~\cite{AAR09,AAR12}, we have discussed the problem of
building a $\C$ compiler which can {\em lift} in a provably correct
way pieces of information on the execution cost of the object code to
cost annotations on the source code.  To this end, we have introduced
a so called {\em labelling} approach and presented its application to
a prototype compiler written in $\ocaml$ from a large fragment of the
$\C$ language
to the assembly languages of $\mips$ and $\eighty$, 
a 32 bits and 8 bits processor, respectively. 

In the following, we are interested in extending the approach to
(higher-order) functional languages. On this issue, a common belief is
well summarized by the following epigram by A.~Perlis \cite{Perlis82}: {\em A Lisp
  programmer knows the value of everything, but the cost of nothing.}
However, we shall show that, with some ingenuity, the methodology
developed for the $\C$ language can be lifted to functional
languages.

\subsection{A standard compilation chain}
Specifically, we shall focus on a rather standard
compilation chain from a call-by-value $\lambda$-calculus to a
register transfer level (RTL) language.  Similar compilation chains
have been explored from a formal viewpoint by Morrisett {\em et
al.}~\cite{M99} (with an emphasis on typing but with no simulation
proofs) and by Chlipala~\cite{Chlipala10} (for type-free languages but
with machine certified simulation proofs).

The compilation chain is described in the lower part of
Table~\ref{chain}. Starting from a standard call-by-value
$\lambda$-calculus with pairs, one performs first a CPS translation,
then a transformation that gives names to values, followed by a closure
conversion, and a hoisting transformation.  All languages considered
are subsets of the initial one though their evaluation mechanism is
refined along the way.   In particular, one moves from an ordinary
substitution to a specialized one where variables can only be replaced
by other variables. One advantage of this approach, as already noted
for instance by Fradet and Le M\'etayer \cite{FM91}, is to have
a homogeneous notation that makes correctness proofs simpler.

Notable differences with respect to Chlipala's
compilation chain~\cite{Chlipala10} is a different choice of the
intermediate languages and the fact that we rely on a small-step
operational semantics.  We also diverge from
Chlipala~\cite{Chlipala10} in that our proofs, following the usual
mathematical tradition, are written to explain to a human why a
certain formula is valid rather than to provide a machine with a
compact witness of the validity of the formula.

The final language of this compilation chain can be directly mapped to a RTL 
language: functions correspond to assembly level routines and
the functions' bodies correspond to sequences of assignments
on pseudo-registers ended by a tail recursive call. 

\begin{table}[t]
{
$$
\xymatrix{
  \lambda^{\cl{M}}
& 
  \lambda^{\ell}
  \ar[l]_{\cal I}
  \ar[r]^{{\cal C}_{{\it cps}}}
  \ar@/^/[d]^{\w{er}}
&
  \lambda_{\w{cps}}^{\ell}
  \ar@/^/[r]^{{\cal C}_{{\it vn}}}
  \ar[d]^{\w{er}}
&
  \lambda_{\w{cps},\w{vn}}^{\ell} 
  \ar@/^/[l]_{\cal R}
  \ar[r]^{{\cal C}_{{\it cc}}}
  \ar[d]^{\w{er}}
&
  \lambda_{\w{cc},\w{vn}}^{\ell}
  \ar[r]^{{\cal C}_{{\it h}}}
  \ar[d]^{\w{er}}
&
  \lambda_{h,\w{vn}}^{\ell}
  \ar[d]^{\w{er}}
\\
&
  \lambda
  \ar@/^/[u]^{\cal L}
  \ar[r]^{{\cal C}_{{\it cps}}}
  \ar[r]^{{\cal C}_{{\it cps}}}
&
  \lambda_{\w{cps}}
  \ar@/^/[r]^{{\cal C}_{{\it vn}}}
&
  \lambda_{\w{cps},\w{vn}}
  \ar@/^/[l]_{\cal R}
  \ar[r]^{{\cal C}_{{\it cc}}}
&
  \lambda_{\w{cc},\w{vn}}
  \ar[r]^{{\cal C}_{{\it h}}}
&
  \lambda_{h,\w{vn}}
}
$$
}
\caption{The compilation chain with its labelling and instrumentation.}
\label{chain}
\end{table}

\subsection{The labelling approach to cost certification}
While the {\em extensional} properties of the compilation chain have
been well studied, we are not aware of previous work focusing on more
{\em intensional} properties relating to the way the compilation
preserves the complexity of the programs.  Specifically, in the
following we will apply to this compilation chain the `labelling
approach' to building certified cost annotations.  In a nutshell the
approach consists in identifying, by means of labels, points in the
source program whose cost is constant and then determining the value
of the constants by propagating the labels along the compilation chain
and analysing small pieces of object code with respect to a target
architecture.

Technically the approach is decomposed in several steps.  First, for
each language considered in the compilation chain, we define an
extended {\em labelled} language and an extended operational semantics
(upper part of Table~\ref{chain}).  The labels are used to mark
certain points of the control. The semantics makes sure that, whenever
we cross a labelled control point, a labelled and observable transition
is produced.

Second, for each labelled language there is an obvious function
$\w{er}$ erasing all labels and producing a program in the
corresponding unlabelled language.  The compilation functions are
extended from the unlabelled to the labelled language so that they
commute with the respective erasure functions.  Moreover, 
the simulation properties of the compilation functions are lifted from the
unlabelled to the labelled languages and transition systems.

Third, assume a {\em labelling} $\cl{L}$ of the source language is 
a right inverse of the respective erasure function.
The evaluation of a labelled source program produces both a value and
a sequence of labels, written $\Lambda$, which intuitively stands for the
sequence of labels crossed during the program's execution.  The central question
we are interested in is whether there is a way of labelling the source
programs so that the sequence $\Lambda$ is a sound and possibly precise
representation of the execution cost of the program.

To answer this question, we observe that the object code 
is some kind of RTL code and that its control flow can be
easily represented as a control flow graph. The fact that we have to
prove the soundness of the compilation function means that we have
plenty of information on the way the control flows in the compiled
code, in particular as far as procedure calls and returns are
concerned.  These pieces of information allow to build a rather
accurate representation of the control flow of the compiled code at
run time.

The idea is then to perform some simple checks on the control flow
graph. The main check consists in verifying that all `loops' go
through a labelled node.  If this is the case then we can associate a
`cost' with every label which over-approximates the actual cost of
running a sequence of instructions.  An optional check amounts to
verify that all paths starting from a label have the same abstract
cost. If this check is successful then we can conclude that the cost
annotations are `precise' in an abstract sense (and possibly concrete too,
depending on the processor considered).

In our previous work~\cite{AAR09,AAR12}, we have showed that it is
possible to produce a sound and precise (in an abstract sense)
labelling for a large class of $\C$ programs with respect to a
moderately optimising compiler.  In the following we show that a
similar result can be obtained for a higher-order functional language
with respect to the standard compilation chain described above.
Specifically we show that there is a simple labelling of the source
program that guarantees that the labelling of the generated object
code is sound and precise.  The labelling of the source program can be
informally described as follows: it associates a distinct label with
every abstraction and with every application which is not `immediately
surrounded' by an abstraction.

In this paper our analysis will stop at the level of an abstract RTL
language, however our previously quoted work~\cite{AAR09,AAR12} shows
that the approach extends to the back-end of a typical moderately
optimising compiler including, {\em e.g.}, dead-code elimination and
register allocation.  Concerning the source language, preliminary
experiments suggest that the approach scales to a larger functional
language such as the one considered in Chlipala's $\coq$
development~\cite{Chlipala10} including fixpoints, sums, exceptions,
and side effects.  Let us also mention that our approach has been
implemented for a simpler compilation chain that bypasses the CPS
translation.  In this case, the function calls are not necessarily
tail-recursive and the compiler generates a $\Cminor$ program which,
roughly speaking, is a type-free, stack aware fragment of $\C$ defined
in the {\sc Compcert} project~\cite{Leroy09}.

\subsection{Reasoning about the certified cost annotations}
If the check described above succeeds every label has a cost which in
general can be taken as an element of a `cost' monoid. Then an {\em
  instrumentation} of the source labelled language is a monadic
transformation $\cl{I}$ (left upper part of Table~\ref{chain}) in the
sense of Gurr's PhD thesis~\cite{Gurr91} that replaces labels with the
associated elements of the cost monoid.  Following this monadic
transformation we are back into the source language (possibly enriched
with a `cost monoid' such as integers with addition).  As a result,
the source program is instrumented so as to monitor its execution cost
with respect to the associated object code. In the end, general logics
developed to reason about functional programs such as the higher-order
Hoare logic co-developed by one of the authors~\cite{RP08} can be
employed to reason about the concrete complexity of source programs by
proving properties on their instrumented versions (see
Table~\ref{code:concat} for an example of a source program with
complexity assertions).

\subsection{Accounting for the cost of memory management}
In a realistic implementation of a functional programming language,
the runtime environment usually includes a garbage collector.  In
spite of considerable progress in {\em real-time garbage
  collectors} (see, {\em e.g.}, the work of Bacon {\em et al.} \cite{BCR03}), 
it seems to us that
this approach does not offer yet a viable path to a certified and
usable WCET prediction of the running time of functional programs.
Instead, the approach we shall adopt, following the seminal work
of  Tofte {\em et al.} \cite{TT97}, is to enrich the last calculus of
the compilation chain described in Table~\ref{chain}, (1) with a notion
of {\em memory region}, (2) with operations to allocate and dispose
memory regions, and (3) with a type and effect system that guarantees
the safety of the dispose operation. This allows to further extend to
the right with one more commuting square the compilation chain
mentioned above and then to include the cost of safe memory management
in our analysis. Actually, because effects are intertwined with types,
what we shall actually do, following the work of Morrisett {\em et al.}
\cite{M99}, is to extend a {\em typed} version of the compilation
chain.

\subsection{Related work}
There is a long tradition starting from the work of Wegbreit \cite{W75}
which reduces the complexity analysis of {\em first-order} functional
programs to the solution of finite difference equations.
Much less is known about higher-order functional programs.
Most previous work on building cost annotations for
higher-order functional programs we are aware of does not take
formally into account the compilation process.  For instance, in an
early work D. Sands~\cite{Sands90} proposes an instrumentation of
call-by-value $\lambda$-calculus in order to describe its execution
cost.  However the notion of cost adopted is essentially the number of
function calls in the source code. In a standard implementation such
as the one considered in this work, different function calls may have
different costs and moreover there are `hidden' function calls which
are not immediately apparent in the source code.  

A more recent work by Bonenfant {\em et al.}~\cite{BFHH06} addresses
the problem of determining the worst case execution time of a
specialised functional language called {\sf Hume}.  The compilation
chain considered consists in first compiling {\sf Hume} to the code of
an intermediate abstract machine, then to $\C$, and finally to
generate the assembly code of the {\sf Resenas M32C/85} processor
using standard $\C$ compilers.  Then for each instruction of the
abstract machine, one computes an upper bound on the worst-case
execution time (WCET) of the instruction relying on a well-known {\sf
  aiT} tool~\cite{AbsInt} that uses abstract interpretation to
determine the WCET of sequences of binary instructions.

While we share common motivations with this work, we differ
significantly in the technical approach.  First, the {\sf Hume}
approach follows a tradition of compiling functional programs to the
instructions of an abstract machine which is then implemented in a
$\C$ like language. In contrast, we have considered a compilation
chain that brings a functional program directly to RTL form. Then the
back-end of a $\C$ like compiler is used to generate binary
instructions.  Second, the cited work \cite{BFHH06} does not address
at all the proof of correctness of the cost annotations; this is left
for future work.  Third, the problem of producing synthetic cost
statements starting from the cost estimations of the abstract
instructions of the {\sf Hume} machine is not considered.  Fourth, the cost
of dynamic memory management, which is crucial in higher-order
functional programs, is not addressed at all.  Fifth, the granularity
of the cost annotations is fixed in Hume~\cite{BFHH06} (the
instructions of the {\sf Hume} abstract machine) while it can vary in
our approach.

We also share with the {\sf Hume} approach one limitation. The precision of our
analyses depends on the possibility of having accurate predictions of
the execution time of relatively short sequences of binary code on a
given processor. Unfortunately, as of today, user interfaces for WCET
systems such as the {\sf aiT} tool mentioned above or 
{\sf Chronos}~\cite{Chronos} do not support modular
reasoning on execution times and therefore experimental work focuses
on processors with simple and predictable architectures.  In a related
direction, another potential loss of precision comes from the introduction of
{\em aggressive} optimisations in the back-end of the compiler such as
loop transformations.
An ongoing work by Tranquilli \cite{T12} addresses this issue
by introducing a refinement of the labelling approach.

\subsection{Paper organisation}
In the following, section \ref{compilation-sec} describes the
certification of the cost-annotations, 
section \ref{instrument-sec} a method to reason about the cost annotations,
section \ref{typing-sec} the typing of the compilation
chain, and section \ref{memory-sec} an extension of the compilation chain to account
for safe memory deallocation.
Proofs are available in the  appendix \ref{proofs-sec}.


\section{The compilation chain: commutation and simulation}\label{compilation-sec}
We describe the intermediate languages and the compilation functions from
an ordinary $\lambda$-calculus to a hoisted, value named $\lambda$-calculus.
For each step we check that: (i) the compilation function commutes with the function
that erases labels and (ii) the object code simulates the source code.

\subsection{Conventions}
The reader is assumed to be acquainted with the type-free 
and typed $\lambda$-calculus,
its evaluation strategies, and its continuation passing style
translations~\cite{W80}.  In the following calculi, all terms are manipulated up
to $\alpha$-renaming of bound names. We denote with~$\equiv$ syntactic
identity up to $\alpha$-renaming.  Whenever
a reduction rule is applied, it is assumed that terms have been
renamed so that all binders use distinct variables and these variables are distinct from
the free ones.  With this assumption, we can omit obvious side conditions 
on binders and free variables.
Similar conventions are applied when reasoning about a substitution,
say $[T/x]T'$, of a term $T$ for a variable~$x$ in a term $T'$.
We denote with $\fv{T}$ the set of variables occurring free in a term~$T$.

Let $C,C_1,C_2,\ldots$ be one hole contexts and $T$ a term.
Then $C[T]$ is the term resulting from the replacement in the
context $C$ of the hole by the term $T$ and $C_1[C_2]$ is 
the one hole context resulting from the replacement in 
the context $C_1$ of the hole by the context $C_2$.

For each calculus, we assume a syntactic category $\w{id}$ of
identifiers with generic elements $x,y,\ldots$ and a syntactic
category $\ell$ of labels with generic elements $\ell,\ell_1,\ldots$
For each calculus, we specify the syntactic categories and the
reduction rules.  For the sake of clarity, the meta-variables of these
syntactic categories are sometimes shared between several calculus:
the context is always sufficiently precise to determine to which
syntax definitions we refer. We let $\alpha$ range over labels and the
empty word~$\epsilon$. We write $M\act{\alpha} N$ if $M$ rewrites to $N$ with a
transition labelled by $\alpha$. We abbreviate $M\act{\epsilon} N$
with $M\arrow N$.  We write~$\trarrow$ for the reflexive and
transitive closure of~$\arrow$.  We also define $M\wact{\alpha} N$ as
$M \trarrow N$ if $\alpha=\epsilon$ and as $M \trarrow \act{\alpha}
\trarrow N$ otherwise.

Given a term $M$ in one of the labelled languages we write $M\eval_{\Lambda} N$
if $M\act{\alpha_1} \cdots \act{\alpha_n} N$, $\Lambda=\alpha_1\cdots \alpha_n$, and
$N$ cannot reduce (in general this does not imply that $N$ is a value). 
We write $M\eval_\Lambda$ for $\xst{N}{M\eval_\Lambda N}$.
Also, if the term $M$ is unlabelled, $\Lambda$ is always the empty word and
we abbreviate $M\eval_{\epsilon} N$ with $M\eval N$.

We shall write $X^+$ (resp. $X^*$) for a non-empty (possibly empty)
finite sequence $X_1,\ldots,X_n$ of symbols.
By extension, $\lambda x^+.M$ stands for $\lambda x_1\ldots x_n.M$,
$[V^+/x^+]M$ stands for  \\\noindent$[V_1/x_1,\ldots,V_n/x_n]M$, and
$\s{let} \ (x=V)^+ \ \s{in} \ M$ stands for 
$\lets{x_{1}}{V_{1}}{\cdots \lets{x_{n}}{V_{n}}{M}}$.

\subsection{The source language}
Table~\ref{lambda} introduces a type-free, left-to-right call-by-value
$\lambda$-calculus.  The calculus includes {\em let-definitions} and
{\em polyadic abstraction} and {\em tupling} with the related
application and projection operators.  Any term $M$ can be {\em
  pre-labelled} by writing $\lb{\ell}{M}$ or {\em post-labelled} by
writing $\plb{\ell}{M}$.  In the pre-labelling, the label $\ell$ is
emitted immediately while in the post-labelling it is emitted after
$M$ has reduced to a value. It is tempting to reduce the
post-labelling to the pre-labelling by writing $\plb{\ell}{M}$ as
$@(\lambda x.\lb{\ell}{x},M)$, however the second notation introduces
an additional abstraction and a related reduction step which is not
actually present in the original code. Roughly speaking, every
$\lambda$-abstraction is a potential starting point for a loop in the
control-flow graph. Thus, we will need the body of every
$\lambda$-abstraction to be pre-labelled so as to maintain the
invariant that all loops go through a labelled node in the
control-flow graph. As the CPS translation introduces new
$\lambda$-abstractions that are not present in the source code but
correspond to the image of some applications, we will also need to
post-label these particular applications so that the freshly introduced
$\lambda$-abstraction can be assigned a label. 

Table~\ref{lambda} also introduces an {\em erasure} function $\w{er}$
from the $\lambda^\ell$-calculus to the $\lambda$-calculus. This
function simply traverses the term and erases all pre and post
labellings. Similar definitions arise in the following calculi of the
compilation chain and are omitted.

\begin{table}[t]
{\footnotesize
\begin{center}
{\sc Syntax}
\end{center}
\[
\begin{array}{lll}

V &::= \w{id} \Alt \lambda \w{id}^+.M \Alt (V^*)                                 &\mbox{(values)} \\
M &::= V \Alt @(M,M^+) \Alt \lets{\w{id}}{M}{M} \Alt (M^*) \Alt \prj{i}{M}\Alt       
  \lb{\ell}{M} \Alt \plb{\ell}{M} \ &\mbox{(terms)} \\
E &::= [~] \Alt @(V^*,E,M^*) \Alt \lets{\w{id}}{E}{M} \Alt (V^*,E,M^*) \Alt \prj{i}{E} \Alt  \plb{\ell}{E}
&\mbox{(eval. cxts.)}

\end{array}
\]
\begin{center}
{\sc Reduction Rules}
\end{center}
\[
\begin{array}{lll}

E[@(\lambda x_1\ldots x_n.M,V_1,\ldots, V_n)] &\arrow &E[[V_1/x_1,\ldots,V_n/x_n]M] \\
E[\lets{x}{V}{M}] &\arrow &E[[V/x]M] \\
E[\prj{i}{V_1,\ldots,V_n}] &\arrow &E[V_i] \qquad 1\leq i \leq n \\
E[\lb{\ell}{M}]            &\act{\ell} &E[M] \\
E[\plb{\ell}{V}]           &\act{\ell} &E[V]

\end{array}
\]
\begin{center}
{\sc Label erasure (selected equations)}
\end{center}
\[
\er{\lb{\ell}{M}} = \er{\plb{\ell}{M}} = \er{M}~ 
\]}
\caption{An ordinary call-by-value $\lambda$-calculus: $\lambda^{\ell}$}\label{lambda}
\end{table}

\subsection{Compilation to CPS form}
Table~\ref{lambda-cps} introduces a fragment of the
$\lambda^\ell$-calculus described in Table~\ref{lambda} and a related
CPS translation.  To avoid all ambiguity, let us assume that
$(V_1,\ldots,V_n)\sco K$ is translated according to the case for
values, but note that if we follow the general case for tuples we
obtain the same result.  We recall that in a CPS translation each
function takes its evaluation context as a fresh additional parameter
(see, {\em e.g.}, the work of Wand \cite{W80}, for an elaboration of
this idea). The results of the evaluation of subterms (of tuples and
of applications) are also named using fresh parameters~$x_0, \ldots,
x_n$. The initial evaluation context is defined relatively to a fresh
variable named '$\w{halt}$'. Then the evaluation context is always
trivial.  Notice that the reduction rules are essentially those of the
$\lambda^{\ell}$-calculus modulo the fact that we drop the rule to
reduce $\plb{\ell}{V}$ since post-labelling does not occur in CPS
terms and the fact that we optimize the rule for the projection to
guarantee that CPS terms are closed under reduction. For instance, the
term $\lets{x}{\prj{1}{V_1,V_2}}{M}$ reduces directly to $[V_1/x]M$
rather than going through the intermediate term $\lets{x}{V_1}{M}$
which does not belong to the CPS terms.

We study next the properties enjoyed by the CPS translation.  In
general, the commutation of the compilation function with the erasure
function only holds up to call-by-value $\eta$-conversion, namely
$\lambda x.@(V,x)=_\eta V$ if $x\notin \fv{V}$.  This is due to the fact
that post-labelling introduces an $\eta$-expansion of the continuation
if and only if the continuation is a variable. To cope with this
problem, we introduce next the notion of {\em well-labelled} term. We will
see later (section \ref{initial-lab}) 
that terms generated by the initial labelling are well-labelled.

\begin{definition}[well-labelling]
We define two predicates $W_i$, $i=0,1$ on the terms of the
 $\lambda^\ell$-calculus as the least sets such that
$W_1$ is contained in $W_0$ and the following conditions hold:

{\footnotesize\[
\begin{array}{c}

\infer{}{x\in W_1}
\qquad
\infer{M\in W_0}{\plb{\ell}{M}\in W_0}
\qquad
\infer{M\in W_1}{\lambda x^+.M\in W_1} \\ \\
\qquad
\infer{M\in W_i\quad i\in \set{0,1}}{\lb{\ell}{M}\in W_i} 
\qquad 
\infer{N\in W_0,M\in W_i\quad i\in \set{0,1}}{\lets{x}{N}{M}\in W_i}
\\ \\
\infer{M_i\in W_0\quad i=1,\ldots,n}{@(M_1,\ldots,M_n)\in W_1}
\qquad
\infer{M_i\in W_0\quad i=1,\ldots,n}{(M_1,\ldots,M_n)\in W_1}
\qquad
\infer{M\in W_0}{\prj{i}{M} \in W_1}~.

\end{array}
\]}
\end{definition}

The intuition is that we want to avoid the situation where a
post-labelling receives as continuation the continuation variable
generated by the translation of a $\lambda$-abstraction. To that end,
we make sure that post-labelling is only applied to terms $M \in W_0$,
that is, terms that are not the immediate body of a
$\lambda$-abstraction (which are in $W_1$).

\begin{example}[labelling and commutation]
Let $M\equiv \lambda x.(\plb{\ell}{@(x,x)})$. Then
$M\notin W_0$ because 
the rule for abstraction requires $\plb{\ell}{@(x,x)}\in W_1$ 
while we can only show $\plb{\ell}{@(x,x)}\in W_0$.
Notice that we have:
\[
\begin{array}{lll}
\er{\cps{M}} &\equiv &@(\w{halt},\lambda x,k.@(x,x,\lambda x.@(k,x))) \\
\cps{\er{M}} &\equiv &@(\w{halt},\lambda x,k.@(x,x,k))~.
\end{array}
\]
So, for $M$, the commutation of the CPS translation and the erasure
function only holds up to~$\eta$.
\end{example}

\begin{proposition}[CPS commutation]\label{cps-commutation}
Let $M\in W_0$ be a term of the $\lambda^\ell$-calculus (Table~\ref{lambda}).
Then: $\er{\cps{M}} \equiv \cps{\er{M}}$. 
\end{proposition}

The proof of the CPS simulation is non-trivial but rather standard
since Plotkin's seminal work \cite{Plotkin75}.  The general idea is
that the CPS translation pre-computes many `administrative' reductions
so that the translation of a term, say $E[@(\lambda x.M,V)]$ is a term
of the shape $@(\psi(\lambda x.M),\psi(V),K_E)$ for a suitable
continuation $K_E$ depending on the evaluation context~$E$.

\begin{proposition}[CPS simulation]\label{cps-simulation}
Let $M$ be a term of the $\lambda^\ell$-calculus. 
If $M \act{\alpha} N$ then  $\cps{M} \wact{\alpha} \cps{N}$.
\end{proposition}

We illustrate this result on the following example.

\begin{example}[CPS]\label{cps-example}
Let $M\equiv @(\lambda x.@(x,@(x,x)),I)$, where $I\equiv \lambda x.x$. Then
\[
\cps{M}\equiv @(\lambda x,k.@(x,x,\lambda y.@(x,y,k)), I', H)
\]
where: $I'\equiv \lambda x,k.@(k,x)$ and $H\equiv \lambda x.@(\w{halt},x)$.
The term $M$ is simulated by $\cps{M}$ as follows:
\[
\begin{array}{ccccccc}

M       &\arrow   &@(I,@(I,I))                  &\arrow   &@(I,I)     &\arrow   &I \\
\cps{M} &\arrow   &@(I',I',\lambda y.@(I',y,H)) &\arrow^+ &@(I',I',H) &\arrow^+ &@(\w{halt},I')~.

\end{array}
\]
\end{example}

\begin{table}[t]
{\footnotesize
\begin{center}
{\sc Syntax CPS terms}
\end{center}
\[
\begin{array}{lll}

V &::= \w{id} \Alt \lambda \w{id}^+.M \Alt (V^*)                                 &\mbox{(values)} \\
M &::= @(V,V^+) \Alt \lets{\w{id}}{\prj{i}{V}}{M} \Alt  \lb{\ell}{M}            &\mbox{(CPS terms)} \\ 
K &::=  \w{id} \Alt \lambda \w{id}.M                                            &\mbox{(continuations)} \\

\end{array}
\]
\begin{center}
{\sc Reduction rules}
\end{center}
\[
\begin{array}{lll}

@(\lambda x_1\ldots x_n.M,V_1,\ldots,V_n) &\arrow &[V_1/x_1,\ldots,V_n/x_n]M \\
\lets{x}{\prj{i}{V_1,\ldots,V_n}}{M}       &\arrow &[V_i/x]M \quad 1\leq i \leq n \\
\lb{\ell}{M}                               &\act{\ell} &M

\end{array}
\]
\begin{center}
{\sc CPS translation}
\end{center}
\[
\begin{array}{lll}

\psi(x) &= & x \\

\psi(\lambda x^+.M) &= &\lambda x^+,k.(M\sco k) \\

\psi((V_1,\ldots,V_n))         &= &(\psi(V_1),\ldots,\psi(V_n)) \\ \\ 

V\sco k    &= &@(k,\psi(V)) \\

V\sco (\lambda x.M) &= &[\psi(V)/x]M  \\

@(M_0,\ldots,M_n)\sco K &= &M_0 \sco \lambda x_0.\ldots (M_n \sco \lambda x_n.@(x_0,\ldots,x_n,K)) \\

\lets{x}{M_1}{M_2}\sco K &= &M_1\sco \lambda x.(M_2\sco K) \\ 

(M_1,\ldots,M_n)\sco K &= &M_1\sco \lambda x_1.\ldots (M_n\sco \lambda x_n.(x_1,\ldots,x_n)\sco K \ ) \\

\prj{i}{M}\sco K      &= &M\sco \lambda x.\lets{y}{\prj{i}{x}}{y\sco K} \\

(\lb{\ell}{M})\sco K    &= &\lb{\ell}{(M\sco K)} \\

(\plb{\ell}{M})\sco K   &= &M\sco (\lambda x.\lb{\ell}{(x\sco K)}) \\ \\ 

\cps{M}           &= &M\sco \lambda x.@(\w{halt},x),\qquad\w{halt} \mbox{ fresh variable}

\end{array}
\]}
\caption{CPS $\lambda$-calculus ($\lambda^{\ell}_{{\it cps}}$) and CPS translation}\label{lambda-cps}
\end{table}

\subsection{Transformation in value named CPS form}
Table~\ref{lambda-adm} introduces a {\em value named} $\lambda$-calculus
in CPS form: $\lambda^{\ell}_{\w{cps},\w{vn}}$. 
In the ordinary $\lambda$-calculus, the application of a $\lambda$-abstraction
to an argument (which is a value) may duplicate the argument as in:
$@(\lambda x.M,V) \arrow [V/x]M$.
In the value named  $\lambda$-calculus, all values are named 
and when we apply the name of a $\lambda$-abstraction to the name of a value
we create a new copy of the body of the function and replace its formal parameter name
with the name of the argument as in:
\[
\lets{y}{V}{\lets{f}{\lambda x.M}{@(f,y)}} \ \arrow  \ 
\lets{y}{V}{\lets{f}{\lambda x.M}{[y/x]M}}~.
\]
We also remark that in the value named $\lambda$-calculus the evaluation contexts are 
a sequence of let definitions associating values to names.
Thus, apart for the fact that the values are not necessarily closed,
the evaluation contexts are similar to the environments of 
abstract machines for functional languages (see, {\em e.g.}, \cite{Curien91}).


\begin{table}[t]
{\footnotesize
\begin{center}
{\sc Syntax}
\end{center}
\[
\begin{array}{lll}

V &::= \lambda \w{id}^+.M \Alt (\w{id}^*)                                                    &\mbox{(values)} \\
C &::= V \Alt \prj{i}{\w{id}}                                                               &\mbox{(let-bindable terms)} \\
M &::= @(\w{id},\w{id}^+) \Alt \lets{\w{id}}{C}{M} \Alt \lb{\ell}{M}
  &\mbox{(CPS terms)} \\
E &::= [~] \Alt \lets{\w{id}}{V}{E}                                                          &\mbox{(evaluation contexts)} \\

\end{array}
\]
\begin{center}
{\sc Reduction Rules}
\end{center}
\[
\begin{array}{cccl}

E[@(x,z_1,\ldots,z_n)]
&\arrow
&E[[z_1/y_1,\ldots,z_n/y_n]M]
&\mbox{if }E(x)=\lambda y_1 \ldots y_n.M \\

E[\lets{z}{\prj{i}x}{M}]
&\arrow
& E[[y_i/z]M]]
&\mbox{if }E(x)=(y_1,\ldots,y_n),  1\leq i \leq n \\

E[\lb{\ell}{M}]
&\act{\ell}
&E[M]

\end{array}
\]
\[
\mbox{where: }E(x)=\left\{
\begin{array}{ll}
V  &\mbox{if }E=E'[\lets{x}{V}{[~]}] \\
E'(x) &\mbox{if }E=E'[\lets{y}{V}{[~]}],x\neq y \\
\mbox{undefined} &\mbox{otherwise}
\end{array}\right.
\]
}
\caption{A value named CPS $\lambda$-calculus: $\lambda^{\ell}_{{\it cps,vn}}$}\label{lambda-adm}
\end{table}

Table~\ref{back-forth} defines the compilation into value named
form along with a readback translation. (Only the case for the local
binding of values is interesting.) The latter is useful to state the
simulation property.  Indeed, it is not true that if $M\arrow M'$ in
$\lambda^{\ell}_{cps}$ then $\vn{M} \trarrow \vn{M'}$ in
$\lambda^{\ell}_{\w{cps},\w{vn}}$. For instance, consider $M\equiv (\lambda
x.xx)I$ where $I\equiv(\lambda y.y)$.  Then $M\arrow II$ but $\vn{M}$
does not reduce to $\vn{II}$ but rather to a term where the `sharing'
of the duplicated value $I$ is explicitly represented.

\begin{example}[value named form]\label{admin-example}
Suppose 
$$N\equiv @(\lambda x,k.@(x,x,\lambda y.@(x,y,k)),I' ,H))$$
where: $I'\equiv \lambda x,k.@(k,x)$ and $H\equiv \lambda x.@(\w{halt},x)$.
(This is the term resulting from the CPS translation in example
\ref{cps-example}.)  The corresponding term in value named form is:
\[
\begin{array}{l}
\s{let} \ z_1 = \lambda x,k. (\lets{z_{11}}{\lambda y.@(x,y,k)}{@(x,x,z_{11})}) \ \s{in} \\
\s{let} \ z_2 = I' \ \s{in} \\
\s{let} \ z_3 = H \ \s{in} \\
@(z_1,z_2,z_3)~.
\end{array}
\]
\end{example}

\begin{table}[t]
{\footnotesize
\begin{center}
{\sc Transformation in value named form (from $\lambda^\ell_{cps}$ to $\lambda^{\ell}_{cps,vn}$)}
\end{center}
\[
\begin{array}{lll}

\vn{@(x_0,\ldots,x_n)} &= &@(x_0,\ldots,x_n) \\

\vn{@(x^*,V,V^*)}      &= &\evn{V}{y}[\vn{@(x^*,y,V^*)}] \quad V\neq \w{id}, y \mbox{ fresh}\\

\vn{\lets{x}{\prj{i}{y}}{M}}  &= &\lets{x}{\prj{i}{y}}{\vn{M}} \\

\vn{\lets{x}{\prj{i}{V}}{M}}  &= 
&\evn{V}{y}[\lets{x}{\prj{i}{y}}{\vn{M}}] \quad V\neq \w{id}, y \mbox{ fresh}\\

\vn{\lb{\ell}{M}}             &= &\lb{\ell}{\vn{M}}  \\ \\

\evn{\lambda x^+.M}{y}            &= &\lets{y}{\lambda x^+.\vn{M}}{[~]} \\

\evn{(x^*)}{y}                   &= &\lets{y}{(x^*)}{[~]} \\ 

\evn{(x^*,V,V^*)}{y}              &= &\evn{V}{z}[\evn{(x^*,z,V^*)}{y}] \quad V\neq \w{id}, z \mbox{ fresh}

\end{array}
\]
\begin{center}
{\sc Readback translation (from $\lambda^{\ell}_{\w{cps},\w{vn}}$ to $\lambda^{\ell}_{\w{cps}}$)} \\
\end{center}
\[
\begin{array}{lll}

\rb{\lambda x^+.M}  &= &\lambda x^+.\rb{M} \\
\rb{x^*}          &= &(x^*) \\

\rb{@(x,x_1,\ldots,x_n)}       &= &@(x,x_1,\ldots,x_n) \\ 

\rb{\lets{x}{\prj{i}{y}}{M}}  &= &\lets{x}{\prj{i}{y}}{\rb{M}} \\

\rb{\lets{x}{V}{M}} &=       &[\rb{V}/x]\rb{M} \\

\rb{\lb{\ell}{M}}  &= &\lb{\ell}{\rb{M}}

\end{array}
\]}
\caption{Transformations in value named CPS form and readback}\label{back-forth}
\end{table}




\begin{proposition}[VN commutation]\label{ad-commutation}
Let $M$ be a  $\lambda$-term in CPS form. Then:

\Defitem{(1)} $\rb{\vn{M}}\equiv M$.

\Defitem{(2)} $\er{\vn{M}} \equiv \vn{\er{M}}$.

\end{proposition}

\begin{proposition}[VN simulation]\label{ad-simulation}
Let $N$ be a $\lambda$-term in CPS value named form.  If
$\rb{N}\equiv M$ and $M\act{\alpha} M'$ then there exists $N'$ such
that $N\act{\alpha} N'$ and $\rb{N'}\equiv M'$.
\end{proposition}

\subsection{Closure conversion}
The next step is called {\em closure conversion}. It consists in
providing each functional value with an additional parameter that
accounts for the names free in the body of the function and in
representing functions using closures. Our closure conversion
implements a closure using a pair whose first component is the code of
the translated function and whose second component is a tuple of the
values of the free variables. 

It will be convenient to write ``$\lets{(y_1,\ldots,y_n)}{x}{M}$'' for
``$\lets{y_1}{\prj{1}{x}}{\cdots\lets{y_n}{\prj{n}{x}}{M}}$'' and
``$\s{let}\ x_1=C_1 \ldots x_n=C_n\ \s{in}\ M$'' for ``$\s{let} \ x_1=C_1
\ \s{in} \ldots \s{let} \ x_n=C_n \ \s{in} \ M$''.  The transformation
is described in Table~\ref{lambda-cc}.  The output of the
transformation is such that all functional values are closed.  In our
opinion, this is the only compilation step where the proofs are rather
straightforward.

\begin{example}[closure conversion]\label{cc-example}
Let $M\equiv \vn{\cps{\lambda x.y}}$, namely
\[
M\equiv \lets{z_1}{\lambda x,k.@(k,y)}{@(\w{halt},z_1)}~.
\]
Then $\cc{M}$ is the following term: 
\[
\begin{array}{l}
\s{let} \ c = \lambda e,x,k.(\s{let} \ (y)=e, (c,e)=k \ \s{in} \ @(c,e,y)) \ \s{in}\\
\s{let} \ e = (y), z_1=(c,e), (c,e) = \w{halt} \ \s{in} \\
@(c,e,z_1)~.

\end{array}
\]
\end{example}

\begin{table}[t]
{\footnotesize
\begin{center}
{\sc Syntactic restrictions on $\lambda^\ell_{cps,\w{vn}}$ after closure conversion} \\
All functional values are closed. ~\\~ \\ 
{\sc Closure Conversion}
\end{center}
\[
\begin{array}{lll}

\cc{@(x,y^+)}                   &=  
\lets{(c,e)}{x}{@(c,e,y^+)} \\ \\

\cc{\lets{x}{C}{M}} &= 
\begin{array}{l}
\s{let} \ c=\lambda e,x^+.\s{let} \ (z_1,\ldots,z_k)=e \ \s{in} \ \cc{N}  \ \s{in} \\
\s{let} \ e=(z_1,\ldots,z_k) \ \s{in} \\
\s{let} \ x = (c,e) \ \s{in} \\
\cc{M} \qquad \qquad (\mbox{if }C=\lambda x^+.N,\fv{C}=\set{z_1,\ldots,z_k})
\end{array} 
 \\ \\

\cc{\lets{x}{C}{M}} &= \lets{x}{C}{\cc{M}} 
\qquad \qquad \quad  (\mbox{if }C \mbox{ not a function})
\\ \\

\cc{\lb{\ell}{M}}&=
\lb{\ell}{\cc{M}} 

\end{array}
\]
}
\caption{Closure conversion  on value named CPS terms}\label{lambda-cc}
\end{table}

\begin{proposition}[CC commutation]\label{cc-commutation}
Let $M$ be a CPS term in value named form. Then
$\er{\cc{M}} \equiv \cc{\er{M}}$.
\end{proposition}

\begin{proposition}[CC simulation]\label{cc-simulation}
Let $M$ be a CPS term in value named form. 
If $M\act{\alpha} M'$ then $\cc{M} \wact{\alpha} \cc{M'}$.
\end{proposition}

\subsection{Hoisting}
The last compilation step consists in moving all functions definitions at top level.
In Table~\ref{lambda-hoist}, we formalise this compilation step as the iteration of a 
set of program transformations that commute with the erasure
function and the reduction relation. Denote with $\lambda z^+.T$ 
a function that does {\em not} contain function definitions.
The transformations consist in hoisting (moving up) the definition 
of a function   $\lambda z^+.T$ with
respect to either a definition of a pair or a projection, or
another including function, or a labelling. 
Note that the hoisting transformations
do not preserve the property that all functions are closed. Therefore the hoisting
transformations are defined on the terms of the $\lambda^{\ell}_{cps,\w{vn}}$-calculus.
As a first step, we analyse the hoisting transformations.

\begin{table}[t]
{\footnotesize
\begin{center}
{\sc Syntax for $\lambda^\ell_h$} \\
{Syntactic restrictions on $\lambda^\ell_{cps,\w{vn}}$ after hoisting} \\
All function definitions are at top level.
\end{center}
\[
\begin{array}{lll}

C &::= (\w{id}^*) \Alt \prj{i}{\w{id}}                                                       &\mbox{(restricted let-bindable terms)} \\
T &::= @(\w{id},\w{id}^+) \Alt \lets{\w{id}}{C}{T} \Alt \lb{\ell}{T}                                          &\mbox{(restricted terms)} \\
P &::= T \Alt  \lets{\w{id}}{\lambda \w{id}^+.T}{P}                                   &\mbox{(programs)} 
\end{array}
\]
\begin{center}
{\sc Specification of the hoisting transformation}
\end{center}
\[
\h{M}=N\mbox{ if }M\leadsto \cdots \leadsto N \not\leadsto, \quad \mbox{where:}
\]
\[
\begin{array}{lll}

D &::= & [~] \Alt \lets{\w{id}}{C}{D} \Alt \lets{\w{id}}{\lambda \w{id}^+.D}{M} \Alt \lb{\ell}{D}
\qquad \mbox{(hoisting contexts)}
\end{array}
\]
\[
\begin{array}{lll}
\\
(h_1)
&D[\lets{x}{C}{\lets{y}{\lambda z^+.T}{M}}] \leadsto \\
&D[\lets{y}{\lambda z^+.T}{\lets{x}{C}{M}}] &\mbox{if }x\notin \fv{\lambda z^+.T} \\\\

(h_2)
&D[\lets{x}{(\lambda w^+.\lets{y}{\lambda z^+.T}{M})}{N}] \leadsto \\
&D[\lets{y}{\lambda z^+.T}{\lets{x}{\lambda w^+.M}{N}}] &\mbox{if }\set{w^+}\inter \fv{\lambda z^+.T} = \emptyset  \\\\

(h_3)
&D[\lb{\ell}{\lets{y}{\lambda z^+.T}{M}}] 
\leadsto\\
&D[\lets{y}{\lambda z^+.T}{\lb{\ell}{M}}] 

\end{array}
\]
}
\caption{Hoisting transformation}\label{lambda-hoist}
\end{table}

\begin{proposition}[on hoisting transformations]\label{h-transform}
The iteration of the hoisting transformation on a term in $\lambda_{cc,\w{vn}}^{\ell}$ 
(all function are closed) 
terminates and produces a term satisfying the syntactic restrictions 
specified in Table~\ref{lambda-hoist}.
\end{proposition}

Next we check that the hoisting transformations commute with the erasure function.

\begin{proposition}[hoisting commutation]\label{h-commutation}
Let $M$ be a term of the $\lambda^\ell_{cps,\w{vn}}$-calculus.

\Defitem{(1)} If $M\leadsto N$ then $\er{M} \leadsto \er{N}$ or $\er{M}\equiv \er{N}$.

\Defitem{(2)} If $M\not\leadsto \cdot$ then $\er{M}\not\leadsto \cdot$.

\Defitem{(3)} $\er{\h{M}} \equiv \h{\er{M}}$.
\end{proposition}

The proof of the simulation property requires some work because to
close the diagram we need to collapse repeated definitions, which may
occur, as illustrated in the example below.

\begin{example}[hoisting transformations and transitions]\label{hoisting-comm-ex}
Let  $$M\equiv \lets{x_{1}}{\lambda y_1.N}{@(x_1,z)}$$
where $N\equiv \lets{x_{2}}{\lambda y_{2}.T_2}{T_1}$ 
and $y_1 \notin \fv{\lambda y_2.T_2}$.
Then we either reduce and then hoist:
\[
\begin{array}{ll}
M  &\arrow \lets{x_{1}}{\lambda y_1.N}{[z/y_1]N}  \\
   &\equiv \lets{x_{1}}{\lambda y_1.N}{\lets{x_{2}}{\lambda y_{2}.T_2}}{[z/y_1]T_1} \\
   &\leadsto \lets{x_{2}}{\lambda y_2.T_2}{\lets{x_{1}}{\lambda y_1.T_1}{\lets{x_{2}}{\lambda y_{2}.T_2}}{[z/y_1]T_1}} \not\leadsto
\end{array}
\]
or hoist and then reduce:
\[
\begin{array}{ll}
M  &\leadsto \lets{x_{2}}{\lambda y_2.T_2}{\lets{x_{1}}{\lambda y_1.T_1}{@(x_1,z)}}  \\
   &\arrow   \lets{x_{2}}{\lambda y_2.T_2}{\lets{x_{1}}{\lambda y_1.T_1}{[z/y_1]T_1}} \quad \not\leadsto
\end{array}
\]
In the first case, we end up duplicating the definition of $x_2$. 
\end{example}

We proceed as follows. First we introduce a relation
$S_h$ that collapses repeated definitions and show that it is a simulation.
Second, we show that the hoisting transformations induce a 
`simulation up to $S_h$'.
Namely if  $M\act{\ell} M'$ and $M\leadsto N$ then there is a $N'$
such that $N\act{\ell} N'$ and $M' \ (\leadsto^* \comp S_h) \  N'$. 
Third, we iterate the previous property to derive the following
one.

\begin{proposition}[hoisting simulation]\label{h-simulation}
There is a simulation relation $\cl{T}_h$ on the terms of the 
$\lambda^\ell_{cps,\w{vn}}$-calculus such that for all terms $M$
of the  $\lambda^\ell_{cc,\w{vn}}$-calculus we have $M \ \cl{T}_h \ \h{M}$.
\end{proposition}

\subsection{Composed commutation and simulation properties}
Let $\cl{C}$ be the composition of the compilation steps we
have considered:
\[
\cl{C} = \cl{C}_h \comp \cl{C}_{cc} \comp \cl{C}_{vn} \comp \cl{C}_{cps}~.
\]
We also define a relation $\cl{R}_{C}$ between terms in 
$\lambda^\ell$ and terms in $\lambda_{h}^{\ell}$ as:
\[
M \cl{R}_{C} P \mbox{ if }
\xst{N}{\cps{M} \equiv \rb{N} \mbox{ and }  \cc{N} \ \cl{T}_h \ P }~.
\]
Notice that for all $M$, $M \ \cl{R}_{C} \  \cmp{M}$.

\begin{theorem}[commutation and simulation]\label{com-sim-thm}
Let $M\in W_0$ be a term of the $\lambda^\ell$-calculus. Then:

\Defitem{(1)} $\er{\cmp{M}} \equiv \cmp{\er{M}}$.

\Defitem{(2)} 
If $M \ \cl{R}_{C} \  N$ and $M \act{\alpha} M'$ then
$N \wact{\alpha} N'$ and 
$M'\  \cl{R}_{C} \  N'$.
\end{theorem}

\section{Reasoning about the cost annotations}\label{instrument-sec}
We describe an initial labelling of the source code leading to a sound
and precise labelling of the object code and an instrumentation of the
labelled source program which produces a source program monitoring its
own execution cost. Then, we explain how to obtain static guarantees
on this execution cost by means of a Hoare logic for purely functional
programs.

\subsection{Initial labelling}\label{initial-lab}
We define a labelling function $\cl{L}$ of the source code (terms of
the $\lambda$-calculus) which guarantees that the associated RTL code
satisfies the conditions necessary for associating a cost with each
label.  We set $\cl{L}(M)=\labi{0}{M}$, where the functions
$\cl{L}_{i}$ are specified in Table~\ref{labelling}. When the
index~$i$ in $\cl{L}_{i}$ is equal to~$1$, it attests that~$M$ is an
immediate body of a $\lambda$-abstraction. In that case, even if~$M$
is an application, it is not post-labelled. Otherwise, when $i$ is
equal to~$0$, the term~$M$ is not an immediate body of
a~$\lambda$-abstraction, and, thus is post-labelled if it is an
application.

\begin{example}[labelling application]\label{label-example}
Let $M\equiv \lambda x.@(x,@(x,x))$. Then 
$\cl{L}(M) \equiv \lambda x.\lb{\ell_0}{@(x,\plb{\ell_{1}}{@(x,x)})}$.
Notice that only the inner application is post-labelled.
\end{example}

\begin{table}[t]
{\footnotesize
\[
\begin{array}{lll}

\cl{L}(M)                    &= &\labi{0}{M} \quad \mbox{ where:} \\ \\

\labi{i}{x}                  &= &x \\

\labi{i}{\lambda x^+.M} &= &\lambda x^+.\lb{\ell}{\labi{1}{M}}  \quad \ell \mbox{ fresh} \\

\labi{i}{(M_1,\ldots,M_n)}    &= &(\labi{0}{M_1},\ldots,\labi{0}{M_n}) \\

\labi{i}{\prj{j}{M}}         &= &\prj{j}{\labi{0}{M}} \\

\labi{i}{@(M,N^+)}           &= &
\left\{
\begin{array}{ll}
\plb{\ell}{@(\labi{0}{M},(\labi{0}{N})^+)} &i=0, \ \ell \mbox{ fresh} \\
@(\labi{0}{M},(\labi{0}{N})^+)             &i=1 
\end{array}
\right. \\

\labi{i}{\lets{x}{M}{N}}     &= 
&\lets{x}{\labi{0}{M}}{\labi{i}{N}}  

\end{array}
\]}
\caption{A sound and precise labelling of the source code}\label{labelling}
\end{table}

\begin{proposition}[labelling properties]\label{labelling-prop}
Let $M$ be a term of the $\lambda$-calculus.

\Defitem{(1)}
The function $\cl{L}$  is a labelling and produces well-labelled
terms, namely:
\[
\er{\labi{i}{M}}\equiv M \mbox{ and } \cl{L}_{i}(M)\in W_i \mbox{ for }i=0,1. 
\]
\Defitemf{(2)} We have: $\cmp{M} \equiv \er{\cmp{\lab{M}}}$.

\Defitem{(3)} Labels occur exactly once in the body of each function definition and nowhere else,
namely, $\cl{C}(\cl{L}(M))$ is a term $P$ specified by the 
following grammar:
\[
\begin{array}{ll}
P &::= T \Alt  \lets{\w{id}}{\lambda \w{id}^+.\w{Tlab}}{P}        \\

\w{Tlab}  &::= \lb{\ell}{T} \Alt \lets{\w{id}}{C}{\w{Tlab}} \\

T         &::= @(\w{id},\w{id}^+) \Alt \lets{\w{id}}{C}{T}      \\

C &::= (\w{id}^*) \Alt \prj{i}{\w{id}} \\

\end{array}
\]
\end{proposition}

Point (2) of the proposition above depends on the commutation property
of the compilation function (theorem \ref{com-sim-thm}(1)).
The point (3) entails that a RTL program 
generated by the compilation function is composed of a set of routines
and that each routine is composed of a sequence of assignments 
on pseudo-registers and a terminal call to another routine.  
Points (2) and (3) entail that the only difference between the compiled code
and the compiled {\em labelled} code is that in the latter, upon
entering a routine, a label uniquely associated with the routine is emitted.

Now suppose we can compute the cost of running once each routine, where the cost
is an element of a suitable commutative monoid $\cl{M}$ with 
binary operation $\oplus$ and identity $\zero$ (the reader may just
think of the natural numbers).
Then we can define a function \s{costof} which associates with 
every label the cost of running once the related routine;
the function $\s{costof}$ 
is extended to words of labels in the standard way.
A run of a terminating program  $M$ corresponds to a finite sequence 
of routine calls which in turn correspond
to the finite sequence of labels that we can observe when running
the labelled program. We summarise this argument in the following
proviso (a modelling hypothesis rather than a mathematical proposition).

\begin{proviso}\label{cost-proviso}
For any term $M$ of the source language,
if $\cl{C}(\cl{L}(M)) \eval_\Lambda$ then
$\s{costof}(\Lambda)$ is the cost of running $M$.
\end{proviso}

We stress that the model at the level of  the RTL programs
is not precise enough to obtain useful predictions on the execution
cost in terms, say, of  CPU cycles. 
However,  the compilation chain of this paper can be composed with
the back-end of a moderately optimising $\C$ compiler
described in our previous work~\cite{AAR09,AAR12}.
For RTL programs such as those characterized by the grammar above, 
the back end produces binary code which satisfies the checks for soundness 
and precision that we outlined in the introduction. 
This remains true even if the source language is enriched with
other constructions such as branching and loops as long as the labelling
function is extended to handle these cases.

\subsection{Instrumentation}
As already mentioned, given a cost monoid $\cl{M}$,
we assume the analysis of the RTL  code associates with each label $\ell$ in the term an element
$m_\ell=\s{costof}(\ell)$ of the cost monoid.  
Table~\ref{instrument} describes a
monadic transformation, extensively analysed in Gurr's
PhD thesis~\cite{Gurr91}, which instruments a program (in our case
$\lambda^\ell$) with the cost of executing its instructions. We are
then back to a standard $\lambda$-calculus (without labels) which
includes a basic data type to represent the cost monoid. 

We assume that the reduction rules of the source language  ($\lambda$)
are extended to account for a call-by-value evaluation of the 
monoidal expressions, where each element of the monoid is regarded as a value.
Then instrumentation and labelling are connected as follows.

\begin{proposition}[instrumentation vs. labelling]\label{inst-lab-prop}
Let $M$ be a term of the source labelled language $\lambda^\ell$.
If $\sem{M}\eval (m,V)$ where $V$ is a value then $M\eval_\Lambda U$,
$\s{costof}(\Lambda)=m$, and $\sem{U}=(\zero,V)$.
\end{proposition}

The following result summarizes the labelling approach 
to certified cost annotations.

\begin{theorem}[certified cost]\label{cost-prediction-thm}
Let $M$ be a term of the source language $\lambda$.
If  $\pi_1 (\sem{\lab{M}}) \eval m$ then the cost
of running $\cmp{M}$ is $m$.
\end{theorem}
\Proof
We take the following steps:
\[
\begin{array}{ll}
\pi_1 (\sem{\lab{M}}) \eval m   \\
\mbox{implies } \quad \lab{M}\eval_\Lambda \mand \s{costof}(\Lambda)=m  
&\mbox{(by proposition \ref{inst-lab-prop} above)} \\
\mbox{implies } \quad \cl{C}(\lab{M})\eval_\Lambda \mand \s{costof}(\Lambda)=m  
&\mbox{(by the simulation theorem \ref{com-sim-thm}(2)).}
\end{array}
\]
By proposition \ref{labelling-prop} and the following proviso \ref{cost-proviso}, we conclude that $m$ is the cost of running the compiled code $\cl{C}(M)$.\qed

\begin{table}[t]
{\footnotesize
\[
\begin{array}{lll}

\psi(x) &= &x \\
\psi(\lambda x^+.M) &= &\lambda x^+.\sem{M} \\
\psi(V_1,\ldots,V_n) &= &(\psi(V_1),\ldots,\psi(V_n)) \\ \\

\sem{V}              &= &(\zero,\psi(V)) \\

\sem{@(M_0,\ldots,M_n)} &= &\s{let} \ (m_0,x_0)=\sem{M_0} \cdots (m_n,x_n)=\sem{M_n}, \\
                       &   &(m_{n+1},x_{n+1})=@(x_0,\ldots,x_n) \ \s{in} \\
                       &&  (m_{0} \oplus m_1 \oplus \cdots\oplus m_{n+1}, x_{n+1}) \\ 

\sem{(M_1,\ldots,M_n)} &= &\s{let} \ (m_1,x_1)=\sem{M_1} \cdots (m_n,x_n)=\sem{M_n} \ \s{in } \\
                      && (m_1 \oplus \cdots \oplus m_n,(x_1,\ldots,x_n)) \quad \mbox{($(M_1,\ldots,M_n)$ not a value)}\\

\sem{\prj{i}{M}}      &= &\lets{(m,x)}{\sem{M}}{(m,\prj{i}{x})} \\

\sem{\lets{x}{M_1}{M_2}} &= &\s{let} \ (m_1,x)=\sem{M_1} \ \s{in} \ (m_2,x_2)=\sem{M_2} \ \s{in } \\
                         &&(m_1\oplus m_2,x_2) \\

\sem{\lb{\ell}{M}}       &= &\lets{(m,x)}{\sem{M}}{(m_\ell \oplus m,x)} \\

\sem{\plb{\ell}{M}}       &= &\lets{(m,x)}{\sem{M}}{(m \oplus m_\ell,x)} \\

\end{array}
\]}
\caption{Instrumentation of labelled $\lambda$-calculus.}\label{instrument}
\end{table}

\subsection{Higher-order Hoare Logic}

\newcommand\reflect[1]{\lceil #1 \rceil}

Many proof systems can be used to obtain static guarantees on the
evaluation of a purely functional program. In our setting, such
systems can also be used to obtain static guarantees on the execution
cost of a functional program by reasoning about its instrumentation.

We illustrate this point using a Hoare logic dedicated to
call-by-value purely functional programs~\cite{RP08}. Given a
well-typed program annotated by logic assertions, this system computes
a set of proof obligations, whose validity ensures the correctness of
the logic assertions with respect to the evaluation of the
functional program.

\begin{table}[t]
\footnotesize
\begin{center}
{\sc Syntax}
\end{center}
\[\begin{array}{lclr}
\footnotesize
F &::=& 

\s{True} 

\Alt \s{False} 

\Alt x 

\Alt F \wedge F 

\Alt F = F 

\Alt (F, F) 
&\mbox{(formulae)} \\
&&\Alt \pi_1 

\Alt \pi_2 

\Alt \lambda (x : \theta). F 

\Alt F\,F  

\Alt F \Rightarrow F

\Alt \forall (x: \theta). F
\\
\theta &::=& \s{prop} \Alt \iota \Alt \theta \times \theta \Alt \theta \arrow \theta
&\mbox{(types)} \\

V &::=& \w{id} \Alt \lambda \w{(id : A)}^+ / F : (id : A)/ F.M \Alt (V^*)
&\mbox{(values)} \\

M &::=& V \Alt @(M,M^+) \Alt \lets{\w{id} : A / F}{M}{M} \Alt (M^*) \Alt \prj{i}{M} 
&\mbox{(terms)} 
\end{array}\]

\begin{center}
{\sc Logical reflection of types}
\end{center}
\[\footnotesize
\begin{array}{rcl}
\reflect{\iota}
& = &
\iota 
\\

\reflect{A_1 \times \ldots \times A_n}
& = &
\reflect{A_1} \times \ldots \reflect{A_n}
\\

\reflect{A_1 \rightarrow A_2}
& = &
(\reflect{A_1} \rightarrow \s{prop}) \times (\reflect{A_1} \times \reflect{A_2} \rightarrow \s{prop})
\\
\end{array}
\]

\begin{center}
{\sc Logical reflection of values}
\end{center}
\[
\begin{array}{rcl}
\reflect{id} 
& = & 
id \\

\reflect{(V_1, \ldots, V_n)} 
& = &
(\reflect{V_1}, \ldots, \reflect{V_n}) \\

\reflect{\lambda (x_1 : A_1) / F_1 : (x_2 : A_2) / F_2.\ M}
& = &
(F_1, F_2)

\end{array}
\]
\caption{The surface language.}\label{fig:surface-language}
\end{table}

Logic assertions are written in a typed higher-order logic whose
syntax is given in Table~\ref{fig:surface-language}. From now on, we
assume that our source language is also typed. The metavariable $A$
ranges over simple types, whose syntax is $A ::= \iota \Alt A \times A
\Alt A \rightarrow A$ where $\iota$ are the basic types including a
data type $\s{cm}$ for the values of the cost monoid. The
metavariable~$\theta$ ranges over logical types. $\s{prop}$ is the
type of propositions. Notice that the inhabitants of arrow types on
the logical side are purely logical (and terminating) functions, while
on the programming language's side they are computational (and
potentially non-terminating) functions. Types are lifted to the
logical level through a logical reflection~$\reflect{\bullet}$ defined
in Table~\ref{fig:surface-language}.

We write ``$\s{let}\ x : A / F = M\ \s{in}\ M$'' to annotate a let
definition by a postcondition $F$ of type $\reflect A \rightarrow
\s{prop}$.  We write ``$\lambda (x_1 : A_1) / F_1 : (x_2 : A_2) / F_2.\ M$''
to ascribe to a $\lambda$-abstraction a precondition~$F_1$ of type
$\reflect{A_1} \rightarrow \s{prop}$ and a postcondition~$F_2$ of type
$\reflect{A_1} \times \reflect{A_2} \rightarrow \s{prop}$. Computational
values are lifted to the logical level using the reflection function defined
in Table~\ref{fig:surface-language}. The key idea of this definition is to
reflect a computational function as a pair of predicates consisting of
its precondition and its postcondition. Given a computational
function~$f$, a formula can refer to the precondition (resp. the
postcondition) of~$f$ using the predicate~$\s{pre}\,f$ (resp. $\s{post}\,f$). 
Thus, $\s{pre}$ (resp. $\s{post}$) is a synonymous for $\pi_1$ (resp.~$\pi_2$). 


To improve the usability of our tool, we define in
Table~\ref{fig:surface-language} a surface language by extending
$\lambda$ with several practical facilities. First, terms are
explicitly typed. Therefore, the labelling~$\cl{L}$ must be extended
to convey type annotations in an explicitly typed version of
$\lambda^\ell$ (the typing system of $\lambda^\ell$ is quite standard
and will be presented formally in the following  section \ref{typing-sec}).
The instrumentation~$\cl{I}$ defined in
Table~\ref{instrument} is extended to types by replacing each type
annotation~$A$ by its monadic interpretation~$\sem{A}$ defined
by $ \sem{A} = \s{cm} \times \overline{A}, \overline{\iota} =
\iota, \overline{A_1 \times A_2} = (\overline{A_1} \times
\overline{A_2})$ and $\overline{A_1 \rightarrow A_2} =
\overline{A_1} \rightarrow \sem{A_2} $.

Second, since the instrumented version of a source program would be
cumbersome to reason about because of the explicit threading of the
cost value, we keep the program in its initial form while allowing
logic assertions to implicitly refer to the instrumented version of
the program. Thus, in the surface language, in the term ``$\s{let}\ x
: A / F = M\ \s{in}\ M$'', $F$ has type $\reflect{\sem{A}}
\rightarrow \s{prop}$, that is to say a predicate over pairs of which
the first component is the execution cost.

Third, we allow labels to be written in source terms as a practical
way of giving names to the labels introduced by the
labelling~$\cl{L}$. By these means, the constant cost assigned to a
label~$\ell$ can be symbolically used in specifications by
writing~$\s{costof}(\ell)$.

Finally, as a convenience, we write ``$x : A /F$'' for ``$x : A
/ \lambda(\textrm{cost} : \s{cm}, x : \reflect{\sem{A}}). F$''.
This improves the conciseness of specifications by automatically
allowing reference to the cost variable in logic assertions without
having to introduce it explicitly. 

\subsection{Prototype implementation}

We implemented a prototype compiler~\cite{YRG11} in $\ocaml$ ($\sim$
3.5Kloc). In addition to the distributed source code, a web
application enables basic experiments without any installation process.

This compiler accepts a program~$P$ written in the surface language
extended with fixpoints and algebraic datatypes. We found no technical
difficulty in handling these extensions and this is the reason why they
are excluded from the core language in the presented formal development.
Specifications are written in the $\coq$ proof
assistant~\cite{Coq}.  A \s{logic} keyword is used to include logical
definitions written in~$\coq$ to the source program.

Type checking is performed on~$P$ and, upon success, it produces a
type annotated program~$P_t$. Then, the labelled
program~$P_\ell = \cl{L}(P_t)$ is generated. Following the same treatment of
branching as in our previous work on imperative programs~\cite{AAR09,AAR12},
the labelling introduces a label at the beginning of each pattern
matching branch.

By erasure of specifications and type annotations, we obtain a
program~$P_\lambda$ of~$\lambda$ (Table~\ref{lambda}). 
Using the compilation chain
presented earlier, $P_\lambda$ is compiled into a
program~$P_{h}$~of~$\lambda_{h, \w{vn}}$ (Table~\ref{lambda-hoist}) . 
The annotating compiler uses the
cost model that counts for each label~$\ell$ the number
of primitive operations that belong to execution paths starting
from~$\ell$ (and ending in another label or in an instruction without
successor). 

Finally, the instrumented version of~$P_\ell$ as well as the actual
cost of each label is given as input to a verification condition
generator to produce a set of proof obligations implying the validity
of the user-written specifications. These proof obligations are either
proved automatically using first-order theorem provers or manually
in~$\coq$.

\subsection{Examples}

In this section, we present two examples that are idiomatic of
functional programming: an inductive function and a higher-order
function. These examples were checked using our prototype
implementation. More involved examples are distributed with the
software. These examples include several standard functions on lists
(fold, map, \ldots), combinators written in continuation-passing style,
and functions over binary search trees.

\paragraph{An inductive function}

Table~\ref{code:concat} contains an example of a simple inductive
function: the standard concatenation of two lists. In the
code, one can distinguish three kinds of toplevel definitions: the
type definitions prefixed by the \textbf{type} keyword, the
definitions at the logical level surrounded by \textbf{logic}\;\{
\ldots \}, and the program definitions introduced by~the~\textbf{let}
keyword.

On lines 1 and 2, the type definitions introduce a type {\sf list} for
lists of natural numbers as well as a type {\sf bool} for
booleans. Between lines 3 and 9, at the logical level, a {\coq}
inductive function defines the length of lists so that we can use this
measure in the cost annotation of {\sf concat}. Notice that the type
definitions are automatically lifted at the {\coq} level, provided
that they respect the strict positivity criterion imposed by {\coq} to
ensure well-foundedness of inductive definitions.

The concatenation function takes two lists {\sf l1} and {\sf l2} as
input, and it is defined, as usual, by induction on {\sf l1}. In order
to write a precise cost annotation, each part of the function body is
labelled so that every piece of code is dominated by a label:
$\ell_{\textrm{match}}$ dominates ~``$\textbf{match } {\sf l1}
\textbf{ with } {\sf Nil } \Rightarrow \bullet \mid {\sf Cons (x, xs)}
\Rightarrow \bullet$'', $\ell_{\textrm{nil}}$ dominates~``${\sf
  Nil}$'', $\ell_{cons}$ dominates ``${\sf Cons (x, \bullet)}$'', and
$\ell_{rec}$ dominates ``${\sf concat (xs, l2)}$''. Looking at the
compiled code in Table~\ref{code:compiledconcat}, it is easy to check
that the covering of the code by the labels is preserved through the
compilation process. One can also check that the computed costs are
{\em correct} with respect to a cost model that simply counts the
number of instructions, \textit{i.e.}, costof($\ell_\textrm{nil}$) =
2, costof($\ell_\textrm{rec}$) = 6, costof($\ell_\textrm{cons}$) = 5
and costof($\ell_{\textrm{match}}$) = 1. Here we are simply assuming
one unit of time per low-level instruction, but a more refined
analysis is possible by propagating the binary instructions till the
binary code (cf. \cite{AAR09,AAR12}).

Finally, the specification says that the cost of executing {\sf concat
  (l1, l2)} is proportional to the size of {\sf l1}. Recall that the
`cost' and `result' variables are implicitly bound in the
post-condition. Notice that the specification is very specific on the
concrete time constants that are involved in that linear
function. Following the proof system of the higher-order Hoare
logic~\cite{RP08}, the verification condition generator produces 37
proof obligations out of this annotated code. All of them are
automatically discharged by~{\coq} (using, in particular, the linear
arithmetic decision procedure {\sf omega}).

\begin{table}
{\footnotesize\bgroup\sf\medskip\begin{flushleft}

\noindent{\textcolor{Gray}{01}\hspace{0.5em}}\hspace*{0.00000em}{\bf type} list = Nil \ensuremath{|} Cons (nat, list)~{\linebreak}
\noindent{\textcolor{Gray}{02}\hspace{0.5em}}\hspace*{0.00000em}{\bf type} bool = BTrue \ensuremath{|} BFalse~{\linebreak}
\noindent{\textcolor{Gray}{03}\hspace{0.5em}}\hspace*{0.00000em}{\bf logic}\; \{~{\linebreak}
\noindent{\textcolor{Gray}{04}\hspace{0.5em}}\hspace*{0.50000em}Fixpoint length (l \ensuremath{\colon} list) \ensuremath{\colon} nat \ensuremath{\colon}= ~{\linebreak}
\noindent{\textcolor{Gray}{05}\hspace{0.5em}}\hspace*{2.00000em}{\bf match} l {\bf with}~{\linebreak}
\noindent{\textcolor{Gray}{06}\hspace{0.5em}}\hspace*{2.00000em}\ensuremath{|} Nil $\Rightarrow$ 0~{\linebreak}
\noindent{\textcolor{Gray}{07}\hspace{0.5em}}\hspace*{2.00000em}\ensuremath{|} Cons (x, xs) $\Rightarrow$ 1 + length (xs)~{\linebreak}
\noindent{\textcolor{Gray}{08}\hspace{0.5em}}\hspace*{2.00000em}{\bf end}.~{\linebreak}
\noindent{\textcolor{Gray}{09}\hspace{0.5em}}\hspace*{0.00000em}\}~{\linebreak}
\noindent{\textcolor{Gray}{10}\hspace{0.5em}}\hspace*{0.00000em}{\bf let} {\bf rec} concat (l1\ensuremath{\colon} list, l2\ensuremath{\colon} list) \ensuremath{\colon} list \{~{\linebreak}
\noindent{\textcolor{Gray}{11}\hspace{0.5em}}\hspace*{1.00000em}cost = costof($\ell_{\textrm{match}}$ )
+ costof($\ell_{\textrm{nil}}$) ~{\linebreak}
\noindent{\textcolor{Gray}{12}\hspace{0.5em}}\hspace*{4.50000em}+ (costof($\ell_\textrm{rec}$) 
+ costof($\ell_{\textrm{match}}$) + costof($\ell_{\textrm{cons}}$)) \ensuremath{\times} length (l1)~{\linebreak}
\noindent{\textcolor{Gray}{13}\hspace{0.5em}}\hspace*{0.00000em}\}~{\linebreak}
\noindent{\textcolor{Gray}{14}\hspace{0.5em}}\hspace*{0.00000em}= ~{\linebreak}
\noindent{\textcolor{Gray}{15}\hspace{0.5em}}\hspace*{1.00000em}$\ell_\textrm{match}$ \ensuremath{>} ~{\linebreak}
\noindent{\textcolor{Gray}{16}\hspace{0.5em}}\hspace*{1.00000em}{\bf match} l1 {\bf with}~{\linebreak}
\noindent{\textcolor{Gray}{17}\hspace{0.5em}}\hspace*{1.00000em}\ensuremath{|} Nil \ensuremath{\rightarrow} 
$\ell_{\textrm{nil}}$ \ensuremath{>} l2~{\linebreak}
\noindent{\textcolor{Gray}{18}\hspace{0.5em}}\hspace*{1.00000em}\ensuremath{|} Cons (x, xs) \ensuremath{\rightarrow} $\ell_{\textrm{cons}}$ \ensuremath{>} Cons (x, concat (xs, l2) \ensuremath{>} $\ell_{\textrm{rec}}$) ~{\linebreak}
\noindent{\textcolor{Gray}{19}\hspace{0.5em}}\hspace*{1.00000em}{\bf end}
\end{flushleft}\medskip\egroup\noindent
with $\left\{
\begin{array}{lcl}
\textsf{costof}(\ell_\textrm{nil}) &=& 2 \\
\textsf{costof}(\ell_\textrm{rec}) &=& 6  \\
\textsf{costof}(\ell_\textrm{cons}) &=& 5  \\
\textsf{costof}(\ell_{\textrm{match}}) &=& 1 \\
\end{array}
\right.$
}
\caption{A function that concatenates two lists, and its cost annotation.}
\label{code:concat}
\end{table}

\begin{table}
{\footnotesize\bgroup\sf\medskip\begin{flushleft}
\noindent{\textcolor{Gray}{01}\hspace{0.5em}}\hspace*{0.00000em}
\textbf{routine} x$_{19}$ (c$_{20}$, x$_7$)
{\linebreak}
\noindent{\textcolor{Gray}{02}\hspace{0.5em}}\hspace*{0.00000em}
  $\ell_{rec}$:
{\linebreak}
\noindent{\textcolor{Gray}{03}\hspace{0.5em}}\hspace*{0.50000em}
    k$_2$ $\leftarrow$ \textbf{proj} 1 c$_{20}$ ;
{\linebreak}
\noindent{\textcolor{Gray}{04}\hspace{0.5em}}\hspace*{0.50000em}
    x $\leftarrow$ \textbf{proj} 2 c$_{20}$ ;
{\linebreak}
\noindent{\textcolor{Gray}{05}\hspace{0.5em}}\hspace*{0.50000em}
    x$_{14}$ $\leftarrow$ \textbf{make\_int} 1 ;
{\linebreak}
\noindent{\textcolor{Gray}{06}\hspace{0.5em}}\hspace*{0.50000em}
    x$_{15}$ $\leftarrow$ \textbf{make\_tuple} (x$_{14}$, x, x$_7$) ;
{\linebreak}
\noindent{\textcolor{Gray}{07}\hspace{0.5em}}\hspace*{0.50000em}
    x$_{22}$ $\leftarrow$ \textbf{proj} 0 k$_2$ ;
{\linebreak}
\noindent{\textcolor{Gray}{08}\hspace{0.5em}}\hspace*{0.50000em}
    \textbf{call} x$_{22}$ (k$_2$, x$_{15}$)
{\linebreak}

\noindent{\textcolor{Gray}{09}\hspace{0.5em}}\hspace*{0.00000em}
\textbf{routine} x$_{16}$ (c$_{17}$, l1, l2, k$_2$)
{\linebreak}
\noindent{\textcolor{Gray}{10}\hspace{0.5em}}\hspace*{0.00000em}
  $\ell_{match}$:
{\linebreak}
\noindent{\textcolor{Gray}{11}\hspace{0.5em}}\hspace*{0.50000em}
    switch l1
{\linebreak}
\noindent{\textcolor{Gray}{12}\hspace{0.5em}}\hspace*{0.50000em}
    0 : $\ell_{nil}:$
{\linebreak}
\noindent{\textcolor{Gray}{13}\hspace{0.5em}}\hspace*{2.00000em}
      x$_{18}$ $\leftarrow$ \textbf{proj} 0 k$_2$ ;
{\linebreak}
\noindent{\textcolor{Gray}{14}\hspace{0.5em}}\hspace*{2.00000em}
      \textbf{call} x$_{18}$ (k$_2$, l2) 
{\linebreak}
\noindent{\textcolor{Gray}{15}\hspace{0.5em}}\hspace*{0.50000em}
    1 : $\ell_{cons}:$
{\linebreak}
\noindent{\textcolor{Gray}{16}\hspace{0.5em}}\hspace*{2.00000em}
      xs $\leftarrow$ \textbf{proj} 2 l1 ;
{\linebreak}
\noindent{\textcolor{Gray}{17}\hspace{0.5em}}\hspace*{2.00000em}
      x $\leftarrow$ \textbf{proj} 1 l1 ;
{\linebreak}
\noindent{\textcolor{Gray}{18}\hspace{0.5em}}\hspace*{2.00000em}
      x$_{13}$ $\leftarrow$ \textbf{make\_tuple} (x$_{19}$, k$_2$, x) ;
{\linebreak}
\noindent{\textcolor{Gray}{19}\hspace{0.5em}}\hspace*{2.00000em}
      x$_{21}$ $\leftarrow$ \textbf{proj} 0 c$_{17}$ ;
{\linebreak}
\noindent{\textcolor{Gray}{20}\hspace{0.5em}}\hspace*{2.00000em}
      \textbf{call} x$_{21}$ (c$_{17}$, xs, l2, x$_{13}$)
{\linebreak}
\end{flushleft}
\medskip\egroup\noindent
}
\caption{The compiled code of {\sf concat}.}
\label{code:compiledconcat}
\end{table}

\paragraph{A higher-order function}
Let us consider a higher-order function $pexists$ that looks for an
integer~$x$ in a list~$l$ such that~$x$ validates a predicate~$p$. In
addition to the functional specification, we want to prove that the
cost of this function is linear in the length~$n$ of the list~$l$. The
corresponding program written in the surface language can be found in
Table~\ref{code:pexists}.

A prelude declares the type and logical definitions used by the
specifications. On lines 1 and 2, two type definitions introduce data
constructors for lists and booleans. Between lines 4 and 5, a $\coq$
definition introduces a predicate $bound$ over the reflection of
computational functions from $nat$ to $nat \times bool$ that ensures
that the cost of a computational function~$p$ is uniformly bounded by
a constant~$k$.

On line 9, the precondition of function $pexists$ requires the
function~$p$ to be total. Between lines 10 and 11, the postcondition
first states a functional specification for $pexists$: the boolean
result witnesses the existence of an element~$x$ of the input
list~$l$ that is related to $BTrue$ by the postcondition of~$p$. The
second part of the postcondition characterizes the cost of $pexists$
in case of a negative result: assuming that the cost of~$p$ is bounded
by a constant~$k$, the cost of $pexists$ is proportional to $k \cdot n$.
Notice that there is no need to add a label in front of $BTrue$ in the
first branch of the inner pattern-matching since the specification only
characterizes the cost of an unsuccessful search. 

The verification condition generator produces 53 proof obligations out of
this annotated program; 46 of these proof obligations are 
automatically discharged and 7 of them are manually proved in $\coq$.

\begin{table}[t]
{\footnotesize
\bgroup\sf\medskip\begin{flushleft}

\noindent\textrm{\textcolor{Gray}{01\;\;\,}}\hspace{0.5em}\hspace*{0.00000em}\textbf{type} list = Nil \ensuremath{|} Cons (nat, list)~\linebreak
\noindent\textrm{\textcolor{Gray}{02\;\;\,}}\hspace{0.5em}\hspace*{0.00000em}\textbf{type} bool = BTrue \ensuremath{|} BFalse~\linebreak
\noindent\textrm{\textcolor{Gray}{03\;\;\,}}\hspace{0.5em}\hspace*{0.00000em}\textbf{logic} \{ ~\linebreak
\noindent\textrm{\textcolor{Gray}{04\;\;\,}}\hspace{0.5em}\hspace*{0.50000em}\textbf{Definition} bound (p \ensuremath{\colon} nat \ensuremath{\longrightarrow} (nat \ensuremath{\ast} bool)) (k \ensuremath{\colon} nat) \ensuremath{\colon} \textbf{Prop} \ensuremath{\colon}=~\linebreak
\noindent\textrm{\textcolor{Gray}{05\;\;\,}}\hspace{0.5em}\hspace*{2.00000em}\ensuremath{\forall} x m\ensuremath{\colon} nat, \ensuremath{\forall} r\ensuremath{\colon} bool, post p x (m, r) \ensuremath{\Rightarrow} m \ensuremath{\le} k.~\linebreak
\noindent\textrm{\textcolor{Gray}{06\;\;\,}}\hspace{0.5em}\hspace*{0.50000em}\textbf{Definition} k0 \ensuremath{\colon}= \ensuremath{\mbox{costof}(\ell_{m})} + \ensuremath{\mbox{costof}(\ell_{nil})}.~\linebreak
\noindent\textrm{\textcolor{Gray}{07\;\;\,}}\hspace{0.5em}\hspace*{0.50000em}\textbf{Definition} k1 \ensuremath{\colon}= \ensuremath{\mbox{costof}(\ell_{m})} + \ensuremath{\mbox{costof}(\ell_{p})} + \ensuremath{\mbox{costof}(\ell_{c})} + \ensuremath{\mbox{costof}(\ell_{f})} + \ensuremath{\mbox{costof}(\ell_{r})}. ~\linebreak
\noindent\textrm{\textcolor{Gray}{08\;\;\,}}\hspace{0.5em}\hspace*{0.00000em}\}~\linebreak
\noindent\textrm{\textcolor{Gray}{09\;\;\,}}\hspace{0.5em}\hspace*{0.00000em}\textbf{let} \textbf{rec} pexists (p \ensuremath{\colon} nat \ensuremath{\rightarrow} bool, l\ensuremath{\colon} list) \{ \ensuremath{\forall} x, pre p x \} \ensuremath{\colon} bool \{~\linebreak
\noindent\textrm{\textcolor{Gray}{10\;\;\,}}\hspace{0.5em}\hspace*{0.50000em}((result = BTrue) \ensuremath{\Leftrightarrow} (\ensuremath{\exists} x c\ensuremath{\colon} nat, mem x l \ensuremath{\wedge} post p x (c, BTrue))) \ensuremath{\wedge}~\linebreak
\noindent\textrm{\textcolor{Gray}{11\;\;\,}}\hspace{0.5em}\hspace*{0.50000em}(\ensuremath{\forall} k\ensuremath{\colon} nat, bound p k \ensuremath{\wedge} (result = BFalse) \ensuremath{\Rightarrow} cost \ensuremath{\le} k0 + (k + k1) \ensuremath{\times} length (l))~\linebreak
\noindent\textrm{\textcolor{Gray}{12\;\;\,}}\hspace{0.5em}\hspace*{0.00000em}\} = \ensuremath{\ell_{m}}\ensuremath{>} \textbf{match} l \textbf{with}~\linebreak
\noindent\textrm{\textcolor{Gray}{13\;\;\,}}\hspace{0.5em}\hspace*{1.00000em}\ensuremath{|} Nil \ensuremath{\rightarrow} \ensuremath{\ell_{nil}}\ensuremath{>} BFalse~\linebreak
\noindent\textrm{\textcolor{Gray}{14\;\;\,}}\hspace{0.5em}\hspace*{1.00000em}\ensuremath{|} Cons (x, xs) \ensuremath{\rightarrow} \ensuremath{\ell_{c}}\ensuremath{>} \textbf{match} p (x) \ensuremath{>} \ensuremath{\ell_{p}} \textbf{with}~\linebreak
\noindent\textrm{\textcolor{Gray}{15\;\;\,}}\hspace{0.5em}\hspace*{10.50000em}\ensuremath{|} BTrue \ensuremath{\rightarrow} BTrue ~\linebreak
\noindent\textrm{\textcolor{Gray}{16\;\;\,}}\hspace{0.5em}\hspace*{10.50000em}\ensuremath{|} BFalse \ensuremath{\rightarrow} \ensuremath{\ell_{f}}\ensuremath{>} (pexists (p, xs) \ensuremath{>} \ensuremath{\ell_{r}})
\end{flushleft}\medskip\egroup\noindent
} 
\caption{A higher-order function and its specification.}
\label{code:pexists}
\end{table}

\section{Typing the compilation chain}\label{typing-sec}
We describe a (simple) typing of the compilation
chain.  Specifically, each $\lambda$-calculus of the compilation chain is equipped
with a type system which enjoys {\em subject reduction}:
if a term has a type then all terms to which it reduces have the same type.
Then the compilation functions are extended to types and are shown
to be {\em type preserving}: if a term has a type then its compilation 
has the corresponding  compiled type. 

Besides providing insight into the compilation chain, typing is used
in two ways. First, the tool for reasoning about cost annotations
presented in section~\ref{instrument-sec} takes as input
a typed $\lambda$-term and second, and more importantly, in
section~\ref{memory-sec}, we rely on an enrichment of a type system
which is expressive enough to type a compiled code with explicit
memory deallocations. 

The two main steps in typing the compilation chain are well studied,
see, {\em e.g.}, the work of Morrisett {\em et al.}~\cite{M99}, and
concern the CPS and the closure conversion steps. In the former, the
basic idea is to type the continuation/the evaluation context of a
term of type $A$ with its negated type $(A\arrow R)$, where $R$ is
traditionally taken as the type of `results'. In the latter, one
relies on existential types to hide the details of the representation
of the `environment' of a function, {\em i.e.} the tuple of variables
occurring free in its body.

\subsection{Typing conventions}
We shall denote with $\w{tid}$  the syntactic category of {\em type variables} 
with generic elements $t,s,\ldots$
and with $A$ the syntactic category of {\em types} with generic elements $A,B,\ldots$
A {\em type context} is denoted with $\Gamma,\Gamma',\ldots$, and it stands for
a finite domain partial function from variables to types. 
To explicit a type context, we write $x_1:A_1,\ldots,x_n:A_n$ where
the variables $x_1,\ldots,x_n$ must be all distinct and the order is irrelevant.
Also we write $x^*:A^*$ for a possibly empty  sequence $x_1:A_1,\ldots,x_n:A_n$, and
$\Gamma,x^*:A^*$ for the context resulting from $\Gamma$ by adding the sequence
$x^*:A^*$. Hence the variables in $x^*$  must not be in the domain of $\Gamma$.
If $A$ is a type, we write $\ftv{A}$ for the set of {\em type variables occurring free} in it
and, by extension, if $\Gamma$ is a type context $\ftv{\Gamma}$ is the union of the sets $\ftv{A}$ 
where $A$ is a type in the codomain of $\Gamma$.
A {\em typing judgement} is typically written as $\Gamma \Gives M:A$ where $M$ is some term.
We shall write $\Gamma \Gives M^*:A^*$ for $\Gamma \Gives M_1:A_1,\ldots,$$\Gamma \Gives M_n:A_n$.
Similar conventions apply if we replace the symbol $`*'$ with the symbol $`+'$ except that in this case the sequence is assumed not-empty.
A type transformation, say $\cl{T}$, is {\em lifted to type contexts} by defining
$\cl{T}(x_1:A_1,\ldots x_n:A_n) = x_1:\cl{T}(A_1),\ldots,x_n:\cl{T}(A_n)$.
Whenever we write:
\begin{quote}
if $\Gamma \Gives^{S_{1}} M:A$ then $\cl{T}(\Gamma) \Gives^{S_{2}} \cl{T}(M):\cl{T}(A)$
\end{quote}
what we actually mean is that if the judgement in the hypothesis is {\em derivable} in a certain `type system $S_1$' then the transformed judgement in derivable in 
the `type system $S_2$'.

\subsection{The type system of the source language}
Table~\ref{type-lambda} describes the typing rules 
for the source language defined in Table~\ref{lambda}.
These rules are standard except those for the labellings,
and, as announced, they are preserved by reduction.

\begin{table}
{\footnotesize
\begin{center}
{\sc Syntax types}
\end{center}
\[
\begin{array}{lll}
A      &::= \w{tid} \Alt A^+\arrow A \Alt \times(A^*)  &\mbox{(types)} 
\end{array}
\]
\begin{center}
{\sc Typing rules}
\end{center}
\[
\begin{array}{cc}

\infer{x:A\in \Gamma}
{\Gamma \Gives x:A}

&\infer{
\begin{array}{c}
\Gamma,x:A\Gives N:B\\
\Gamma \Gives M:A
\end{array}}
{\Gamma \Gives \lets{x}{M}{N}:B} \\ \\

\infer{\Gamma,x^+:A^+ \Gives M:B}
{\Gamma \Gives \lambda x^+.M:A^+\arrow B} 

&\infer{\begin{array}{c}
\Gamma \Gives M:A^+\arrow B \\
\Gamma \Gives N^+:A^+
\end{array}}
{\Gamma \Gives @(M,N^+):B}  \\ \\

\infer{\Gamma \Gives M^*:A^*}
{\Gamma \Gives (M^*):\times(A^*)}

&\infer{\Gamma \Gives M:\times(A_1,\ldots,A_n)\quad 1\leq i  \leq n}
{\Gamma \Gives \prj{i}{M}:A_i} \\ \\



\infer{\Gamma \Gives M:A}
{\Gamma \Gives \lb{\ell}{M}:A} 

&\infer{\Gamma \Gives M:A}
{\Gamma \Gives \plb{\ell}{M}:A}

\end{array}
\]
\begin{center}
{\sc Restricted syntax CPS types, $R$ type of results}
\end{center}
\[
\begin{array}{lll}
A &::= \w{tid} \Alt A^+\arrow R \Alt \times(A^*)  &\mbox{(CPS types)}
\end{array}
\]
\begin{center}
{\sc CPS type compilation}
\end{center}
\[
\begin{array}{ll}

\cps{t} &= t \\
\cps{\times(A^*)}  &= \times(\cps{A}^*) \\
\cps{A^+ \arrow B} &=(\cps{A})^+, \neg \cps{B} \arrow R \\
                   &\mbox{where: }\neg A \equiv (A\arrow R)
\end{array}
\]
}
\caption{Type system for $\lambda^\ell$ and $\lambda^\ell_{\w{cps}}$}\label{type-lambda}
\end{table}

\begin{proposition}[subject reduction]\label{subj-red-prop}
If $M$ is a term of the $\lambda^\ell$ calculus,
$\Gamma \Gives M:A$ and $M\arrow N$ then $\Gamma \Gives N:A$.
\end{proposition}

The typing rules described in Table~\ref{type-lambda} 
apply to the CPS $\lambda$-calculus too. Table~\ref{type-lambda}
describes  the restricted syntax of the CPS types
and the CPS type translation. Then the CPS term translation
defined in Table~\ref{lambda-cps} preserves typing in the following
sense.

\begin{proposition}[type CPS]\label{type-cps-prop}
If $\Gamma \Gives M:A$ then 
$\cps{\Gamma},\w{halt}:\neg \cps{A} \Gives \cps{M}:R$.
\end{proposition}

\subsection{Type system for the value named calculi}
Table~\ref{types-adm} describes the typing rules for the
value named calculi.  
For the sake of brevity, we shall omit the type of a 
term since this type is always the
type of results $R$ and write $\Gamma \aGives M$ rather than 
$\Gamma \aGives M:R$.
The first 6 typing rules are just a
specialization of the corresponding rules in Table~\ref{type-lambda}.
The last two rules allow for the introduction and elimination of {\em
 existential types}; we shall see shortly how they 
are utilised in typing closure conversion. 

In the proposed formalisation, we rely on the tuple constructor to
introduce an existential type and the first projection to eliminate
it.  This has the advantage of leaving unchanged the syntax and the
operational semantics of the value named $\lambda$-calculus. An
alternative presentation consists in introducing specific operators to
introduce and eliminate existential types which are often denoted with
\s{pack} and \s{unpack}, respectively.  The reader who is familiar
with this notation may just read $(x)$ as $\pack{x}$ and $\prj{1}{x}$
as $\unpack{x}$ when $x$ has an existential type. With this
convention, the rewriting rule which allows to {\em unpack} a {\em
  packed} value is just a special case of the rule for projection.

As in the previous system, typing is preserved by reduction.

\begin{proposition}[subject reduction, value named]\label{subj-red-adm-prop}
If $M$ is a term of the $\lambda^\ell_{\w{cps},\w{vn}}$-calculus,
$\Gamma \aGives M$ and $M\arrow N$ then 
$\Gamma \aGives N$.
\end{proposition}

\begin{table}
{\footnotesize
\begin{center}
{\sc Syntax types}
\end{center}
\[
\begin{array}{lll}

A&::= \w{tid} \Alt (A^+\arrow R) \Alt \times(A^*) \Alt \exists \w{tid}.A
\end{array}
\]
\begin{center}
{\sc Typing rules}
\end{center}
\[
\begin{array}{cc}

\infer{\Gamma,x^+:A^+\aGives M}
{\Gamma \aGives \lambda x^+.M:A^+\arrow R} 

&\infer{x:A^+\arrow R,y^+:A^+\in \Gamma}
{\Gamma \aGives @(x,y^+)} \\ \\

\infer{x^*:A^*\in \Gamma}
{\Gamma \aGives (x^*):\times(A^*)}

&\infer{\begin{array}{c}
y:\times(A_1,\ldots,A_n)\in \Gamma\quad 1\leq i\leq n\quad\\
\Gamma,x:A_i\aGives M
\end{array}}
{\Gamma \aGives \lets{x}{\prj{i}{y}}{M}}  \\ \\

\infer{\Gamma \aGives V:A\quad \Gamma,x:A\aGives M}
{\Gamma \aGives \lets{x}{V}{M}} 

&\infer{\Gamma \aGives M}
{\Gamma \aGives \lb{\ell}{M}}   \\ \\

\infer{x:[B/t]A\in \Gamma}
{\Gamma \aGives (x) :\exists t.A}

&\infer{y:\exists t.A \in \Gamma\quad
\Gamma,x:A\aGives M\quad t\notin \ftv{\Gamma}}
{\Gamma \aGives \lets{x}{\prj{1}{y}}{M}}



\end{array}
\]
\begin{center}
{\sc Closure conversion type compilation}
\end{center}
\[
\begin{array}{ll}

\cc{t} &=t \\
\cc{\times(A^*)} &=\times(\cc{A}^*) \\
\cc{A^+\arrow R} &= \exists t.\times((t,\cc{A}^+ \arrow R),\ t) \\
\cc{\exists t.A} &= \exists t.\cc{A}

\end{array}
\]
}
\caption{Type system for the value named calculi and closure conversion}\label{types-adm}
\end{table}

The transformation into value named CPS form specified in Table~\ref{back-forth} affects the terms but not the types.

\begin{proposition}[type value named]\label{type-adm-prop}
If $M$ is a term of the $\lambda^{\ell}_{\w{cps}}$-calculus
and $\Gamma \Gives M:R$ then $\Gamma \aGives \vn{M}$.
\end{proposition}

On the other hand, to type the closure conversion we rely on
existential types to abstract/hide the type of the environment as
specified in Table~\ref{types-adm}.  Then the term translation of the
function definition and application given in Table~\ref{lambda-cc} has to be
slightly modified to account for the introduction and elimination of
existential types.  The revised definition is as follows:

{\footnotesize
\[
\begin{array}{ll}

\cc{@(x,y^+)}                   &=  
\begin{array}{l}
\s{let} \ x =\prj{1}{x} \ \s{in}  \qquad \mbox{($\leftarrow$ {\sc existential elimination})}\\
\lets{(c,e)}{x}{@(c,e,y^+)} 
\end{array}
\\ \\

\cc{\lets{x}{C}{M}} &= 
\begin{array}{l}
\s{let} \ c=\lambda e,x^+.\s{let} \ (z_1,\ldots,z_k)=e \ \s{in} \ \cc{N}  \ \s{in} \\
\s{let} \ e=(z_1,\ldots,z_k) \ \s{in} \\
\s{let} \ x = (c,e) \ \s{in} \\
\s{let} \ x = (x) \ \s{in}  \qquad \mbox{($\leftarrow$ {\sc existential introduction})}\\
\cc{M} \qquad \qquad (\mbox{if }C=\lambda x^+.N,\fv{C}=\set{z_1,\ldots,z_k})
\end{array} 

\end{array}
\]
}

This modified closure conversion does not affect the commutation and simulation properties 
stated in propositions \ref{cc-commutation} and \ref{cc-simulation} and moreover 
it  preserves typing  as follows. 

\begin{proposition}[type closure conversion]\label{type-cc-prop}
If $M$ is a term in $\lambda^{\ell}_{\w{cps},\w{vn}}$  and  $\Gamma \aGives M$ 
then  $\cc{\Gamma}\aGives \cc{M}$.
\end{proposition}

Similarly to the transformation in value named form, 
the hoisting transformations  affect the terms but not the types.

\begin{proposition}[type hoisting]\label{type-hoisting-prop}
If $M$ is a term in $\lambda^{\ell}_{\w{cps},\w{vn}}$, 
$\Gamma \aGives M$,  and  $M\leadsto N$ then $\Gamma \aGives N$.
\end{proposition}

\subsection{Typing the compiled code}
We can now extend the compilation function to types by defining:
\[
\cmp{A} = \cc{\cps{A}}
\]
and by composing the previous results we derive the following type preservation 
property of the compilation function.

\begin{theorem}[type preserving compilation]\label{type-compil-thm}
If $M$ is a term of the  $\lambda^\ell$-calculus
and $\Gamma \Gives M:A$  then 
\[
\cmp{\Gamma},\w{halt}:\exists t.\times(t,\cmp{A}\arrow R,t) \aGives \cmp{M}~.
\]
\end{theorem}

\begin{remark}\label{halt-rmk}
The `\w{halt}' variable introduced by the CPS translation 
can occur only in a subterm of the shape $@(\w{halt},x)$ in the intermediate
code prior to closure conversion. 
Then in the closure conversion translation, it suffices that $\cc{@(\w{halt},x)}=@(\w{halt},x)$ 
and give to $`\w{halt}'$
a functional rather than an existential type. With this proviso,
theorem \ref{type-compil-thm} above can be restated as follows:
\begin{quote}
If $M$ is a term of the $\lambda^\ell$-calculus and 
$\Gamma \Gives M:A$ then $\cmp{\Gamma},\w{halt}:\neg \cmp{A} \aGives \cmp{M}$.
\end{quote}
\end{remark}

\begin{example}[typing the compiled code]\label{type-example}
We consider again the compilation of the term $\lambda x.y$ (cf. example \ref{cc-example}) which can be typed, {\em e.g.}, as follows:
\[
y:t_1 \Gives \lambda x.y:(t_2\arrow t_1)~.
\]
Its CPS translation is then typed as:
\[
y:t_1,\w{halt}:\neg \cps{t_2\arrow t_1} \Gives @(\w{halt},\lambda x,k.@(k,y)): R~.
\]
The value named translation does not affect the types:
\[
y:t_1,\w{halt}:\neg \cps{t_2\arrow t_1} \aGives \lets{z_1}{\lambda x,k.@(k,y)}{@(\w{halt},z_1)}~.
\]
After closure conversion we obtain the following term $M$:
\[
\begin{array}{l}
\s{let} \ c = \lambda e,x,k.\s{let} \ y=\prj{1}{e}, k=\prj{1}{k},c=\prj{1}{k},e=\prj{2}{k} \ \s{in} \ @(c,e,y) \ \s{in}\\
\s{let} \ e = (y), z_1=(c,e), z_1=(z_1),\w{halt} =\prj{1}{\w{halt}},c=\prj{1}{\w{halt}}, e = \prj{2}{\w{halt}} \ \s{in} \\
@(c,e,z_1)
\end{array}
\]
which is typed as follows:
\[
y:t_1,\w{halt}: \exists t.\times(t,\cmp{t_2\arrow t_1} \arrow R,t) \aGives M~.
\]
In this case no further hoisting transformation applies.
If we adopt the optimised compilation strategy sketched in remark \ref{halt-rmk}
then after closure conversion we obtain the following term $M'$:
\[
\begin{array}{l}
\s{let} \ c = \lambda e,x,k.\s{let} \ y=\prj{1}{e}, k=\prj{1}{k},c=\prj{1}{k},e=\prj{2}{k} \ \s{in} \ @(c,e,y) \ \s{in}\\
\s{let} \ e = (y), z_1=(c,e), z_1=(z_1), \ \s{in} \\
@(\w{halt},z_1)
\end{array}
\]
which is typed as follows:
\[
y:t_1,\w{halt}:\cmp{t_2\arrow  t_1} \arrow R \aGives M'~.
\]

\end{example}

\section{Memory management}\label{memory-sec}
We describe an enrichment of the
$\lambda^{\ell}_{h,\w{vn}}$-calculus called
$\lambda^{\ell,r}_{h,\w{vn}}$-calculus which explicitly handles allocation
and deallocation of {\em memory regions}. 
At our level of abstraction, the memory regions are just names
$r,r',\ldots$ of a countable set.  Intuitively, the `live' locations
of a memory are partitioned into regions.  The three new operations
the enriched calculus may perform are: (1) allocate a new region, (2)
allocate a value (in our case a non-empty tuple) in a region, and (3)
dispose a region. The additional operation of reading a value from a
region is implicit in the projection
operation which is already
available in the non-enriched calculus $\lambda^{\ell}_{h,\w{vn}}$.  In
order to gain some expressivity we shall also allow a function to be
parametric in a collection of region names which are provided as
arguments at the moment of the function call.

From our point of view, the important property of this approach to
memory management is both its cost predictability and the possibility
of formalising and certifying it using techniques similar to those
presented in section \ref{compilation-sec}.  Indeed the operations
(1-3) inject short sequences of instructions in the compiled code that
can be executed in constant time as stressed by Tofte and Talpin~\cite{TT97}
(more on this at the end of  section \ref{reg-enriched-sec}). 
 
Because of the operation (3) described above (dispose), the following
memory errors may arise at run-time: (i) write a value in a disposed
region, (ii) access (project) a value in a disposed region, and (iii)
dispose an already disposed region.  To avoid these errors, we
formulate a {\em type and effect} system in the sense of Lucassen and
  Gifford~\cite{LG88} that over-approximates the visible set of
  regions and guarantees safe memory disposal, following Tofte and
  Talpin~\cite{TT97}.  This allows to further extend to the right with
  one more commuting square a typed version of the compilation chain
  described in Table~\ref{chain} and then to include the cost of safe
  memory management in our analysis.

\subsection{Region conventions}
We introduce a syntactic category of {\em regions} $\w{rid}$ with
generic elements $r,r',\ldots$ and a syntactic category of {\em effects} $e$
with generic elements $e,e',\ldots$ An effect is a 
finite set of region variables. 
We keep denoting {\em types} with $A,B,\ldots$ However types may depend
on both regions and effects. Regions can be {\em bound} when occurring either in types
or in terms. In the first case, the binder is a universal quantifier
$\forall r.A$, while in the second it is either a new region allocation operator
$\letall{r}{T}$ or a region  $\lambda$-abstraction.
On the other hand, we stress that the disposal operator $\dis{r}{T}$ is
{\em not} a binder.
Because of the universal quantification and the $\lambda$-abstraction 
both the untyped and the typed region enriched calculi include a notion of 
{\em region substitution}. Note that such a substitution operates on the effects 
contained in the types too and that, as a result, it may reduce the cardinality
of the set of regions which composes an effect. The change of cardinality however, 
can only arise in the untyped calculus. In the typed calculus, 
all the region substitutions are guaranteed to be injective.
We denote with $\frv{A}$ the set of regions occurring free in the type $A$, and
$\frv{\Gamma}$ denotes the obvious extension to type contexts.

\subsection{A region enriched calculus}\label{reg-enriched-sec}
A formalisation of the operations and the related memory errors is given through
the region enriched calculus presented in Tables~\ref{region-predicates} and \ref{lambda-reg}. Notice that an empty tuple is stored in a local variable rather than in a region and that a similar stategy would be adopted for basic data values such as booleans or integers.
The usual formalisation of the operational
semantics relies on a rather concrete
representation of a heap as a (finite domain) function 
mapping regions to stores which
are (finite domain) functions from locations to values 
satisfying some coherence conditions (see, {\em e.g.} \cite{TT97,AFL95,B08}).
In the following, we will take a slightly more abstract 
approach by representing the heap implicitly
as a {\em heap context} $H$. The latter is simply a list of region allocations,
value allocations at a region, and region disposals.

It turns out that it is possible to formulate the coherence conditions
on the memory directly on this list so that we  do not have to
commit to a concrete representation of the heap as the one sketched
above.  A first consequence of this design choice, is that we can
dispense with the introduction of additional syntactic entities like
that of `location' or `address' and consequently avoid the
non-deterministic rules that choose fresh regions or locations (as in,
say, the $\pi$-calculus, $\alpha$-renaming will take care of that).  A
second consequence is that the proof of the simulation property of the
standard calculus by the region-enriched calculus is rather direct.

Continuing the comparison with formalisations found in the literature,
we notice that the fact that region disposal is 
decoupled from allocation avoids the introduction of a special
`disposed' or `free' region which is sometimes used in the operational
semantics to represent the situation where a region becomes
inaccessible  (see, {\em e.g.}, \cite{TT97,AFL95}). 
What we do instead is to keep track of the disposal operation in
the {\em heap context}. 

Finally, let us notice that we certainly take advantage of the fact
that our formalisation of region management targets an intermediate
RTL language where the execution order and the operations of writing
and reading a value from memory are completely explicit.  The
formalisation of region management at the level of the source
language, {\em e.g.}, the $\lambda$-calculus, {\em appears} a bit more
involved because one has to enrich the language with operations that
really refer to the way the language is compiled. For instance, one
has to distinguish between the act of storing a value in memory and
the act of referring to it without exploring its internal
structure. 

Table~\ref{region-predicates} specifies the coherence predicate $\w{Coh}(H,L)$ of the
heap context $H$ relatively to a set of `live' regions $L$. Briefly, a heap context is {\em coherent}
if whenever the context contains an allocation for a tuple in a region, or a disposal of a region, the region in question is alive.
This is defined by induction on the structure of the heap context $H$.

The reduction rules in  Table~\ref{lambda-reg} are a refinement of those
of the value named $\lambda$-calculus described in Table~\ref{lambda-adm}.
The main novelties are that a transition can be fired only if the heap context
is coherent relatively to an empty set of regions in the sense described above 
and moreover that a tuple can be projected 
only if it is allocated in a region which has not been disposed. To formalise this
last property we have refined the definition of the function $E(x)$ which looks for the
value bound to a variable in an evaluation context. The refined function, upon success,
returns both the value and the part of the heap context, say $H$, which has been explored.
Then the predicate $\w{NDis}(r,H)$ defined in Table~\ref{region-predicates} checks
that the region $r$ where the tuple is allocated is not disposed by $H$.
Again, this predicate is defined by induction on the structure of the heap context $H$.

\begin{table}
{\footnotesize
\begin{center}
{\sc Coherent heap context relative to live regions}
\end{center}
\[
\begin{array}{c}

\infer{}{\w{Coh}([~],L)}

\qquad
\infer{\w{Coh}(H,L)}
{\w{Coh}(\lets{x}{()}{H},L)}
\qquad

\infer{\w{Coh}(H,L)\quad r\in L}
{\w{Coh}(\lets{x}{(y^+)\at{r}}{H},L)}  \\ \\

\infer{\w{Coh}(H,L\union\set{r})}
{\w{Coh}(\letall{r}{H},L)}

\qquad \infer{\w{Coh}(H,L\minus\set{r})\quad r\in L}
{\w{Coh}(\dis{r}{H},L)}

\end{array}
\]
\begin{center}
{\sc Region not-disposed in a heap context}
\end{center}
\[
\begin{array}{c}

\infer{}
{\w{NDis}(r,[~])}

\qquad

\infer{\w{NDis}(r,H)}
{\w{NDis}(r,\lets{x}{()}{H})}   

\qquad 
\infer{\w{NDis}(r,H)}
{\w{NDis}(r,\lets{x}{(y^+)\at{r'}}{H})}   \\ \\

\infer{(r=r') \mbox{ or }(r\neq r' \mbox{ and }\w{NDis}(r,H))}
{\w{NDis}(r,\letall{r'}{H})}

\qquad \infer{r\neq r' \quad \w{NDis}(r,H)}
{\w{NDis}(r,\dis{r'}{H})}

\end{array}
\]}
\caption{Coherence predicate on heap contexts}\label{region-predicates}
\end{table}

\begin{table}
{\footnotesize

\begin{center}
{\sc Syntax}
\end{center}
\[
\begin{array}{lll}

\w{rid} &::= r \Alt r' \Alt \cdots   &\mbox{(region identifiers)}\\

C &::= () \Alt (\w{id}^+)\at{\w{rid}} \Alt \prj{i}{\w{id}}                                                       &\mbox{(restricted let-bindable terms)} \\
T &::= @(\w{id},\w{rid}^*,\w{id}^+) \Alt \lets{\w{id}}{C}{T} \Alt \lb{\ell}{T} \Alt \\
       &\qquad \letall{\w{rid}}{T} \Alt \dis{\w{rid}}{T} 
                                         &\mbox{(restricted terms)} \\
P &::= T \Alt  \lets{\w{id}}{\lambda \w{rid}^*,\w{id}^+.T}{P}                                   &\mbox{(programs)} \\
F &::= [~] \Alt \lets{\w{id}}{\lambda \w{rid}^*,\w{id}^+.T}{F}  &\mbox{(function contexts)} \\

H &::= [~] \Alt  \lets{\w{id}}{()}{H} \Alt
\lets{\w{id}}{(\w{id}^+)\at{\w{rid}}}{H} \Alt \\
  &\qquad               \letall{\w{rid}}{H} \Alt
                 \dis{\w{rid}}{H} &\mbox{(heap contexts)}\\

E &::= F[H]       &\mbox{(evaluation contexts)}

\end{array}
\]
\begin{center}
{\sc Reduction rules}
\end{center}
\[
\begin{array}{l}

E[@(x,r'_1,\ldots,r'_m,z_1,\ldots,z_n)]
\arrow
E[[r'_1/r_1,\ldots,r'_m/r_m,z_1/y_1,\ldots,z_n/y_n]T] \\
\qquad \mbox{if }\pi_1(E(x))\equiv \lambda r_1,\ldots,r_m,y_1,\ldots,y_n.T, E\equiv F[H], 
\ \w{Coh}(H,\emptyset)\\ \\

E[\lets{z}{\prj{i}x}{T}]
\arrow
 E[[y_i/z]T]]\\
\qquad \mbox{if }E(x)=((y_1,\ldots,y_n)\at{r},H'), \ 1\leq i \leq n, \ E\equiv F[H],\ \w{Coh}(H,\emptyset),\ \w{NDis}(r,H')
\\\\

E[\lb{\ell}{T}]
\act{\ell}
E[T] 
\qquad \mbox{if }E\equiv F[H], \w{Coh}(H,\emptyset)

\end{array}
\]
\[
\mbox{where: }
\left[
\begin{array}{l}

E(x)=\left\{
\begin{array}{ll}
(V,[~])  &\mbox{if }E=E'[\lets{x}{V}{[~]}] \\
(V,E''[\w{El}]) &\mbox{otherwise if }E=E'[\w{El}], E'(x)=(V,E'') \\
\mbox{undefined} &\mbox{otherwise}
\end{array}\right.\\ \\ 

V ::=() \Alt (\w{id}^*)\at{\w{rid}} \Alt \lambda \w{rid}^*,\w{id}^+.T \\

\w{El} ::= \lets{\w{id}}{V}{[~]} \Alt \letall{\w{rid}}{[~]} \Alt \dis{\w{rid}}{[~]} 
\end{array}
\right]
\]
}
\caption{The region-enriched calculus: $\lambda^{\ell,r}_{h,\w{vn}}$}\label{lambda-reg}
\end{table}

We remark the following  decomposition property of region-enriched programs.

\begin{proposition}[decomposition]\label{decom-prop}
A program $P$ in the region enriched $\lambda$-calculus can be uniquely decomposed
as $F[H[\Delta]]$ where $F$ is a function context, $H$ a heap context, and $\Delta$
is either an application of the shape $@(x,r^*,y^+)$ or a projection of the 
shape $\lets{x}{\prj{i}{y}}{T}$, or a labelling of the shape $\lb{\ell}{T}$.
\end{proposition}

We define an obvious {\em erasure function} on the region-enriched
types, values, and terms that just erases all the region related
pieces of information (please refer to the formal definition in
Table~\ref{region-erasure} of the appendix for details).

Because of the possible memory errors described above, a region
enriched program does not necessarily simulate its region erasure. 

\begin{example}[memory errors]\label{memory-fault-ex}
Consider the following program~$P$ in $\lambda_{h,\w{vn}}^{\ell}$ (not necessarily the result of a compilation):
\[
\begin{array}{lll}

P\equiv &F[@(\w{pair},v_1,v_2)]\\

F \equiv 
        &\s{let} \ \w{prj1}= \lambda x.\lets{y}{\prj{1}{x}}{@(\w{halt},y)} \ \s{in} \\
        &\s{let} \ \w{pair} = \lambda x_1,x_2.\lets{y}{(x_1,x_2)}{@(\w{prj1},y)} \ \s{in} ~[~]~.
\end{array}
\]
One strategy to manage memory regions in $P$ is to allocate a region upon entering the
\w{pair} function and to dispose it just before calling the \w{prj1} function
as in the following program~$P_1$ in $\lambda_{h,\w{vn}}^{\ell,r}$.
\[
\begin{array}{lll}
P_1  &\equiv &F_1[@(\w{pair},v_1,v_2)] \\
F_1    & \equiv     &\s{let} \ \w{prj1}= \lambda x.\lets{z}{\prj{1}{x}}{@(\w{halt},z)} \ \s{in}  \\
       &            &\s{let} \ \w{pair} = \lambda x_1,x_2.\letall{r}{\lets{y}{(x_1,x_2)\at{r}}{\dis{r}{@(\w{prj1},y)}}} \ \s{in} ~[~]~.
      
\end{array}
\]
Unfortunately this strategy leads to a memory error as:
\[
\begin{array}{lll}
P_1 &\trarrow &F_1[H_1[\lets{z}{\prj{1}{y}}{@(\w{halt},z)}]]  \\
H_1   &\equiv   & \letall{r}{ \lets{y}{(v_1,v_2)\at{r}}{\dis{r}{[~]}}}
\end{array}
\]
Formally, $F_1[H_1](y) = ((v_1,v_2)\at{r}, H_2)$, $H_2 = \dis{r}{[~]}$, and 
the predicate $\w{NDis}(r,H_2)$ does {\em not} hold. In plain words,  the problem 
with this strategy is that it disposes the region $r$ before the value $(v_1,v_2)$ allocated
into it is projected. 
A better strategy is to pass the region created in the function
$\w{pair}$ to the function $\w{prj1}$ and let this function dispose the region once the value 
$(v_1,v_2)$ has been projected. This strategy is described by the following program~$P_2$
in $\lambda_{h,\w{vn}}^{\ell,r}$.
\[
\begin{array}{lll}
P_2  &\equiv &F_2[@(\w{pair},v_1,v_2)] \\
F_2   &\equiv   &\s{let} \ \w{prj1}= \lambda r,x.\lets{z}{\prj{1}{x}}{\dis{r}{@(\w{halt},z)}} \ \s{in}  \\
          &    &\s{let} \ \w{pair} = \lambda x_1,x_2.\letall{r}{\lets{y}{(x_1,x_2)\at{r}}{@(\w{prj1},r,y)}} \ \s{in} ~[~]~.
\end{array}
\]
This time the reduction leads to a normal termination:
\[
\begin{array}{lll}
P_2 &\trarrow &F_2[H_2[@(\w{halt},v_1)]] \\
H_2 &\equiv   &\letall{r}{ \lets{y}{(v_1,v_2)\at{r}}{\dis{r}{[~]}}}~.
\end{array}
\]
\end{example}

We conclude this section with an overview of 
a rather standard {\em implementation scheme} of region based memory 
management (see, {\em e.g.}, \cite{M03} for more details).  Initially,
the available memory is partitioned in pages which constitute a free
list. A region is a pointer to a `region descriptor' that contains a
pointer to the beginning and the end of a list of pages and a counter
which gives the amount of memory available in the last page of the
list. A value (a non-empty tuple in our case) is just a pointer to 
a memory address and an access to a value is direct. 
Storing a value in a region means storing the value in the last page
of the list related to the region and updating the region descriptor. 
If the space available is not sufficient, then one or more pages are
taken from the free list and appended to the end of the region 
list and again the region descriptor is updated. This operation 
can be executed in constant time as long as the size of the values
to be allocated can be determined at compile time (which is 
obviously true in our case). Deallocating a region 
means concatenating the list related to the region to the free list.
We refrain from going into the details of the
implementation scheme mentioned above which really belong to the
backend of the compiler. Indeed, the scheme is rather independent 
from the source language (for instance, Christiansen {\em et al.} \cite{CHNV98}
rely on it to implement an object-oriented language) while depending
for its efficiency on the memory organisation of the processor and
possibly the operating system.

\subsection{A type and effect system}
In order to have the simulation property, we require that the
region-enriched program is typable with respect to an enhanced {\em
  type and effect} system described in Table~\ref{type-effect} whose
purpose is precisely to avoid memory errors at run time.  The system
defines judgment (i) $\Gamma \rGives T : e$ read ``restricted term
$T$ has effect $e$ under $\Gamma$'' ; (ii) $\Gamma \rGives C : A$ read
``restricted let-bindable term $C$ has type $A$ under $\Gamma$'' and
(iii) $\Gamma \rGives P : e$ read ``program $P$ has effect $e$ under
$\Gamma$''. The formalisation follows the work of Aiken {\em et al.}
\cite{AFL95} in that allocation and disposal of a region are decoupled
(see also Henglein {\em et al.} \cite{HHN05} for a survey and Boudol
\cite{B08} for a discussion).
Then a region can be disposed only if  the following computation neither accesses nor  disposes it.
Note that in typing values we omit the effect (which is always empty) and
in typing terms we omit the type (which is always the type of results $R$).
The typing rules are designed to maintain several invariants. First,
if a program $P$ has effect $e$ then the set of regions $e$
over-approximates the {\em visible} regions that the program $P$ may dispose 
or access for allocating or reading a value.
Second,  all the region names have been allocated 
(and possibly disposed afterwards).
Third, distinct region names in the program correspond at run time to
different regions, {\em i.e.}, all region substitutions are {\em injective}.
With respect to the system described in Table~\ref{types-adm}, we notice that
we distinguish the  rules for typing an empty and a non-empty tuple as
the former has no effect on the heap. For similar reasons, we split the rule
for typing a value definition in three depending on whether the
value is an empty tuple, a function, or a non-empty tuple (possibly of existential type). Only in the
last case an effect on the heap is recorded. As already mentioned, 
empty tuples do not affect the heap and function
definitions eventually become sequences of assembly language instructions
which are  stored in a statically allocated and read-only 
zone of memory separated by the data memory.

\begin{table}
{\footnotesize
\begin{center}
{\sc Types and effects syntax}
\end{center}
\[
\begin{array}{lll}

e &::=\set{\w{rid},\ldots,\w{rid}}    &\mbox{(effects)} \\

A &::= \w{tid} \Alt \forall \w{rid}^*.A^+ \act{e} R \Alt 
       \times() \Alt \times(\w{A}^+)\at{\w{rid}} \Alt 
        (\exists \w{tid}.A)\at{\w{rid}}          &\mbox{(types)} \\

\end{array}
\]
\begin{center}
{\sc Typing rules}
\end{center}
\[
\begin{array}{cc}

\infer{
\begin{array}{c}
\Gamma,y^+:A^+ \rGives T:e \\
\set{r^*}\inter\frv{\Gamma}=\emptyset \quad 
\frv{\lambda r^*,y^+.T}=\emptyset
\end{array}}
{\Gamma \rGives \lambda r^*,y^+.T: \forall r^*. A^+\act{e} R} 

&\infer{\begin{array}{c} 
      B\equiv \forall r^*_1.A^+\act{e} R \quad
      x:B \in \Gamma \quad  y^+:[r^*/r^*_1]A^+ \in \Gamma \\
       r^* \mbox{ distinct}
\quad \frv{B}=\emptyset
      \end{array}}
{\Gamma \rGives @(x,r^*,y^+):[r^*/r^*_{1}]e}   \\ \\

\infer{x^+:A^+\in \Gamma}
{\Gamma \rGives (x^+)\at{r}: \times(A^+)\at{r} }   

&\infer{\begin{array}{c}
y:\times(A_1,\ldots,A_n)\at{r} \in \Gamma\\ 
1\leq i \leq n\quad \Gamma,x:A_i \rGives T: e 
\end{array}}
{\Gamma \rGives \lets{x}{\prj{i}{y}}{T}: e\union\set{r}}  \\ \\

\infer{x:[B/t]A\in \Gamma}
{\Gamma \rGives (x)\at{r} :(\exists t.A)\at{r}}

&\infer{y:(\exists t.A)\at{r} \in \Gamma\quad
\Gamma,x:A\rGives T:e\quad t\notin \ftv{\Gamma}}
{\Gamma \rGives \lets{x}{\prj{1}{y}}{T}:e\union \set{r}} \\ \\

\infer{}
{\Gamma \rGives (): \times()}

&\infer{\begin{array}{c}
\Gamma \rGives ():A \\
\Gamma,x:A \rGives T:e
\end{array}}
{\Gamma \rGives \lets{x}{()}{T}:e} \\ \\

\infer{\begin{array}{c}
\Gamma \rGives  \lambda r^*,y^+.T:A \\
\Gamma,x:A \rGives P:e
\end{array}}
{\Gamma \rGives \lets{x}{\lambda r^*,y^+.T}{P}:e}

&\infer{\begin{array}{c}
\Gamma \rGives (y^+)\at{r}:A \\
\Gamma,x:A \rGives T:e
\end{array}}
{\Gamma \rGives \lets{x}{(y^+)\at{r}}{T}:e\union\set{r}} \\ \\

\infer{\Gamma \rGives T: e\union\set{r}\quad r\notin \frv{\Gamma},e}
{\Gamma \rGives \letall{r}{T}:e} 

&\infer{\Gamma \rGives T:e\quad r\notin e}
{\Gamma \rGives \dis{r}{T}:e\union\set{r}} \\ \\



\infer{\Gamma \rGives T:e}
{\Gamma \rGives \lb{\ell}{T}: e} 

&\infer{\Gamma \rGives P:e \quad e\subseteq e'}
{\Gamma \rGives P:e'} 

\end{array}
\]
}
\caption{Type and effect system for the region-enriched calculus}\label{type-effect}
\end{table}

\begin{example}[types and effects]\label{type-effect-ex}
Going back to example \ref{memory-fault-ex}, let us assume the types
$t_i:v_i$,  $i=1,2$ and $\w{halt}:t_1\act{\emptyset} R$.
Then the reader may check that the program $P_2$ is typable (has an effect)
assuming the following types for the functions:
\[
\begin{array}{ll}

\w{pair} \ : \  t_1,t_2 \act{\emptyset} R \qquad
&\w{prj1} \ : \ \forall r.\times(t_1,t_2)\at{r} \act{\set{r}} R

\end{array}
\]
On the other hand, any attempt at typing $P_1$ fails trivially because 
the type of the function $\w{prj1}$ must be of the shape $\times(t_1,\ldots)\at{r}\act{\set{r}} R$
and it cannot match the type of the pair allocated by the function $\w{pair}$ in a {\em new} region.
If we fix this problem by abstracting the function $\w{prj1}$ w.r.t. a region so that it has the type
$\forall r.\times(t_1,\ldots)\at{r}\act{\set{r}} R$ we stumble on the main problem, 
namely the function $\w{pair}$ disposes a region which is used in the continuation; 
this is forbidden  by the typing rule for 
region disposal. 
\end{example}

\begin{example}[injective region substitutions]
The soundness and relative simplicity of the type and effect system bear
on the fact that region substitutions are injective. Technically this property
is enforced by the rules for typing an application and an abstraction.
In an application, say $@(x,r^*,y^+)$, 
the region variables $r^*$ are {\em distinct} and the type of $x$ is 
region closed.
In an abstraction, say $\lambda r^*,x^+.T$, the function is region closed.
It is instructive to see what can go wrong if we drop these conditions.
Consider a function $x$ of the following shape with its possible (region closed) type:
\[
x=\lambda r_1,r_2,y.\dis{r_1}{\lets{z}{(y)\at{r_2}}{T}}:
\forall r_1,r_2.\times() \act{\set{r_1,r_2}} R~.
\]
An application $@(x,r,r,y)$ where we pass twice the same region name $r$ will produce
a memory error since we dispose $r$ before writing into it.
A similar phenomenon arises with a function of the following shape and related (region open!) type:
\[
x=\lambda r_1,y.\dis{r_1}{\lets{z}{(y)\at{r_2}}{T}}:
\forall r_1.\times() \act{\set{r_1,r_2}} R~.
\]
Then an application $@(x,r_2,y)$ where we pass a region name 
which is free in the type of the function will also produce a memory error.
As a final example, consider a function
\[
x=\lambda r_1,y. \lets{z}{()}{@(y,r_1,r_2,z)}
\] 
with a free region variable $r_2$. 
Note that $r_1,r_2$ may not appear in the type of the
function $x$ because, {\em e.g.}, $y$ makes no use of them.
Then  if we apply $x$ as in $@(x,r_2,y)$ we end up with an application
$@(y,r_2,r_2,z)$ which is not typable because it violates the condition
that all the region variables passed as arguments are distinct.
\end{example}

\subsection{Properties of the type and effect system}
We notice that the region erasing function preserves typing.

\begin{proposition}[region erasure]\label{reg-erasure-prop}
 If $\Gamma \rGives P:e$ then $\rer{\Gamma}\aGives \rer{P}$.
\end{proposition}

In the other direction, it is always possible to insert region
annotations in a typable program of $\lambda_{h,\w{vn}}^{\ell}$ so as to
produce a typable region-enriched program. A simple but {\em not} very
interesting way to do this is to allocate one region at the very
beginning of the computation which is never disposed and which is
shared by all functions.

\begin{proposition}[region enrichment]\label{reg-enrich-prop}
Let $\Gamma_0$ be a type context such that if $x:A\in \Gamma_0$
then $A$ is not a type of the shape $\times(B^+)$ or $\exists t.B$.
If $\Gamma_0 \aGives P$ then it is always possible to find
a region enriched typable program $P'$ such $\rer{P'}\equiv P$.
\end{proposition}

Fortunately, more interesting strategies are available; we refer to
Aiken {\em et al.} \cite{AFL95} for their description and for an encouraging experimental
evaluation and to Henglein {\em et al.} \cite{HHN05} for a survey of region inference techniques.  For our purposes, it is enough to know that it is always
possible to define a compilation function $\cl{C}_{\w{rg}}$ from the
typed $\lambda^{\ell}_{h,\w{vn}}$-calculus to the typed
$\lambda^{\ell,r}_{h,\w{vn}}$-calculus which is a right inverse of the
region erasing function, {\em i.e.}, $\rer{\rg{P}}\equiv P$, and which
commutes with the label erasure functions, {\em i.e.}, $\er{\rg{P}} \equiv
\rg{\er{P}}$. 
Also, following remark \ref{halt-rmk}, we notice that the typing context of the
compiled code satisfies the conditions in proposition \ref{reg-enrich-prop}
provided that if $\Gamma$ is the typing context of the source code and 
$x:A\in \Gamma$ then the type $A$ is not of the shape $\times(B^+)$.

Next we remark that the type system entails the coherence
of the heap context of a program; this leads to the following {\em progress} property.

\begin{proposition}[progress]\label{progress-prop}
Let $P$ be a typable program in the region enriched calculus such that $\frv{P}=\emptyset$.
Then $P$ decomposes as $F[H[\Delta]]$  (proposition \ref{decom-prop})
and 
either  (i) $P$ reduces or (ii) $\Delta$ has the shape
$@(x,r^*,y^+)$ or $\lets{y}{\prj{i}{x}}{T}$, where $x\in \fv{P}$.
\end{proposition}

Of course, we must also prove that the region enriched types are preserved by reduction.

\begin{proposition}[subject reduction, types and effects]\label{type-eff-subjred-prop}
If $\Gamma \rGives P:e$ and $P\arrow P'$ then $\Gamma \rGives P':e$.
\end{proposition}

Finally, we can show that a well-typed region enriched program does indeed simulate
its region erasure.

\begin{theorem}[region simulation] \label{reg-sim-thm}
If $\Gamma \rGives P:e$, $\frv{P}=\emptyset$, and $\rer{P} \act{\alpha} Q$ then 
$P\act{\alpha} P'$ and $\rer{P'}\equiv Q$.
\end{theorem}


\section{Conclusion}
We have shown that our approach, that we call the `labelling'
approach, can be used to obtain certified execution costs on
functional programs following a standard compilation chain which
composes well with the back-end of a moderately optimising $\C$
compiler. The technique allows to compute the cost of the compiled
code while reasoning abstractly at the level of the source language
and it accounts precisely for the cost of memory management for a
particular memory management strategy that uses regions.  To provide
technical evidence for this claim has required to have an in-depth and
sometimes novel look at the formal properties of the compilation
chain; notable examples are the commutation property of the CPS
transformation and the simulation property for the hoisting and the
region aware transformations.


\paragraph{Acknowledgements}
The authors would like to thank the anonymous reviewers for their
valuable comments and suggestions that significantly helped to improve
this paper, as well as Anindya Banerjee and Olivier Danvy for their 
efficient work in the edition process.

\appendix

\section{Proofs}\label{proofs-sec}
We outline the proofs of the results we have stated.

\subsection*{Proof of  proposition \ref{cps-commutation} [CPS commutation]}
The proof takes the following steps:

\begin{enumerate}

\item  We remark that if $V$ is a value in $\lambda^\ell$ and 
$K$ a continuation in $\lambda^\ell_{cps}$ then so are $\er{V}$ and $\er{K}$.
The proof is a direct induction on the structure of $V$ and $K$, respectively.

\item For all values $V$ and terms $M$ of the $\lambda^\ell$-calculus, we check that:
\[
\er{[V/x]M}\equiv [\er{V}/x]\er{M}~.
\]
The proof proceeds by induction on the structure of $M$.

\item We notice that $\lambda x.(x\sco K) \equiv K$ holds,
for all continuations $K$ such that $K$ is an abstraction.

\item For all terms $M$ and continuations $K$ such that either
$M\in W_0$ and $K$ is an abstraction or $M\in W_1$ the following holds:
\[
\er{M\sco K} \equiv \er{M}\sco \er{K}~.
\]
We proceed by induction on $M$. 

\begin{description}

\item[$x$] We expand the definition of $x\sco K$ depending on whether 
$K$ is a variable or a function and we rely on step 2.

\item[$\lambda x^+.M$]
We have $\lambda x^+.M\in W_1$ and $M\in W_1$. 
We analyse $\lambda x^+.M\sco K$ depending on whether $K$ is a variable or a function
and we apply the inductive hypothesis on $M$ and step 2. Notice that
it is essential that $M\in W_1$ to apply the inductive hypothesis.

\item[$@(M_0,\ldots,M_n)$]
We know $M_0,\ldots,M_n\in W_0$. We apply the inductive hypothesis on $M_n,\ldots, M_0$
to conclude that:
\[
\begin{array}{l}

\er{@(M_0,\ldots,M_n)}\sco \er{K} \\
\equiv \er{M_0}\sco \lambda x_0.\ldots \er{M_n} \sco \lambda x_n.@(x_0,\ldots,x_n,\er{K}) \\
\equiv \er{M_0}\sco \lambda x_0.\ldots \er{M_n\sco \lambda x_n.@(x_0,\ldots,x_n,K)} \\
\equiv \cdots \\
\equiv \er{M_0\sco \lambda x_0.\ldots M_n\sco \lambda x_n.@(x_0,\ldots,x_n,K)} \\
\equiv \er{@(M_0,\ldots,M_N)\sco K}~.

\end{array}
\]

\item[$\lb{\ell}{M}$]
We know that if $\lb{\ell}{M}\in W_i$ then $M\in W_i$ and 
we apply the inductive hypothesis on $M$.

\item[$\plb{\ell}{M}$]
By definition, we must have $\plb{\ell}{M}\in W_0$.
Hence $K$ is a function and $M\in W_0$.
Then we apply the inductive hypothesis on $M$ and step 3.

\item[$(M_1,\ldots,M_n)$]
We know that $M_i \in W_0$ for $i=1,\ldots,n$.
First we notice that:
\[
\er{\lambda x_n.(x_1,\ldots,x_n)\sco K} \equiv  \lambda x_n.(x_1,\ldots,x_n)\sco \er{K}~.
\]
Then we apply the inductive hypothesis on $M_n,\ldots,M_0$ to conclude that:
\[
\begin{array}{ll}

\er{(M_1,\ldots,M_n)}\sco \er{K} \\
\equiv \er{M_1}\sco \lambda x_1\ldots \er{M_n}\sco \lambda x_n.(x_1,\ldots,x_n)\sco \er{K} \\
\equiv  \er{M_1}\sco \lambda x_1\ldots \er{M_n}\sco  \er{\lambda x_n.(x_1,\ldots,x_n)\sco K} \\
\equiv \er{M_1}\sco \lambda x_1\ldots \er{M_n\sco \lambda x_n.(x_1,\ldots,x_n)\sco K} \\
\equiv \cdots \\
\equiv \er{M_1\sco \lambda x_1\ldots M_n\sco \lambda x_n.(x_1,\ldots,x_n)\sco K} \\
\equiv \er{(M_1,\ldots,M_n)\sco K}~.
\end{array}
\]

\item[$\prj{i}{M}$]
We know $M\in W_0$.
We observe that $\er{y\sco K} \equiv y\sco \er{K}$.
Then we apply the inductive hypothesis on $M$ to conclude that:
\[
\begin{array}{l}

\er{\prj{i}{M}}\sco \er{K} \\
\equiv \prj{i}{\er{M}}\sco \er{K} \\
\equiv \er{M}\sco \lambda x.\lets{y}{\prj{i}{x}}{y\sco \er{K}} \\
\equiv \er{M}\sco \er{\lambda x.\lets{y}{\prj{i}{x}}{y\sco K}} \\
\equiv \er{M\sco \lambda x.\lets{y}{\prj{i}{x}}{y\sco K}} \\
\equiv \er{\prj{i}{M}\sco K}~.

\end{array}
\]

\item[$\lets{x}{N}{M}$]
If $\lets{x}{N}{M}\in W_i$ then we know $N\in W_0$ and $M\in W_i$.
We apply the inductive hypothesis on $N$ and $M$ to conclude that:
\[
\begin{array}{l}
\er{\lets{x}{N}{M}\sco K} \\
\equiv \er{N\sco \lambda x.(M\sco K)} \\
\equiv \er{N}\sco \lambda x.\er{M\sco K} \\
\equiv \er{N}\sco \lambda x.\er{M}\sco \er{K} \\
\equiv \er{\lets{x}{N}{M}}\sco \er{K}~.
\end{array}
\]
\end{description}


\item Then we prove the assertion for $M\in W_0$ as follows:
\[
\begin{array}{lll}

\er{\cps{M}} &\equiv \er{M\sco \lambda x.@(\w{halt},x)} &\mbox{(by definition)}\\
             &\equiv \er{M}\sco \lambda x.@(\w{halt},x)  &\mbox{(by point 4)}\\
             &\equiv \cps{\er{M}}                  &\mbox{(by definition).} 
\end{array}
\]
\qed
\end{enumerate}

\subsection*{Proof of  proposition \ref{cps-simulation} [CPS simulation]}
The proof takes the following steps.

\begin{enumerate}

\item We show that for all values $V$, terms $M$, and continuations $K\neq x$:
\[
[V/x]M\sco [\psi(V)/x]K \equiv [\psi(V)/x](M\sco K)~.
\]
We proceed by induction on $M$.

\begin{description}

\item[$M$ is a variable.] By case analysis: $M\equiv x$ or $M\equiv y\neq x$.

\item[$\lambda z^+.M$] By case analysis on $K$ which is either a variable or a function.
We develop the second case with $K\equiv \lambda y.N$. 
We observe:
\[
\begin{array}{l}

[V/x](\lambda z^+.M)\sco [\psi(V)/x]K \\
\equiv [\lambda z^+,k.([V/x]M\sco k)/y][\psi(V)/x]N \\
\equiv [\lambda z^+,k.[\psi(V)/x](M\sco k)/y][\psi(V)/x]N \\
\equiv [\psi(V)/x][\lambda z^+,k.(M\sco k)/y]N \\
\equiv [\psi(V)/x]((\lambda z^+.M)\sco K)~.

\end{array}
\]

\item[$@(M_0,\ldots,M_n)$]
We apply the inductive hypothesis on $M_0,\ldots,M_n$ as follows:
\[
\begin{array}{l}

[\psi(V)/x](@(M_0,\ldots,M_n)\sco K) \\
\equiv [\psi(V)/x](M_0\sco \lambda x_0\ldots M_n\sco \lambda x_n.@(x_0,\ldots,x_n,K)) \\
\cdots \\
\equiv [V/x]M_0\sco \lambda x_0\ldots [\psi(V)/x](M_n\sco \lambda x_n.@(x_0,\ldots,x_n,K)) \\
\equiv [V/x]M_0\sco \lambda x_0\ldots [V/x]M_n\sco \lambda x_n.@(x_0,\ldots,x_n,[\psi(V)/x]K) \\
\equiv [V/x]@(M_0,\ldots,M_n)\sco [\psi(V)/x]K~.

\end{array}
\]
Note that in this case the substitution $[\psi(V)/x]$ may operate on the continuation.
The remaining cases (pairing, projection, let, pre and post labelling) 
follow a similar pattern and are omitted.
\end{description}

\item
The evaluation contexts for the $\lambda^\ell$-calculus 
described in Table~\ref{lambda}
can also be specified `bottom up' as follows:
\[
\begin{array}{lll}
E  &::= &[~] \Alt E[@(V^*,[~],M^*)]  \Alt E[\lets{\w{id}}{[~]}{M}] \Alt
 E[(V^*,[~],M^*)] \Alt \\ 
&&E[\prj{i}{[~]}] \Alt E[\plb{\ell}{[~]}]~.
\end{array}
\]
Following this specification, we associate a continuation $K_E$ with an evaluation context as follows:
\[
\begin{array}{lll}

K_{[~]} &= &\lambda x.@(\w{halt},x) \\

K_{E[@(V^*,[~],M^*)]} 
&= 
&\lambda x.M^*\sco \lambda y^*.@(\psi(V)^*,x,y^*,K_E) \\

K_{E[\lets{x}{[~]}{N}]}
&=
&\lambda x.N\sco K_E \\

K_{E[(V^*,[~],M^*)]}
&=
&\lambda x.M^*\sco \lambda y^*.(\psi(V)^*,x,y^*)\sco K_E \\

K_{E[\prj{i}{[~]}}
&=
&\lambda x.\lets{y}{\prj{i}{x}}{y\sco K_E} \\

K_{E[\plb{\ell}{[~]}]} 
&=
&\lambda x.\lb{\ell}{x\sco K_E}

\end{array}
\]
where $M^*\sco \lambda x^*.N$ stands for $M_0\sco \lambda x_0\ldots M_n\sco \lambda x_n.N$ with $n\geq 0$.

\item For all terms $M$ and evaluation contexts $E, E'$ we prove by induction
on the evaluation context $E$ that the following holds:
\[
E[M]\sco K_{E'} \equiv M\sco K_{E'[E]}~.
\]
For instance, we detail the case where the context has the shape $E[@(V^*,[~],M^*)]$.
\[
\begin{array}{ll}

E[@(V^*,[M],M^*)]\sco K_{E'} \\
\equiv @(V^*,[M],M^*)\sco  K_{E'[E]}  &\mbox{(by inductive hypothesis)} \\
\equiv M\sco \lambda x.M^*\sco \lambda x^*.@(\psi(V)^*,x,x^*,K_{E'[E]}) \\
\equiv M\sco  K_{E'[E[@(V^*,[~],M^*)]]}~.
\end{array}
\] 

\item For all terms $M$, continuations $K, K'$, and variable $x\notin \fv{M}$ 
we prove by induction on $M$ and case analysis that the following  holds:
\[
\begin{array}{c}

[K/x](M\sco K') \left\{
\begin{array}{ll}
\arrow  M\sco K'  &\mbox{if }K \mbox{ abstraction}, M\mbox{ value}, K'=x \\
\equiv (M\sco [K/x]K') &\mbox{ otherwise.}
\end{array}\right.
\end{array}
\]

\item Finally, we prove the assertion by proceeding by case analysis 
on the reduction rule.

\begin{itemize}

\item $E[@(\lambda x^+.M,V^+)] \arrow E[[V^+/x^+]M]$.
We have:
\[
\begin{array}{l}
E[@(\lambda x^+.M,V^+)]\sco K_{[~]}  \\
\equiv @(\lambda x^+.M,V^+)\sco K_E  \\
\equiv @(\lambda x^+,k.M\sco k,\psi(V)^+,K_E) \\
\arrow [K_E/k,\psi(V)^+/x^+](M\sco k) \\
\equiv [K_E/k]([V^+/x^+]M\sco k) \\
\trarrow [V^+/x^+]M\sco  K_E \\
\equiv E[[V^+/x^+]M]\sco K_{[~]}~.
\end{array}
\]

\item $E[\lets{x}{V}{M}] \arrow E[[V/x]M]$.
We have:
\[
\begin{array}{l}
E[\lets{x}{V}{M}]\sco K_{[~]} \\
\equiv \lets{x}{V}{M}\sco K_E \\
\equiv V\sco \lambda x.(M\sco K_E) \\
\equiv [\psi(V)/x](M\sco K_E) \\
\equiv [V/x]M\sco K_E \\
\equiv E[[V/x]M]\sco K_{[~]}~.
\end{array}
\]

\item $E[\prj{i}{V}] \arrow E[V_i]$, where $V\equiv (V_1,\ldots,V_n)$
and $1\leq i \leq n$. We have:
\[
\begin{array}{l}
E[\prj{i}{V}]\sco K_{[~]} \\
\equiv \prj{i}{V}\sco K_E \\
\equiv V\sco \lambda x.\lets{y}{\prj{i}{x}}{y\sco K_E} \\
\equiv \lets{y}{\prj{i}{\psi(V_1),\ldots,\psi(V_n)}}{y\sco K_E} \\
\arrow [\psi(V_i)/y](y\sco K_E) \\
\equiv  V_i\sco K_E \\
\equiv E[V_i]\sco K_{[~]}~.
\end{array}
\]

\item $E[\lb{\ell}{M}] \act{\ell} E[M]$. We have:
\[
\begin{array}{l}
E[\lb{\ell}{M}]\sco K_{[~]} \\
\equiv \lb{\ell}{M}\sco  K_{E} \\
\equiv \lb{\ell}{(M\sco K_E)} \\
\act{\ell} (M\sco K_E) \\
\equiv E[M]\sco K_{[~]}~.
\end{array}
\]

\item $E[\plb{\ell}{V}] \act{\ell} E[V]$. We have:
\[
\begin{array}{l}
E[\plb{\ell}{V}]\sco K_{[~]} \\
\equiv \plb{\ell}{V}\sco  K_{E} \\
\equiv V\sco \lambda x.\lb{\ell}{x\sco K_E} \\
\equiv \lb{\ell}{(V\sco K_E)} \\
\act{\ell} V\sco K_E \\
\equiv E[V]\sco K_{[~]}~.

\end{array}
\]
\end{itemize}
\qed
\end{enumerate}

\subsection*{Proof of  proposition \ref{ad-commutation} [VN commutation]}
 \Proofitemf{(1)}
We show that for every $P$ which is either a term or a value of the $\lambda^{\ell}_{cps}$-calculus
the following properties hold:

\begin{description}

\item[A] If $P$ is a term then  $\rb{\vn{P}}\equiv P$.

\item[B] If $P$ is a value then for any term $N$, $\rb{\evn{P}{x}[N]}\equiv [P/x]\rb{N}$.

\end{description}

We prove the two properties at once by induction on the structure of $P$.

\begin{description}

\item[$@(x,x^+)$] We are in case A and by definition we have:
\[
\rb{\vn{@(x,x^+)}} \equiv  \rb{@(x,x^+)} \equiv  @(x,x^+)~.
\]

\item[$@(x^*,V,V^*),V\neq \w{id}$] Again in case A. We have:
\[
\begin{array}{ll}
\rb{\vn{@(x^*,V,V^*)}} \\
\equiv \rb{\evn{V}{y}[\vn{@(x^*,y,V^*)}]}     \\
\equiv [V/y]\rb{\vn{@(x^*,y,V^*)}}       &\mbox{(by ind. hyp. on B)} \\
\equiv [V/y]@(x^*,y,V^*)                 &\mbox{(by ind. hyp. on A)} \\
\equiv @(x^*,V,V^*)~.
\end{array}
\]

\item[$\lets{x}{\prj{i}{z}}{M}$] Again in case A. We have:
\[
\begin{array}{ll}
\rb{\vn{\lets{x}{\prj{i}{z}}{M}}} \\
\equiv \rb{\lets{x}{\prj{i}{z}}{\vn{M}}} \\
\equiv \lets{x}{\prj{i}{z}}{\rb{\vn{M}}} \\
\equiv \lets{x}{\prj{i}{z}}{M} &\mbox{(by ind. hyp. on A)}~.
\end{array}
\]

\item[$\lets{x}{\prj{i}{V}}{M}, V\neq \w{id}$] Again in case A. We have:
\[
\begin{array}{ll}
\rb{\vn{\lets{x}{\prj{i}{V}}{M}}} \\
\equiv \rb{\evn{V}{y}[\lets{x}{\prj{i}{y}}{\vn{M}}]} \\
\equiv [V/y]\rb{\lets{x}{\prj{i}{y}}{\vn{M}}}  &\mbox{(by ind. hyp. on B)}\\
\equiv [V/y]\lets{x}{\prj{i}{y}}{\rb{\vn{M}}}  \\
\equiv [V/y]\lets{x}{\prj{i}{y}}{M} &\mbox{(by ind. hyp. on A)} \\
\equiv \lets{x}{\prj{i}{V}}{M}~.
\end{array}
\]

\item[$\lb{\ell}{M}$] Last case for A. We have:
\[
\begin{array}{ll}
\rb{\vn{\lb{\ell}{M}}}  \\
\equiv \rb{\lb{\ell}{\vn{M}}}  \\
\equiv \lb{\ell}{\rb{\vn{M}}} \\
\equiv \lb{\ell}{M} &\mbox{(by ind. hyp. on A)}~.
\end{array}
\]

\item[$\lambda y^+.M$] We now turn to case B. We have:
\[
\begin{array}{ll}
\rb{\evn{\lambda y^+.M}{x}[N]} \\
\equiv \rb{\lets{x}{\lambda y^+.\vn{M}}{N}} \\
\equiv [\rb{\lambda y^+.\vn{M}}/x]\rb{N} \\
\equiv [\lambda y^+.\rb{\vn{M}}/x]\rb{N} \\
\equiv [\lambda y^+.M/x]\rb{N} &\mbox{(by ind. hyp. on A)}~.

\end{array}
\]

\item[$(y^*)$] Again in case B. We have:
\[
\begin{array}{ll}
\rb{\evn{(y^*)}{x}[N]} \\
\equiv \rb{\lets{x}{(y^*)}{N}} \\
\equiv [(y^*)/x]\rb{N}~.
\end{array}
\]

\item[$(y^*,V,V^*), V\neq \w{id}$] Last case for B. We have:
\[
\begin{array}{ll}
\rb{\evn{(y^*,V,V^*)}{x}[N]} \\
\equiv \rb{\evn{V}{z}[\evn{(y^*,z,V^*)}{x}[N]]} \\
\equiv [V/z]\rb{\evn{(y^*,z,V^*)}{x}[N]} &\mbox{(by ind. hyp. on B)} \\
\equiv [V/z]([(y^*,z,V^*)/x]\rb{N})   &\mbox{(by ind. hyp. on B)} \\
\equiv  [(y^*,V,V^*)/x]\rb{N}~.
\end{array}
\]
\end{description}

\Proofitem{(2)} The proof is similar to the previous one.
We show that for every $P$ which is either a term or a value of the $\lambda^{\ell}_{cps}$-calculus
the following properties hold:

\begin{description}

\item[A] If $P$ is a term then  $\er{\vn{P}}\equiv \vn{\er{P}}$.

\item[B] If $P$ is a value then for any term $N$, $\er{\evn{P}{x}[N]}\equiv \evn{\er{P}}{x}[\er{N}]$.

\end{description}

We prove the two properties at once by induction on the structure of $P$.

\begin{description}

\item[$@(x,x^+)$] We are in case A and by definition we have:
\[
\er{\vn{@(x,x^+)}} \equiv \er{@(x,x^+)} \equiv  @(x,x^+) \equiv \vn{\er{@(x,x^+)}}~.
\]

\item[$@(x^*,V,V^*),V\neq \w{id}$] Again in case A. We have:
\[
\begin{array}{ll}

\er{\vn{@(x^*,V,V^*)}} \\
\equiv \er{\evn{V}{y}[\vn{@(x^*,y,V^*)}]} \\
\equiv \evn{\er{V}}{y}[\er{\vn{@(x^*,y,V^*)}}] &\mbox{(by ind. hyp. on B)} \\
\equiv \evn{\er{V}}{y}[\vn{\er{@(x^*,y,V^*)}}] &\mbox{(by ind. hyp. on A)} \\
\equiv \vn{\er{@(x^*,V,V^*)}}~.

\end{array}
\]

\item[$\lets{x}{\prj{i}{z}}{M}$] Again in case A. We have:
\[
\begin{array}{ll}

\er{\vn{\lets{x}{\prj{i}{z}}{M}}} \\
\equiv \er{\lets{x}{\prj{i}{z}}{\vn{M}}} \\
\equiv \lets{x}{\prj{i}{z}}{\er{\vn{M}}} \\
\equiv  \lets{x}{\prj{i}{z}}{\vn{\er{M}}}  &\mbox{(by ind. hyp. on A)} \\
\equiv \vn{\er{ \lets{x}{\prj{i}{z}}{M}}}~.

\end{array}
\]

\item[$\lets{x}{\prj{i}{V}}{M}, V\neq \w{id}$] Again in case A. We have:
\[
\begin{array}{ll}

\er{\vn{\lets{x}{\prj{i}{V}}{M}}} \\
\equiv \er{\evn{V}{z}[\lets{x}{\prj{i}{z}}{\vn{M}}]} \\
\equiv  \evn{\er{V}}{z}[\lets{x}{\prj{i}{z}}{\er{\vn{M}}}]  &\mbox{(by ind. hyp. on B)}\\
\equiv  \evn{\er{V}}{z}[\lets{x}{\prj{i}{z}}{\vn{\er{M}}}]   &\mbox{(by ind. hyp. on A)} \\
\equiv  \vn{\er{\lets{x}{\prj{i}{V}}{M}}}~.

\end{array}
\]

\item[$\lb{\ell}{M}$] Last case for A. We have:
\[
\begin{array}{ll}
\er{\vn{\lb{\ell}{M}}} \\
\equiv \er{\lb{\ell}{\vn{M}}} \\
\equiv \er{\vn{M}} \\
\equiv \vn{\er{M}}  &\mbox{(by ind. hyp. on A)} \\
\equiv \vn{\er{\lb{\ell}{M}}}~.

\end{array}
\]

\item[$\lambda y^+.M$] We now turn to case $B$. We have:
\[
\begin{array}{ll}

\er{\evn{\lambda y^+.M}{x}[N]} \\
\equiv \er{\lets{x}{\lambda y^+.\vn{M}}{N}} \\
\equiv \lets{x}{\lambda y^+.\er{\vn{M}}}{\er{N}} \\
\equiv \lets{x}{\lambda y^+.\vn{\er{M}}}{\er{N}}  &\mbox{(by ind. hyp. on A)} \\
\equiv \evn{\er{\lambda y^+.M}}{x}[\er{N}]~.

\end{array}
\]

\item[$(y^*)$] Again in case B. We have:
\[
\begin{array}{ll}

\er{\evn{(y^*)}{x}[N]} \\
\equiv \er{\lets{x}{(y^*)}{N}} \\
\equiv \lets{x}{(y^*)}{\er{N}} \\
\equiv \evn{\er{(y^*)}}{x}[\er{N}]~.

\end{array}
\]

\item[$(y^*,V,V^*), V\neq \w{id}$] Last case for B. We have:
\[
\begin{array}{ll}
\er{\evn{(y^*,V,V^*)}{x}[N]} \\
\equiv \er{\evn{V}{z}[\evn{(y^*,z,V^*)}{x}[N]]} \\
\equiv \evn{\er{V}}{x}[\er{\evn{(y^*,z,V^*)}{x}[N]}]  &\mbox{(by ind. hyp. on B)} \\
\equiv \evn{\er{V}}{x}[\evn{\er{(y^*,z,V^*)}}{x}[\er{N}]]  &\mbox{(by ind. hyp. on B)} \\
\equiv \evn{\er{(y^*,V,V^*)}}{x}[\er{N}]~.

\end{array}
\] 
~\qed
\end{description}

\subsection*{Proof of  proposition \ref{ad-simulation} [VN simulation]}
First we fix some notation. 
We associate a substitution $\sigma_E$ with an evaluation context $E$ of the $\lambda^\ell_{cps,\w{vn}}$-calculus
as follows:
\[
\begin{array}{ll}
\sigma_{[~]} = \w{Id}
&\sigma_{\lets{x}{V}{E}} = [\rb{V}/x]\comp \sigma_E~.
\end{array}
\]
Then we prove the property by case analysis.

\begin{itemize}

\item If $\rb{N}\equiv @(\lambda y^+.M,V^+) \arrow [V^+/y^+]M$ 
then $N\equiv E[@(x,x^+)]$, $\sigma_E(x)\equiv \lambda y^+.M$, and 
$\sigma_E(x^+)\equiv V^+$. 

Moreover, $E\equiv E_1[\lets{x}{\lambda y^+.M'}{E_2}]$ and
$\sigma_{E_{1}}(\lambda y^+.M') \equiv \lambda y^+.M$.

Therefore, $N\arrow E[[x^+/y^+]M']\equiv N'$ and we check that
$\rb{N'}\equiv \sigma_E([x^+/y^+]M')\equiv [V^+/y^+]M$.

\item  If $\rb{N}\equiv \lets{x}{\prj{i}{(V_1,\ldots,V_n)}}{M} \arrow [V_i/x]M$ 
then $N\equiv E[\lets{x}{\prj{i}{y}}{N''}]$, $\sigma_E(y)\equiv (V_1,\ldots,V_n)$, and 
$\sigma_E(N'')\equiv M$. 

Moreover, $E\equiv E_1[\lets{y}{(z_1,\ldots,z_n)}{E_2}]$ and
$\sigma_{E_{1}}(z_1,\ldots,z_n) \equiv (V_1,\ldots,V_n)$.

Therefore, $N\arrow E[[z_i/x]N'']\equiv N'$ and we check that
$\rb{N'} \equiv \sigma_E([z_i/x]N'')\equiv [V_i/x]M$.

\item If $\rb{N}\equiv \lb{\ell}{M} \act{\ell} M$ then
$N\equiv E[\lb{\ell}{N''}]$ and $\sigma_E(N'')\equiv M$.
We conclude by observing that
$N\act{\ell}E[N'']$. \qed
\end{itemize}

\subsection*{Proof of  proposition \ref{cc-commutation} [CC commutation]}
 This is a simple induction on the structure of the term $M$.

\begin{description}

\item[$@(x,y^+)$] We have:
\[
\begin{array}{ll}

\er{\cc{@(x,y^+)}} \\
\equiv \er{\lets{(c,e)}{x}{@(c,e,y^+)}} \\
\equiv \lets{(c,e)}{x}{@(c,e,y^+)} \\
\equiv \cc{@(x,y^+)}\\
\equiv \er{ \cc{@(x,y^+)}}~.
\end{array}
\]

\item[$\lets{x}{C}{M}$, $C$ not a function] We have:
\[
\begin{array}{ll}

\er{\cc{\lets{x}{C}{M}}} \\
\equiv \er{\lets{x}{C}{\cc{M}}} \\
\equiv \lets{x}{C}{\er{\cc{M}}} \\
\equiv  \lets{x}{C}{\cc{\er{M}}} &\mbox{(by ind. hyp.)} \\
\equiv \cc{\er{ \lets{x}{C}{M}}}~.

\end{array}
\]

\item[$\lets{x}{\lambda x^+.N}{M}, \fv{\lambda x^+.N}=\set{z_1,\ldots,z_k}$] We have:
\[
\begin{array}{ll}

\er{\cc{\lets{x}{\lambda x^+.N}{M}}} \\
\equiv 
\w{er}( \ \s{let} \ c=\lambda e,x^+.\s{let} \ (z_1,\ldots,z_k)=e \ \s{in} \ \cc{N}  \ \s{in} \\
\qquad \s{let} \ e = (z_1,\ldots,z_k), x=(c,e) \ \s{in} \ \cc{M} \ ) \\
\equiv 
\s{let} \ c=\lambda e,x^+.\s{let}\ (z_1,\ldots,z_k)=e \ \s{in} \ \er{\cc{N}}  \ \s{in} \\
\qquad \s{let} \ e = (z_1,\ldots,z_k),x=(c,e) \ \s{in} \ \er{\cc{M}} \\
\equiv 
\s{let} \ c=\lambda e,x^+.\s{let} \ (z_1,\ldots,z_k)=e \ \s{in} \ \cc{\er{N}}  \ \s{in} \\
\qquad \s{let} \ e = (z_1,\ldots,z_k),x=(c,e) \ \s{in} \ \cc{\er{M}} &\mbox{(by ind. hyp.)} \\
\equiv \cc{\er{\lets{x}{\lambda x^+.N}{M}}}~.
\end{array}
\]

\item[$\lb{\ell}{M}$] We have:
\[
\begin{array}{ll}

\er{\cc{\lb{\ell}{M}}} \\
\equiv \er{\lb{\ell}{\cc{M}}} \\
\equiv \er{\cc{M}} \\
\equiv \cc{\er{M}} &\mbox{(by ind. hyp.)} \\
\equiv \cc{\er{\lb{\ell}{M}}}~.

\end{array}
\]
~\qed
\end{description}

\subsection*{Proof of  proposition \ref{cc-simulation} [CC simulation]}
As a first step we check that  the closure conversion function commutes with name 
substitution:
\[
\cc{[x/y]M} \equiv [x/y]\cc{M}~.
\]
This is a direct induction on the structure of the term $M$.
Then we extend the closure conversion function to contexts as follows:
\[
\begin{array}{ll}

\cc{[~]} &=[~] \\

\cc{\lets{x}{(y^*)}{E}} &= \lets{x}{(y^*)}{\cc{E}} \\

\cc{\lets{x}{\lambda x^+.M}{E}} &=      \s{let} \ c=\lambda e,x^+.\s{let} \ (z_1,\ldots,z_k)=e \ \s{in} \ \cc{M}  \ \s{in} \\
                                &\quad \s{let} \ e = (z_1,\ldots,z_k),x=(c,e) \ \s{in} \ \cc{E} \\
                                &\mbox{where: }\fv{\lambda x^+.M}=\set{z_1,\ldots,z_k} ~.
\end{array}
\]
We note that for any evaluation context $E$, $\cc{E}$ is again an evaluation context, and 
moreover for any term $M$ we have:
\[
\cc{E[M]} \equiv \cc{E}[\cc{M}]~.
\]
Finally we prove the simulation property by case analysis of the reduction rule being 
applied.

\begin{itemize}

\item Suppose $M\equiv E[@(x,y^+)] \arrow E[[y^+/x^+]M]$ where $E(x)=\lambda x^+.M$ 
and $\fv{\lambda x^+.M}=\set{z_1,\ldots,z_k}$.
Then:
\[
\cc{E[@(x,y^+)]} \equiv \cc{E}[\lets{(c,e)}{x}{@(c,e,y^+)}]
\]
with $\cc{E}(x)=(c,e)$,
$\cc{E}(c)=\lambda e,x^+.\s{let} \ (z_1,\ldots,z_k)=e \ \s{in} \ \cc{M}$ and
$\cc{E}(e)=(z_1,\ldots,z_k)$. Therefore:
\[
\begin{array}{l}
\cc{E}[\lets{(c',e')}{x}{@(c',e',y^+)}] \\
\trarrow \cc{E}[\s{let} \ (z_1,\ldots,z_k)=e \ \s{in} \ [y^+/x^+]\cc{M}] \\
\trarrow \cc{E}[[y^+/x^+]\cc{M}] \\
\equiv \cc{E}[\cc{[y^+/x^+]M}]  \qquad \mbox{(by substitution commutation)} \\
\equiv \cc{E[[y^+/x^+]M]}~.
\end{array}
\]

\item Suppose $M \equiv E[\lets{x}{\prj{i}{y}}{M}] \arrow E[[z_i/x]M]$ 
where $E(y)=(z_1,\ldots,z_k)$, $1\leq i \leq k$.
Then:
\[
\cc{E[\lets{x}{\prj{i}{y}}{M}]} \equiv \cc{E}[\lets{x}{\prj{i}{y}}{\cc{M}}]
\]
with $\cc{E}(y)=(z_1,\ldots,z_k)$. 
Therefore:
\[
\begin{array}{l}
 \cc{E}[\lets{x}{\prj{i}{y}}{\cc{M}}] \\
\arrow \cc{E}[[z_i/x]\cc{M}] \\
\equiv \cc{E}[\cc{[z_i/x]M}] \qquad \mbox{(by substitution commutation)} \\
\equiv \cc{E[[z_i/x]M]}~.
\end{array}
\]

\item Suppose $M\equiv E[\lb{\ell}{M}] \act{\ell} E[M]$.
Then:
\[
\begin{array}{ll}

\cc{E[\lb{\ell}{M}]} \\
\equiv \cc{E}[\cc{\lb{\ell}{M}}] \\
\equiv \cc{E}[\lb{\ell}{\cc{M}}] \\
\act{\ell} \cc{E}[\cc{M}] \\
\equiv \cc{E[M]}~.
\end{array}
\]
~\qed
\end{itemize}

\subsection*{Proof of  proposition \ref{h-transform} [on hoisting transformations]}
As a preliminary remark, note that the hoisting contexts $D$ can be defined
in an equivalent way as follows:
\[
D::= [~] \Alt D[\lets{x}{C}{[~]}] \Alt D[\lets{x}{\lambda y^+.[~]}{M}] \Alt D[\lb{\ell}{[~]}]
\]
If $D$ is a hoisting context and $x$ is a variable we define $D(x)$ as follows:
\[
D(x)=
\left\{
\begin{array}{ll}

\lambda z^+.T    &\mbox{if }D=D'[\lets{x}{\lambda z^+.T}{[~]}] \\
D'(x)            &\mbox{o.w. if }D=D'[\lets{y}{C}{[~]}], x\neq y \\
D'(x)            &\mbox{o.w. if }D=D'[\lets{y}{\lambda y^+.[~]}{M}], x\notin\set{y^+} \\
\mbox{undefined} &\mbox{o.w.}
\end{array}
\right.
\]
The intuition is that $D(x)$ checks whether $D$ binds $x$ to a simple function $\lambda z^+.T$.
If this is the case it returns the simple function as a result, otherwise the result is undefined.

Let $I$ be the set of terms of the  $\lambda_{cps,\w{vn}}^{\ell}$ such that 
if $M\equiv D[\lets{x}{\lambda y^+.T}{N}]$ and $z\in \fv{\lambda y^+.T}$
then $D(z)=\lambda z^+.T'$. Thus a  name free in a simple function must
be bound to another simple function.
We prove the following properties:

\begin{enumerate}

\item The hoisting transformations terminate.

\item The hoisting transformations are confluent (hence the result of the hoisting transformations
is unique).

\item If a term $M$ of the $\lambda_{cps,\w{vn}}^{\ell}$-calculus contains a function
definition then $M\equiv D[\lets{x}{\lambda y^+.T}{N}]$ for some $D,T,N$.

\item All terms in  $\lambda_{cc,\w{vn}}^{\ell}$ belong to the set $I$ (trivially).

\item The set $I$ is an invariant of the hoisting transformations, {\em i.e.},
if $M\in I$ and $M\leadsto N$ then $N\in I$. 

\item If a term satisfying the invariant above is not a program then 
a hoisting transformation applies. 

\end{enumerate}

\Proofitem{(1)} 
To prove the termination of the hoisting transformations
we introduce a size function from terms to positive natural numbers as follows:
\[
\begin{array}{ll}
\ms{@(x,x^+)} &= 1 \\
\ms{\lets{x}{\lambda y^+.M}{N}} &= 2\cdot \ms{M}+\ms{N} \\
\ms{\lets{x}{C}{N}}             &=2\cdot \ms{N} \\
\ms{\lb{\ell}{N}}               &=2\cdot \ms{N}~.
\end{array}
\]
Then we check that if $M\leadsto N$ then $\ms{M}>\ms{N}$.
Note that the hoisting context $D$ induces a function which is
strictly monotone on natural numbers. Thus it is enough to check
that the size of the redex term is larger than the size of the reduced term.

\begin{description}
\item[$(h_1)$]~
\[
\begin{array}{l}

\ms{\lets{x}{C}{\lets{y}{\lambda z^+.T}{M}}}\\
=2\cdot (2 \cdot \ms{T}+ \ms{M}) \\
>2 \cdot \ms{T} + 2 \cdot \ms{M} \\
=\ms{\lets{y}{\lambda z^+.T}{\lets{x}{C}{M}}}~.

\end{array}
\]
\item[$(h_2)$]~
\[
\begin{array}{l}

\ms{\lets{x}{\lambda w^+.\lets{y}{\lambda z^+.T}{M}}{N}} \\
= 2\cdot (2 \cdot \ms{T}+\ms{M})+ \ms{N} \\
> 2 \cdot \ms{T} + 2\cdot \ms{M} + \ms{N} \\
= \ms{\lets{y}{\lambda z^+.T}{\lets{x}{\lambda w^+.M}{N}}}~.

\end{array}
\]
\item[$(h_3)$]~
\[
\begin{array}{l}

\ms{\lb{\ell}{\lets{y}{\lambda z^+.T}{M}}} \\
= 2 \cdot (2\cdot \ms{T}+\ms{M}) \\
> 2 \cdot \ms{T} + 2 \cdot \ms{M} \\
= \ms{\lets{y}{\lambda z^+.T}{\lb{\ell}{M}}}~.

\end{array}
\]
\end{description}

\Proofitem{(2)} Since the hoisting transformation is terminating, by Newman's lemma it is enough
to prove local confluence. There are $9=3\cdot 3$ cases to consider.
In each case one checks that the two redexes cannot superpose. 
Moreover, since the hoisting transformations neither duplicate nor erase terms,
one can close the diagrams in one step. 

For instance,  suppose the term $D[\lets{x}{\lambda w^+.\lets{y}{\lambda z^+.T}{M}}{N}]$
contains a distinct redex $\Delta$ of the same type (a function definition containing 
a {\em simple} function definition).
Then the root of this redex can be in the subterms $M$ or $N$ or in the context
$D$. Moreover if it is in $D$, then  either it is disjoint from the first redex
or it contains it strictly. Indeed, the second let of the second redex cannot be the first
let of the first redex since the latter is not defining a simple function.

\Proofitem{(3)} 
By induction on $M$. Let $F$ be an abbreviation for $\lets{x}{\lambda y^+.T}{N}$.

\begin{description}

\item[$@(x,x^+)$] The property holds trivially.

\item[$\lets{y}{C}{M}$] Then $M$ must contain a function definition. Then by inductive hypothesis,
$M\equiv D'[F]$. We conclude by taking $D\equiv\lets{y}{C}{D'}$.

\item[$\lets{y}{\lambda x^+.M'}{M}$] If $M$ is a restricted term then we take $D\equiv[~]$.
Otherwise, $M'$ must contain a function definition and by inductive hypothesis,
$M'\equiv D'[F]$. Then we take $D\equiv\lets{y}{\lambda x^+.D'}{M}$.

\item[$\lb{\ell}{M}$] Then $M$ contains a function definition and by inductive hypothesis
$M\equiv D'[F]$. We conclude by taking $D\equiv \lb{\ell}{D'}$.

\end{description}

\Proofitem{(4)} In the terms of the $\lambda_{cc,\w{vn}}^{\ell}$ calculus all functions are closed and
therefore the condition is vacuously satisfied.

\Proofitem{(5)} We proceed by case analysis on the hoisting transformations.

\Proofitem{(6)} We proceed by induction on the structure of the term $M$.

\begin{description}

\item[$@(x,y^+)$] This is a program.

\item[$\lets{x}{C}{M'}$] There are two cases:
\begin{itemize}

\item If $M'$ is not a program then by inductive hypothesis a hoisting transformation applies and 
the same transformation can be applied to $M$.

\item If $M'$ is a program then it has a function definition on top (otherwise $M$ is a program). 
Because $M$ belongs to $I$ the side condition of $(h_1)$ is satisfied. 

\end{itemize}

\item[$\lets{x}{\lambda y^+.M'}{M''}$]  Again there are two cases:

\begin{itemize}

\item If $M'$ or $M''$ are not programs then by inductive hypothesis a hoisting transformation applies and 
the same transformation can be applied to $M$.

\item Otherwise, $M'$ is a program with a function definition on top (otherwise $M$ is a program). Because $M$ belongs to
$I$ the side condition of $(h_2)$ is satisfied. 

\end{itemize}

\item[$\lb{\ell}{M'}$] Again there are two cases:

\begin{itemize}

\item  If $M'$ is not a program then by inductive hypothesis a hoisting transformation applies and 
the same transformation can be applied to $M$.

\item If $M'$ is a program then it has a function definition on top (otherwise $M$ is a program) and 
$(h_3)$ applies to $M$. \qed
\end{itemize}
\end{description}

\subsection*{Proof of  proposition \ref{h-commutation} [hoisting commutation]}
As a preliminary step, extend the erasure function to the hoisting contexts in the
obvious way and notice that (i) if $D$ is a hoisting context then
$\er{D}$ is a hoisting context too, and (ii)
$\er{D[M]} \equiv \er{D}[\er{M}]$.

\Proofitem{(1)} We proceed by case analysis on the hoisting transformation applied to $M$.
The case where $\er{M}\equiv \er{N}$ arises in $(h_3)$:
\[
\begin{array}{ccc}
D[\lb{\ell}{\lets{x}{\lambda y^+.T}{M}}] &\leadsto &D[\lets{x}{\lambda y^+.T}{\lb{\ell}{M}}] \\

\er{D[\lb{\ell}{\lets{x}{\lambda y^+.T}{M}}]} &\equiv &\er{D[\lets{x}{\lambda y^+.T}{\lb{\ell}{M}}]}

\end{array}
\]
\Proofitemf{(2)} We show that $\er{M}\leadsto$ entails that $M\leadsto $.
Since $\er{M}$ has no labels, either $(h_1)$ or $(h_2)$ apply.
Then $M$ is a term that is derived from $\er{M}$ by inserting (possibly empty) sequences of pre-labelling
before each subterm. We check that either the hoisting transformation applied to $\er{M}$ can
be applied to $M$ too or $(h_3)$ applies.

\Proofitem{(3)} If $\h{M}\equiv N$ then by definition we have $M \leadsto^* N \not\leadsto$.
By (1) $\er{M}\leadsto^* \er{N}$, and by (2) $\er{N} \not\leadsto$.
Hence $\h{\er{M}}\equiv \er{N} \equiv \er{\h{M}}$. \qed

\subsection*{Proof of  proposition \ref{h-simulation} [hoisting simulation]}

\begin{definition}
A (strong) simulation on the terms of the $\lambda^{\ell}_{cps,\w{vn}}$-calculus is a binary relation $R$
such that if $M \rl{R} N$ and $M\act{\alpha} M'$ then there is $N'$ such that
$N\act{\alpha} N'$ and $M'\rl{R} N'$.
\end{definition}

\begin{definition}
A (pre-)congruence on the terms of the  $\lambda^{\ell}_{cps,\w{vn}}$-calculus
is an equivalence relation (a pre-order) which is preserved by the operators of the calculus.
\end{definition}

\begin{definition}
Let $\simeq$ be the smallest congruence on terms of the $\lambda^{\ell}_{cps,\w{vn}}$-calculus
which is induced by structural equivalence and the following commutation of let-definitions:
\[
\lets{x_{1}}{V_{1}}{\lets{x_{2}}{V_{2}}{M}} 
\simeq \lets{x_{2}}{V_{2}}{\lets{x_{1}}{V_{1}}{M}}
\]
where: $x_1\neq x_2, x_1\notin \fv{V_2}, x_2\notin \fv{V_1}$.
\end{definition}

The relation $\simeq$ is preserved by name substitution and it is a simulation.

\begin{definition}
Let $\succeq$ the smallest pre-congruence on terms of the $\lambda^{\ell}_{cps,\w{vn}}$-calculus
which is induced by structural equivalence and the following collapse of let-definitions:
\[
\lets{x}{V}{\lets{x}{V}{M}} 
\simeq \lets{x}{V}{M} 
\]
where: $x\notin \fv{V}$.
\end{definition}

The relation $\succeq$ is preserved by name substitution and it is a simulation.

\begin{definition}
Let $S_h$ be the relation  $\simeq \comp \succeq$.
\end{definition}

Note that $S_h$ is a simulation too. Then we can state the main lemma.

\begin{lemma}
Let $M$ be a term of the $\lambda^{\ell}_{cps,\w{vn}}$-calculus.
If $M \act{\alpha} M'$ and $M\leadsto N$ then there is $N'$
such that $N\act{\alpha} N'$ and $M' \rl{(\leadsto^*) \comp S_h} N'$.
\end{lemma}
\Proof
As a preliminary remark we notice that the hoisting transformations
are preserved by name substitution. Namely if $M\leadsto N$ then
$[y^+/x^+]M \leadsto [y^+/x^+]N$.

There are three reduction rules and three hoisting transformations
hence there are $9$ cases to consider and for each case 
we have to analyse how the two redexes can superpose.

As usual a term can be regarded as a tree and an occurrence
in the tree is identified by a path $\pi$ which is a sequence of
natural numbers. 

\begin{itemize}

\item The reduction rule is 
\[
E[@(x,y^+)] \arrow E[[y^+/z^+]M]
\]
where $E(x)=\lambda z^+.M$.
We suppose that $\pi$ is the path which corresponds to
the let-definition of the variable $x$ and $\pi'$ is that
path that determines the redex of the hoisting transformation.

\begin{description}

\item[$(h_1)$]
There are two critical cases.

\begin{enumerate}

\item The let-definition that defines a function of the
hoisting transformation coincides with the let-definition of
$x$. In this case $M$ is actually a restricted term $T$.
The diagram is closed in one step.

\item The path $\pi'$ determines a subterm of $M$.
If we reduce first then we have to apply the hoisting
transformation twice to close the diagram using the
fact that these transformations are preserved by name substitution.

\end{enumerate}

\item[$(h_2)$]
Again there are two critical situations.

\begin{enumerate}

\item The top level let-definition of the hoisting transformation 
coincides with the let-definition of the variable $x$ in the reduction.
This is  the case illustrated by the example \ref{hoisting-comm-ex}.
If we reduce first then we have to apply the hoisting transformation
twice (again using preservation under name substitution).
After this we have to commute the let-definitions and finally collapse
two identical ones.

\item The path $\pi'$ determines a subterm of $M$. 
If we reduce first then we have to apply the hoisting
transformation twice to close the diagram using the
fact that these transformations are preserved by name substitution.

\end{enumerate}

\item[$(h_3)$]
There are two critical cases.

\begin{enumerate}

\item The function let-definition in the hoisting transformation
coincides with the let-definition of the variable $x$ in the reduction.
We close the diagram in one step.

\item The path $\pi'$ determines a subterm of $M$.
If we reduce first then we have to apply the hoisting
transformation twice to close the diagram using the
fact that these transformations are preserved by name substitution.

\end{enumerate}
\end{description}

\item The reduction rule is
\[
E[\lets{x}{\prj{i}{y}}{M}] \arrow E[[z_i/x]M]
\]
where $E(y)=(z_1,\ldots z_n)$ and $1\leq i \leq n$.

\begin{description}

\item[$(h_1)$]
There are two critical cases.

\begin{enumerate}

\item The first let-definition in the hoisting transformation
coincides with the let-definition of the tuple in the reduction.
We close the diagram in one step.

\item The first let-definition in the  hoisting transformation
coincides with the projection in the reduction.
If we reduce first then there is no need to apply a hoisting
transformation to close the diagram because the projection disappears.

\end{enumerate}

\item[$(h_2)$]
The only critical case arises when the redex for the  hoisting transformation
is contained in $M$. We close the diagram in one step using the fact that
the transformations are preserved by name substitution.

\item[$(h_3)$]
Same argument as in the previous case.

\end{description}

\item The reduction rule is
\[
E[\lb{\ell}{M}] \act{\ell} E[M]
\]
The hoisting transformations can be either in $E$ or in $M$.
In both cases we close the diagram in one step. \qed

\end{itemize}

We conclude by proving by diagram chasing the following proposition.
We rely on the previous lemma and the fact that $S_h$ is a simulation.

\begin{proposition}
The relation $T_h= ((\leadsto^*) \comp S_h)^*$ is a simulation and
for all terms of the  $\lambda^{\ell}_{cc,\w{vn}}$-calculus,
$M \rl{T_h} \h{M}$.
\end{proposition}

\subsection*{Proof of theorem \ref{com-sim-thm} [commutation and simulation]}
By composition of the commutation and simulation properties of the four
compilation steps.

\subsection*{Proof of proposition \ref{labelling-prop} [labelling properties]}
\Proofitemf{(1)}
Both properties are proven by induction on $M$.
The first is immediate. We spell out the second.

\begin{description}

\item[$x$] Then $\labi{i}{x}=x\in W_1 \subseteq W_0$.

\item[$\lambda x^+.M$] Then 
$\labi{i}{\lambda x^+.M} = \lambda x^+.\lb{\ell}{\labi{1}{M}}$ and
by inductive hypothesis $\labi{1}{M}\in W_1$. 

Hence, $\lb{\ell}{\labi{1}{M}}\in W_1$ and $\lambda x^+.\lb{\ell}{\labi{1}{M}}\in W_1$.

\item[$(M_1,\ldots,M_n)$]
Then $\labi{i}{(M_1,\ldots,M_n)} = (\labi{0}{M_1},\ldots,\labi{0}{M_n})$ and
by inductive hypothesis $\labi{0}{M_j}\in W_0$ for $j=1,\ldots,n$.

Hence,  $(\labi{0}{M_1},\ldots,\labi{0}{M_n})\in W_1\subseteq W_0$.

\item[$\prj{j}{M}$]
Same argument as for the pairing.

\item[$\lets{x}{M}{N}$]
Then $\labi{i}{\lets{x}{M}{N}} = \lets{x}{\labi{0}{M}}{\labi{i}{N}}$ and by
inductive hypothesis $\labi{0}{M}\in W_0$ and $\labi{i}{N}\in W_i$.
Hence $\lets{x}{\labi{0}{M}}{\labi{i}{N}}\in W_i$.

\item[$@(M_1,\ldots,M_n)$ and $i=0$]
Then $\labi{0}{@(M_1,\ldots,M_n)} = \plb{\ell}{@(\labi{0}{M_1},\ldots,\labi{0}{M_n})}$
and by inductive hypothesis $\labi{0}{M_j}\in W_0$ for $j=1,\ldots,n$. 
Hence  $\plb{\ell}{@(\labi{0}{M_1},\ldots,\labi{0}{M_n})}\in W_0$.

\item[$@(M_1,\ldots,M_n)$ and $i=1$]
Same argument as in the previous case to conclude that  \\\noindent
$@(\labi{0}{M_1},\ldots,\labi{0}{M_n})\in W_1$.

\end{description}

\Proofitem{(2)}
By (1) we know that $\er{\lab{M}}\equiv M$ and $\lab{M}\in W_0$. Then:
\[
\begin{array}{lll}
P &\equiv \cmp{M}  \\
  &\equiv \cmp{\er{\lab{M}}}  \\
  &\equiv \er{\cmp{\lab{M}}} &\mbox{(by theorem \ref{com-sim-thm}(1))}~.
\end{array}
\]
\Proofitemf{(3)}
The main point is to show that the CPS compilation of a labelled term is
a term where a pre-labelling appears exactly after each $\lambda$-abstraction.
The following compilation steps (value named, closure conversion, hoisting)
neither destroy nor introduce new $\lambda$-abstractions while maintaining the
invariant that the body of each function definition contains exactly 
one pre-labelling.

As a preliminary step, we define a restricted syntax for the $\lambda^\ell_{cps}$-calculus
where labels occur exactly after each $\lambda$-abstraction.
\[
\begin{array}{lll}

V &::= \w{id} \Alt \lambda \w{id}^+.\lb{\ell}{M} \Alt (V^*)   &\mbox{(restricted values)} \\
M &::= @(V,V^+) \Alt \lets{\w{id}}{\prj{i}{V}}{M}             &\mbox{(restricted CPS terms)} \\ 
K &::=  \w{id} \Alt \lambda \w{id}.M                          &\mbox{(restricted continuations)} 

\end{array}
\]
Let us call this language $\lambda^{\ell}_{cps,r}$ ($r$ for restricted).
First we remark that if $V$ is a restricted value and $M$ is a restricted CPS term then
$[V/x]M$ is again a restricted CPS term.
Then we show the following property.

\begin{quote}
For all terms $M$ of the $\lambda$-calculus and all continuations
$K$ of the $\lambda^{\ell}_{cps,r}$-calculus the term $\labi{i}{M}\sco K$ is 
again a term of the  $\lambda^{\ell}_{cps,r}$-calculus provided that 
$i=0$ if $K$ is a function and $i=1$ if $K$ is a variable.
\end{quote}

Notice that the initial continuation $K_0=\lambda x.@(\w{halt},x)$ is
a functional continuation in the restricted calculus and recall that
by definition $\cps{\lab{M}}=\labi{0}{M}\sco K_0$.
We proceed by induction on $M$ and case analysis assuming that if $i=0$ then
$K=\lambda y.N$.

\begin{description}

\item[$x$, $i=0$]
We have: $\labi{0}{x}\sco K = x\sco K = [x/y]N$.

\item[$x$, $i=1$]
We have: $\labi{i}{x}\sco k = x\sco k = @(k,x)$.

\item[$\lambda x^+.M$, $i=0$] We have:
\[
\labi{0}{\lambda x^+.M}\sco K
= \lambda x^+.\lb{\ell}{\labi{1}{M}}\sco K
=[\lambda x^+,k.\lb{\ell}{\labi{1}{M}\sco k}/y]N
\]
and we apply the inductive hypothesis on $\labi{1}{M}\sco k$ and closure under value substitution.

\item[$\lambda x^+.M$, $i=1$] We have:
\[
\labi{1}{\lambda x^+.M}\sco k
= \lambda x^+.\lb{\ell}{\labi{1}{M}}\sco k
=@(k,\lambda x^+,k.\lb{\ell}{\labi{1}{M}\sco k})
\]
and we apply the inductive hypothesis on $\labi{1}{M}\sco k$.

\item[$@(M_1,\ldots,M_n)$, $i=0$]
We have:
\[
\begin{array}{l}
\labi{i}{@(M_1,\ldots,M_n)}\sco K  \\
\equiv  \plb{\ell}{@(\labi{0}{M_1},\ldots,\labi{0}{M_n})}\sco K  \\
\equiv @(\labi{0}{M_1},\ldots,\labi{0}{M_n})\sco K' \\
\equiv \labi{0}{M_1}\sco \lambda x_1\ldots \labi{0}{M_n}\sco \lambda x_n.@(x_1,\ldots,x_n,K')
\end{array}
\]
where $K'= \lambda y.\lb{\ell}{N}$. Then we apply the inductive hypothesis on 
$M_n,\ldots,M_1$ with the suitable functional continuations.

\item[$@(M_1,\ldots,M_n)$, $i=1$]
We have:
\[
\begin{array}{l}
\labi{i}{@(M_1,\ldots,M_n)}\sco K  \\
\equiv  @(\labi{0}{M_1},\ldots,\labi{0}{M_n})\sco K  \\
\equiv \labi{0}{M_1}\sco \lambda x_1\ldots \labi{0}{M_n}\sco \lambda x_n.@(x_1,\ldots,x_n,K)~.
\end{array}
\]
Again we apply the inductive hypothesis on 
$M_n,\ldots,M_1$ with the suitable functional continuations.

\item[$(M_1,\ldots,M_n)$]
We have:
\[
\begin{array}{l}
\labi{i}{(M_1,\ldots,M_n)}\sco K  \\
\equiv (\labi{0}{M_1},\ldots,\labi{0}{M_n})\sco K  \\
\equiv  \labi{0}{M_1}\sco \lambda x_1\ldots \labi{0}{M_n}\sco \lambda x_n.(x_1,\ldots,x_n)\sco K ~.
\end{array}
\]
We apply  the inductive hypothesis on 
$M_n,\ldots,M_1$ with the suitable functional continuations.

\item[$\prj{j}{M}$]
We have:
\[
\begin{array}{l}
\labi{i}{\prj{j}{M}}\sco K  \\
\equiv \prj{j}{\labi{0}{M}}\sco K  \\
\equiv \labi{0}{M}\sco \lambda x.\lets{y}{\prj{j}{x}}{y\sco K}~.
\end{array}
\]
We apply the inductive hypothesis on $M$ with a functional continuation.

\item[$\lets{x}{N}{M}$]
We have:
\[
\begin{array}{l}
\labi{i}{\lets{x}{N}{M}} \sco K \\
\equiv \lets{x}{\labi{0}{N}}{\labi{i}{M}}\sco K \\
\equiv \labi{0}{N}\sco \lambda x.\labi{i}{M}\sco K~.
\end{array}
\]
We apply the inductive hypothesis on $M$ and then on $N$
with a functional continuation. \qed

\end{description}

\subsection*{Proof of proposition \ref{inst-lab-prop} [instrumentation vs. labelling]}
As a preliminary step, we show that for all terms $M$ and values $V,V'$ of the $\lambda^\ell$-calculus
the following (mutually dependent) properties hold.
\begin{enumerate}

\item $[\psi(V)/x]\psi(V') \equiv \psi([V/x]V')$.

\item $[\psi(V)/x]\sem{M} \equiv \sem{[V/x]M}$.
\end{enumerate}
Let $S=[V_1/x_1,\ldots,V_n/x_n]$ denote a substitution in the $\lambda^\ell$-calculus.
Then let $\psi(S)$ be the substitution $[\psi(V_1)/x_1,\ldots,\psi(V_n)/x_n]$.
We prove the following generalisation of the proposition.
\begin{quote}
For all terms $M$ and substitutions $S$, if $\psi(S)\sem{M}\eval (m,V)$ then
$\sem{SM}\eval_\Lambda V'$, $\s{costof}(\Lambda)=m$ and $\psi(V')\equiv V$.
\end{quote}
We proceed by induction on the length of the derivation of the judgement $\psi(S)\sem{M}\eval (m,V)$
and case analysis on $M$. 

We consider the case for application which explains the need for the generalisation.
Suppose $\psi(S)\sem{@(M_0,M_1,\ldots,M_n)}\eval (m,V)$. 
By the shape of $\sem{@(M_0,M_1,\ldots,M_n)}$ this entails that 
$\psi(S)\sem{M_i}\eval (m_i,V_i)$ for $i=0,\ldots,n$.
By induction hypothesis, $\sem{SM_i} \eval_{\Lambda_{i}} V'_i$, 
$\s{costof}(\Lambda_i)=m_i$, and $\psi(V'_i)=V_i$, for $i=0,\ldots,n$.
We also have $@(V_0,V_1,\ldots,V_n) \eval (m_{n+1},V)$ and 
$m=m_0\oplus \cdots \oplus m_{n+1}$.
Since $\psi(V'_0)=V_0$, this requires $V'_0 =\lambda x_1\cdots x_n.M'$ 
and $V_0=\lambda x_1\cdots x_n.\sem{M'}$. 
So we must have $[V_1/x_1,\ldots,V_n/x_n]\sem{M'} \eval (m_{n+1},V)$.
Again by inductive hypothesis, this entails that 
$\sem{[V'_1/x_1,\ldots,V'_n/x_n]M'} \eval_{\Lambda_{n+1}} V'$, 
$\s{costof}(\Lambda_{n+1}) = m_{n+1}$, and  $\psi(V')=V$.
We conclude that $\sem{S(@(M_0,M_1,\ldots,M_n))} \eval_\Lambda V'$ with
$\Lambda = \Lambda_0\cdots\Lambda_{n+1}$. \qed


\subsection*{Proof of proposition \ref{subj-red-prop} [subject reduction]}
The proof of this result is standard, so we just recall the main steps. 

\begin{enumerate}

\item Prove a weakening lemma: $\Gamma \Gives M:A$ implies $\Gamma,x:B \Gives M:A$ for $x$ fresh.

\item Prove a substitution lemma: $\Gamma,x:A\Gives M:B$ and $\Gamma \Gives N:A$ implies
$\Gamma \Gives [N/x]M:B$; by induction on the proof of $M$.

\item Derive by iteration the following substitution lemma: $\Gamma,x_1:A_1,\ldots,x_n:A_n \Gives M:B$ and
$\Gamma \Gives N_i:A_i$ for $i=1,\ldots,n$ implies $\Gamma \Gives [N_1/x_1,\ldots,N_n/x_n]M:B$.

\item With reference to Table~\ref{lambda}, notice that in an evaluation context one does not
cross any binder. Then we have that $E[M]\equiv [M/x]E[x]$ for $x$ fresh variable.
Moreover, if $\Gamma \Gives E[M]:A$ then for some $B$, $\Gamma \Gives M:B$.

\item Now examine the 5 possibilities for reduction specified in Table~\ref{lambda}. They
all have the shape $E[\Delta] \arrow E[\Delta']$. By the previous remark, it suffices
to show that if $\Gamma \Gives \Delta:B$ then $\Gamma \Gives \Delta':B$. 
Note that the typing rules in Table~\ref{type-lambda} are driven by the syntax of the term.
Then the property is checked by case analysis while appealing,
for the first two rewriting rules, to the substitution properties mentioned above.

\end{enumerate}

\subsection*{Proof of proposition \ref{type-cps-prop} [type CPS]}
First, we prove the following properties at once by induction on the
structure of the term (possibly a value).
\begin{enumerate}

\item If $\Gamma \Gives V:A$ then $\cps{\Gamma}\Gives \psi(V):\cps{A}$.

\item If $\Gamma \Gives M:A$ then
$\cps{\Gamma},k:\neg \cps{A} \Gives (M\sco k):R$.

\item If $\Gamma \Gives M:A$ and $\cps{\Gamma},\Gamma',x:\cps{A} \Gives N:R$ then
$\cps{\Gamma},\Gamma' \Gives (M\sco (\lambda x.N)):R$.

\end{enumerate}

We illustrate the analysis for the cases of abstraction and application.

\begin{description}

\item[Abstraction] Suppose 
$\Gamma \Gives \lambda x^+.M:A^+\arrow B$ is derived from
$\Gamma,x^+:A^+ \Gives M:B$. We prove the $3$ properties above.

\begin{description}

\item[(1)] By induction hypothesis (property $2$), we know:
\[
\cps{\Gamma},x^+:\cps{A}^+, k:\neg \cps{B} \Gives (M\sco k):R~.
\]
Then, recalling that:
\[
\psi(\lambda x^+.M)\equiv \lambda x^+,k.(M\sco k) \mbox{ and }
\cps{A^+\arrow B}=\cps{A}^+,\neg\cps{B}\arrow R~,
\]
we derive:
\[
\cps{\Gamma} \Gives \psi(\lambda x^+.M):\cps{A^+\arrow B}~.
\]

\item[(2)] 
Recall that $(\lambda x^+.M)\sco k\equiv @(k,\psi(\lambda x^+.M))$.
By property 1, we derive:
\[
\cps{\Gamma} \Gives \psi(\lambda x^+.M) : \cps{A^+\arrow B}
\]
Then by weakening and substitution we derive 
\[
\cps{\Gamma},k:\neg \cps{A^+\arrow B} \Gives \psi(\lambda x^+.M) : \cps{A^+\arrow B}
\]
Finally the application rule gives
\[
\cps{\Gamma},k:\neg \cps{A^+\arrow B} \Gives ((\lambda x^+.M):k):R~.
\]

\item[(3)] 
Suppose additionally that $\cps{\Gamma},\Gamma',y:\cps{A^+\arrow B} \Gives N:R$.
Recall that $(\lambda x^+.M\sco \lambda y.N) \equiv [\psi(\lambda x^+.M)/y]N$.
By property 1, we know that:
\[
\cps{\Gamma} \Gives \psi(\lambda x^+.M):\cps{A^+\arrow B}~.
\]
Then by weakening and substitution we derive that:
\[
\cps{\Gamma},\Gamma' \Gives [\psi(\lambda x^+.M)/y]N:R~.
\]
\end{description}

\item[Application] 
Suppose $\Gamma \Gives @(M_0,\ldots,M_n):B$ is derived
from $\Gamma \Gives M_0:A$, $A\equiv A_1,\ldots,A_n\arrow B$, and 
$\Gamma \Gives M_i:A_i$ for $i=1,\ldots,n$.
In this case, we just look at the last
two properties since an application cannot be a value.

\begin{description}

\item[(2)]
Clearly: 
\[
\cps{\Gamma},\Gamma',x_n:\cps{A_n} \Gives @(x_0,x_1,\ldots,x_n,k):R ~,
\]
where $\Gamma'\equiv x_0:\cps{A},\ldots,x_{n-1}:\cps{A_{n-1}},k:\neg \cps{B}$.
By induction hypothesis (property 3) on $M_n$, we derive:
\[
\cps{\Gamma},\Gamma' \Gives (M_n:\lambda x_n.@(x_0,x_1,\ldots,x_n,k)):R~.
\]
Then by applying the inductive hypothesis (property 3) on $M_{n-1},\ldots,M_0$ we obtain:
\[
\cps{\Gamma}, k:\neg \cps{B} \Gives (M_0\sco \lambda x_0.\cdots M_n\sco \lambda x_n.@(x_0,x_1,\ldots,x_n,k)):R~.
\]

\item[(3)]
Suppose additionally that $\cps{\Gamma},\Gamma'',y:\cps{B} \Gives N:R$.
Then we have:
\[
\cps{\Gamma},\Gamma',\Gamma'',x_n:\cps{A_n} \Gives @(x_0,x_1,\ldots,x_n,\lambda y.N):R ~,
\]
where $\Gamma'\equiv x_0:\cps{A},\ldots,x_{n-1}:\cps{A_{n-1}}$. 
Then proceed as in the previous case by applying the inductive hypothesis (property 3)
on $M_n,\ldots,M_0$.
\end{description}

\end{description}
The proof of the omitted cases follows a similar pattern.
Now to derive proposition \ref{type-cps-prop}, 
recall that $\cps{M}\equiv M\sco \lambda x.@(\w{halt},x)$.
Then we obtain the desired statement from the property~3 above
observing that if $\Gamma \Gives M:A$ and
$\cps{\Gamma},\w{halt}:\neg \cps{A},x:\cps{A} \Gives @(\w{halt},x):R$
then $\cps{\Gamma},\w{halt}:\neg \cps{A} \Gives \cps{M}:R$.

\subsection*{Proof of proposition \ref{subj-red-adm-prop} [subject reduction, value named]}
First, we prove some standard properties for the type system described in Table~\ref{types-adm}.
\begin{description}

\item[Weakening]
If $\Gamma \aGives M$ then $\Gamma,x:A \aGives M$ with $x$ fresh.

\item[Variable substitution]
If $\Gamma,x:A \aGives M$ and $y:A\in \Gamma$ then $\Gamma \aGives [y/x]M$.
This property generalizes to $\Gamma,x^+:A^+ \aGives M$ and $y^+:A^+\in \Gamma$ implies
$\Gamma \aGives [y^+/x^+]M$.

\item[Type substitution]
If $\Gamma \aGives M$ then $[B/t]\Gamma \aGives M$.

\end{description}

Next, suppose $\Gamma \aGives M$ and $M\arrow N$ according to the rules specified in
Table~\ref{lambda-adm}.
This means $M\equiv E[\Delta]$ where 
for some $\Gamma'$, we have 
$\Gamma,\Gamma'\aGives \Delta$ and  $\Delta$ is either an application, 
or a projection or a labelling. We consider each case in turn.

\begin{description}

\item[$\Delta \equiv @(x,y^+)$.] Then $y^+:A^+\in \Gamma,\Gamma'$, \ 
$\Gamma'\equiv \Gamma_1,x:A^+\arrow R,\Gamma_2$, $x$ is bound to some
function $\lambda z^+.M'$, and  
$\Gamma,\Gamma_1,z^+:A^+ \aGives M'$. By weakening, we have
$\Gamma,\Gamma',z^+:A^+ \aGives M'$ and by substitution $\Gamma,\Gamma' \aGives [y^+/z^+]M'$.
Then we derive $\Gamma \aGives E[[y^+/z^+]M']$ as required.

\item[$\Delta \equiv \lets{x}{\prj{i}{y}}{M'}$.] This case splits in two sub-cases:
the first for product types and the second for existential types.

\begin{description}

\item[Product] $\Gamma'\equiv \Gamma_1,y:\times(A_1,\ldots,A_n),\Gamma_2$, $1\leq i \leq n$,
$\Gamma,\Gamma',x:A_i \aGives M'$, 
and for some $z_1,\ldots,z_n$, $z_1:A_1,\ldots,z_n:A_n\in \Gamma,\Gamma_1$.
By substitution, $\Gamma,\Gamma' \aGives [z_i/x]M'$.
Then we derive $\Gamma \aGives E[[z_i/x]M']$ as required.

\item[Existential] $i=1$, $\Gamma'\equiv \Gamma_1,y:\exists t.A, \Gamma_2$,  \ 
$\Gamma,\Gamma',x:A \aGives M'$ with $t\notin \ftv{\Gamma,\Gamma'}$,
and for some $z,B$, we have $z:[B/t]A \in \Gamma,\Gamma_1$.
By type substitution, $\Gamma,\Gamma',x:[B/t]A \aGives M'$ and 
by substitution, $\Gamma,\Gamma' \aGives [z/x]M'$.
Then we derive $\Gamma \aGives E[[z/x]M']$ as required.

\end{description}

\item[$\Delta \equiv \lb{\ell}{M'}$.]
Then $\Gamma,\Gamma'\aGives M'$ and we derive 
$\Gamma \aGives E[M']$ as required.

\end{description}

\subsection*{Proof of proposition \ref{type-adm-prop} [type value named]}
We prove at once the following two properties:
\begin{enumerate}

\item If $\Gamma \Gives M:R$ then $\Gamma \aGives \vn{M}$.

\item If $\Gamma \Gives V:A$, $V\neq \w{id}$ and $\Gamma,y:A \aGives N:R$ then 
$\Gamma \aGives \evn{V}{y}[N]$.

\end{enumerate}

We proceed by induction on the structure of $M$ and $V$ along the pattern 
of the definition of the value named translation in Table~\ref{back-forth}.
We spell out two typical cases.

\begin{description}

\item[$M\equiv @(x^*,V,V^*)$, $V\neq \w{id}$.]
Suppose $\Gamma \Gives @(x^*,V,V^*):R$. 
This entails $\Gamma \Gives V:A$ for some type $A$.
We also have $\Gamma,y:A \Gives @(x^*,y,V^*):R$ and
by inductive hypothesis (property 1) 
$\Gamma,y:A \aGives \vn{@(x^*,y,V^*)}$.
Then, we apply the inductive hypothesis on $V$ 
(property 2) to derive that:
$\Gamma \aGives \evn{V}{y}[\vn{@(x^*,y,V^*)}]$,
and this last term equals  $\vn{ @(x^*,V,V^*)}$.

\item[$V\equiv (x^*,V',V^*)$, $V'\neq \w{id}$.]
Suppose  $\Gamma,y:A \aGives N$ and 
$\Gamma \Gives (x^*,V',V^*):A$.
This entails $\Gamma \Gives V':B$ for some type $B$ and 
$\Gamma,z:B \Gives (x^*,z,V^*):A$.
By weakening and inductive hypothesis (property 2) on $(x^*,z,V^*)$ we derive
$\Gamma,z:B \aGives \evn{(x^*,z,V^*)}{y}[N]$.
Then by inductive hypothesis on $V'$ (again by property 2) we derive:
\[
\Gamma \aGives \evn{V'}{z}[\evn{(x^*,z,V^*)}{y}[N]]~,
\]
and this last term equals $\evn{(x^*,V',V^*)}{y}[N]$.

\end{description}

\subsection*{Proof of proposition \ref{type-cc-prop} [type closure conversion]}
By induction on the typing of $\Gamma \aGives M$ according to the rules 
specified in Table~\ref{types-adm}. We detail the cases of abstraction and application.

\begin{description}

\item[Abstraction] 
Suppose $\Gamma \aGives \lets{x}{\lambda y^+.M}{N}$ is derived from
$\Gamma,y^+ : A^+ \aGives M$ and $\Gamma,x:A^+\arrow R \aGives N$.
Let us pose $\set{z^*}=\fv{\lambda y^+.M}$. Then for some $C^*$
we have $z^*:C^* \in \Gamma$.
We have to show that:
\[
\begin{array}{ll}
\cc{\Gamma} \aGives  &\s{let } \ c=\lambda e,y^+.\lets{(z^*)}{e}{\cc{M}}  \ \s{in} \\
                     &\s{let}  \  e=(z^*)   \ \s{in} \\
                     &\s{let}  \  x'=(c,e)  \ \s{in} \\
                     &\s{let}  \  x=(x')    \ \s{in} \ \cc{N}~.
\end{array}
\]
By inductive hypothesis on $M$, variable substitution, and weakening we derive:
$\cc{\Gamma},\Gamma' \aGives \cc{M}$,
with $\Gamma'\equiv  c:\times(C^*),\cc{A}^+\arrow R, e:\times(C^*)$.

Also, by inductive hypothesis on $N$ and weakening we derive:
\[
\cc{\Gamma}, \Gamma',\Gamma'' \aGives \cc{N}~,
\]
with $\Gamma''\equiv x':[\times(C^*)/t]B, x:\exists t.B$ and 
$B\equiv \times((t,\cc{A}^+\arrow R),t)$. 

\item[Application]
Suppose $\Gamma \aGives @(x,y^+)$ is derived from $x:A^+\arrow R,y^+:A^+\in \Gamma$.
We have to show:
\[
\begin{array}{ll}

\cc{\Gamma} \aGives &\s{let}  \  x'=\prj{1}{x}    \ \s{in} \\
                    &\s{let}  \  (c,e)=x'    \ \s{in} \
                    @(c,e,y^+)~.

\end{array}
\]
Since $\cc{A^+\arrow R} \equiv \exists t.\times((t,\cc{A}^+\arrow R),t)$,
the judgement above is derived from:
\[
\cc{\Gamma},x':\times((t,\cc{A}^+\arrow R),t), c: (t,\cc{A}^+\arrow R), e:t \aGives @(c,e,y^+)~.
\]
\end{description}

\subsection*{Proof of proposition \ref{type-hoisting-prop} [type hoisting]}
First, we show the following property.
\begin{description}

\item[Strengthening] If $\Gamma,x:A \aGives P$ and $x\notin \fv{P}$ then $\Gamma \aGives P$. 

\end{description}

Then we proceed by case analysis ($3$ cases) on the hoisting transformations
specified in Table~\ref{lambda-hoist}.
They all have the shape $D[\Delta]\leadsto D[\Delta']$, so it suffices to show
that if $\Gamma \aGives \Delta$ then $\Gamma \aGives \Delta'$.
We detail the analysis for the transformation $(h_2)$.
Suppose:
\[
\Gamma \aGives \lets{x}{\lambda w^+.\lets{y}{\lambda z^+.T}{M}}{N},
\]
with $\set{w^+}\inter \fv{\lambda z^+.T}=\emptyset$, is derived from:
\[
\begin{array}{ll}

(1) &\Gamma, x:A^+\arrow R \aGives N ~,\\
(2) &\Gamma, w^+:A^+,z^+:B^+ \aGives T ~,\\
(3) &\Gamma, w^+:A^+,y:B^+\arrow R \aGives M~.
\end{array}
\]
Then we derive:
\[
\Gamma \aGives \lets{y}{\lambda z^+.T}{\lets{x}{\lambda w^+.M}{N}}~.
\]
as follows:
\[
\begin{array}{lll}

(1') &\Gamma, x:A^+\arrow R,y:B^+\arrow R \aGives N  &\mbox{(by (1) and weakening)}\\
(2') &\Gamma, z^+:B^+ \aGives T                      &\mbox{(by (2) and strengthening)}\\
(3') &\Gamma, w^+:A^+,y:B^+\arrow R \aGives M        &\mbox{(by (3)) .}
\end{array}
\]

\subsection*{Proof theorem \ref{type-compil-thm} [type preserving compilation]}
Suppose $M$ term of the $\lambda^\ell$-calculus and $\Gamma \Gives M:A$. Then:
\[
\begin{array}{ll}

\cps{\Gamma},\w{halt}:\neg \cps{A} \Gives \cps{M}:R 
&\mbox{(by proposition \ref{type-cps-prop})} \\

\cps{\Gamma},\w{halt}:\neg \cps{A} \aGives \vn{\cps{M}}
&\mbox{(by proposition \ref{type-adm-prop})} \\

\cmp{\Gamma},\w{halt}:\exists t.\times((t,\cmp{A}\arrow R),t) \aGives \cc{\vn{\cps{M}}}
&\mbox{(by proposition \ref{type-cc-prop})} 

\end{array}
\]
Next recall that the compiled term $\cmp{M}$ is the result of iterating the hoisting transformations
on the term $\cc{\vn{\cps{M}}}$ a finite number of times. Hence, by proposition
\ref{type-hoisting-prop} we conclude:
\[
\cmp{\Gamma},\w{halt}:\exists t.\times((t,\cmp{A}\arrow R),t) \aGives \cmp{M}
~.
\]


\begin{table}
{\footnotesize
\[
\begin{array}{ll}
\rer{t} &= t\\
\rer{A^+\act{e} R} &=\rer{A}^+\arrow R \\
\rer{\times()}         &=\times() \\
\rer{\times(A^+)\at{r}} &= \times(\rer{A}^+) \\
\rer{(\exists t.A)\at{r}}  &= \exists t.\rer{A} \\ \\ 

\rer{\lambda r^*,x^+.T} &=\lambda x^+.\rer{T} \\
\rer{()}               &=()\\
\rer{(x^+)\at{r}}           &=(x^+) \\ \\

\rer{\lets{x}{V}{P}}   &=\lets{x}{\rer{V}}{\rer{P}} \\
\rer{@(x,r^*,y^+)}         &=@(x,y^+) \\
\rer{\lets{x}{\prj{i}{y}}{T}} &= \lets{x}{\prj{i}{y}}{\rer{T}}

\end{array}
\]}
\caption{Region erasure for types, values and terms.}\label{region-erasure}
\end{table}

\subsection*{Proof of proposition \ref{decom-prop} [decomposition]}
With reference to Table~\ref{lambda-reg}, we know that a program $P$ is a list
of function definitions, determining the function context $F$, followed
by a term $T$. The latter is a list of value definitions and region allocations and disposals,
determining the heap context $H$, and 
ending either in an application or a projection  or a labelling. 
This last part of the program corresponds to the redex $\Delta$.

\subsection*{Proof of proposition \ref{reg-erasure-prop} [region erasure]}
First, we notice that the region erasure function is invariant under region substitutions:
$\rer{[r'/r]A}=\rer{A}$. 
Then we prove at once the following two properties
where it is intended that the judgements on the left are derivable in
the type and effect system described in Table~\ref{type-effect} and
the ones on the right in the type system described in Table~\ref{types-adm}.
\begin{enumerate}

\item If $\Gamma \rGives P:e$  then $\rer{\Gamma} \aGives \rer{P}$.

\item If $\Gamma \rGives V:A$ then $\rer{\Gamma} \aGives \rer{V}:\rer{A}$.

\end{enumerate}
We detail the cases of abstraction and application.

\begin{description}

\item[Abstraction]
Suppose $\Gamma \rGives \lambda r^*,y^+.T: \forall r^*.A^+\act{e} R$ is derived from
$\Gamma, y^+:A^+ \rGives T:e$. Then by inductive hypothesis,
$\rer{\Gamma}, y^+:\rer{A}^+ \aGives \rer{T}$. And we conclude:
$\rer{\Gamma} \aGives \lambda y^+.T:\rer{A}^+\arrow R$ as required.

\item[Application]
Suppose $\Gamma \rGives @(x,r^+,y^+)$ is derived from $x:B\in \Gamma$,
$B\equiv \forall r_{1}^{*}.A^+\act{e}R$, $y^+:[r^*/r^*_1]A^+\in \Gamma$.
Then, by the invariance property of the region erasure function noticed above,
$x:\rer{A}^+\arrow R,y^+:\rer{A}^+\in \rer{\Gamma}$. So we conclude:
$\rer{\Gamma} \aGives @(x,y^+)$ as required.

\end{description}

\subsection*{Proof of proposition \ref{reg-enrich-prop} [region enrichment]}
We define a region enrichment function ${\it ren}$ from the programs of the $\lambda_{h,\w{vn}}^{\ell}$-calculus to those
of the $\lambda_{h,\w{vn}}^{\ell,r}$-calculus. With reference to Table~\ref{lambda-hoist},
we recall  that a program $P$ of the $\lambda_{h,\w{vn}}^{\ell}$ is composed of a list of 
function definitions and a term. Thus $P$ is decomposed uniquely as $F[T]$ where $F$ is 
a functional context defined as follows
\[
F::=[~] \Alt \lets{\w{id}}{\lambda \w{id}^+.T}{F}~.
\]
We fix one region variable $r$ and define the region enrichment function relatively to it
as follows:

{\footnotesize
\[
\begin{array}{lll}

\ren{F[T]} &= \ren{F}[\letall{r}{\ren{T}}]  &\mbox{({\sc Programs})} \\ \\ 

\ren{[~]} &=[~]                             &\mbox{({\sc Function contexts})} \\
\ren{\lets{x}{\lambda y^+.T}{F}} &= \lets{x}{\lambda r,y^+.\ren{T}}{\ren{F}} \\ \\ 

\ren{@(x,y^+)}  &= @(x,r,y^+)               &\mbox{({\sc Restricted terms})} \\
\ren{\lets{x}{C}{T}} &= \lets{x}{\ren{C}}{\ren{T}} \\
\ren{\lb{\ell}{T}}   &=\lb{\ell}{\ren{T}} \\ \\

\ren{()}      &=()                          &\mbox{({\sc Restricted let-bindable terms})} \\
\ren{(x^+)}   &=(x^+)\at{r}         \\
\ren{\prj{i}{x}} &=\prj{i}{x} 

\end{array}
\]}

The intuition is that a region $r$ is created initially and never disposed, 
that all tuples are allocated in this region, and that at every function call
we pass this region as a parameter. Then all functions when
applied will produce (at most) an effect $\set{r}$. 

Next we extend the region enrichment function to types as follows:

{\footnotesize
\[
\begin{array}{ll}

\ren{t} &=t \\
\ren{A^+\arrow R} &= \forall r.\ren{A}^+ \act{\set{r}} R \\
\ren{\times()}    &= \times() \\
\ren{\times(A^+)} &=\times(\ren{A}^+)\at{r} \\
\ren{(\exists t.A)} &=(\exists t.\ren{A})\at{r}

\end{array}
\]}

We notice that function definitions are region closed and so are
the functional types in the image of the function $\w{ren}$.
Let us denote with $\Gamma_0$ a type context such that if $x:A\in \Gamma_0$
then $A$ is not a type of the shape $\times(B^+)$ or $\exists t.B$.
It follows that $\frv{\ren{\Gamma_0}}=\emptyset$.

We show the following {\em enrichment} property:
\begin{quote}
If $\Gamma_0,\Gamma \aGives T$ then
$\ren{\Gamma_0,\Gamma} \rGives \ren{T}:\set{r}$.
\end{quote}

We detail three cases.

\begin{description}

\item[Tuple construction]
Suppose $\Gamma_0,\Gamma \aGives \lets{x}{(y^+)}{T}$ is derived from
\[
\begin{array}{lll} 
\Gamma_0,\Gamma \aGives (y^+):A^+  &\mbox{and} 
&\Gamma_0,\Gamma, x:\times(A^+)\aGives T~.
\end{array}
\]
Then we derive:
\[
\ren{\Gamma_0,\Gamma} \rGives \lets{x}{(y^+)\at{r}}{\ren{T}}:\set{r}
\] 
from:
\[
\begin{array}{ll}
\ren{\Gamma_0,\Gamma} \rGives (y^+)\at{r}:\times(\ren{A}^+)\at{r} & \mbox{ and}\\ 
\ren{\Gamma_0,\Gamma},x:\times(\ren{A}^+)\at{r} \rGives \ren{T}:\set{r} &\mbox{(inductive hypothesis)}.
\end{array}
\]

\item[Projection]
Suppose $\Gamma_0,\Gamma \aGives \lets{x}{\prj{i}{y}}{T}$  is derived from
$y:\times(A_1,\ldots,A_n)\in \Gamma$, $1\leq i \leq n$, and 
$\Gamma_0,\Gamma,x:A_i \aGives T$.
Then: 
\[
y:\times(\ren{A_1},\ldots,\ren{A_n})\at{r}\in \ren{\Gamma} \mbox{ and }
\ren{\Gamma_0,\Gamma,x:A_i} \aGives \ren{T}:\set{r} ~.
\]
Hence:
\[
\ren{\Gamma_0,\Gamma} \aGives \ren{\lets{x}{\prj{i}{y}}{T}}:\set{r}~.
\]

\item[Application]
Suppose $\Gamma_0,\Gamma \aGives @(x,y^+)$ is derived from
$x:A^+\arrow R,y^+:A^+\in \Gamma_0,\Gamma$. 
Then we derive $\ren{\Gamma_0,\Gamma} \rGives @(x,r,y^+):\set{r}$
from:
\[
x:\forall r.\ren{A}^+\act{\set{r}} R,\ y^+:\ren{A}^+\in \ren{\Gamma_0,\Gamma}~.
\]

\end{description}

Finally, we derive from the enrichment property above 
the following two properties which suffice to
derive the statement:

\begin{itemize}

\item If  $\Gamma_0 \aGives \lambda x^+.T:A^+\arrow R$ then 
$\ren{\Gamma_0} \rGives \ren{\lambda x^+.T} : \ren{A^+\arrow R}$.

\item If $\Gamma_0 \aGives T$ then 
$\ren{\Gamma_0} \rGives \letall{r}{\ren{T}}:\emptyset$.

\end{itemize}

\subsection*{Proof of proposition \ref{progress-prop} [progress]}
First we prove by induction on the structure of a heap context
the following monotonicity property of the coherence predicate:
\begin{quote}
if $\w{Coh}(H,L)$ and $L\subseteq L'$ then $\w{Coh}(H,L')$. 
\end{quote}

Let $\frv{H}$ denote the set of region variables free in a heap context.
If the program $P\equiv F[H[\Delta]]$ is typable then a judgement
of the form $\Gamma \rGives H[\Delta]:e$ is derivable. We show
by induction on the typing of such judgement the following two properties:

\begin{enumerate}

\item $\w{Coh}(H,\frv{H})$.

\item If $r\in \frv{H}$ then $r\in e$.

\end{enumerate}

Because we assumed $\frv{P}=\emptyset$ we must have $\frv{H}=\emptyset$
and by the first property we derive $\w{Coh}(H,\emptyset)$. In other
terms, in a typable program without free region variables 
the heap context is coherent relatively to the empty set.
We look at the shape of $\Delta$.

\begin{description}

\item[Labelling] If $\Delta$ is a labelling then the program may reduce.

\item[Application] If $\Delta$ is an application $@(x,r^*,y^+)$ then either 
the variable $x$ is not bound in the function context or it is bound
to a function value. The fact that 
the number of parameters matches the number of 
arguments is forced (as usual) by typing. Then a reduction is possible.

\item[Projection]
The last case is when the redex is a projection $\lets{x}{\prj{i}{y}}{T}$.
Similarly to the previous case, either $y$ is not bound or it is bound
to a tuple allocated at a region $r$.
Then we must be able to type a term of the shape:
\begin{equation}\label{ndis-type}
\Gamma, y:(A)\at{r} \rGives H[\lets{x}{\prj{i}{y}}{T}]
\end{equation}
The fact that the projection is in the right range is forced (as usual) by typing.
To fire the transition we need to check that $\w{NDis}(r,H)$ holds.
In fact let us argue that if the predicate does not hold then 
the judgement (\ref{ndis-type}) above cannot be typed.
By inspecting the definition of $\w{NDis}(r,H)$ we see that 
for the predicate to fail, $H$ must have the shape $H_1[\dis{r}[H_2]]$ 
for a heap context  $H_1$ which contains 
neither allocations nor disposals on the region $r$.
But then $H_2[\lets{x}{\prj{i}{y}}{T}]$ must produce a visible effect on $r$.
Indeed the typing system records an effect on $r$ when projecting $y$ and 
this effect cannot be hidden by an allocation because the region variable $r$
is free in the context $\Gamma,y:(A)\at{r}$. 
Then the typing of $\dis{r}[H_2[\lets{x}{\prj{i}{y}}{T}]]$ fails because
the typing forbids disposing a region which is in the effect of the continuation.

\end{description}

\subsection*{Proof of proposition \ref{type-eff-subjred-prop} [subject reduction, types and effects]}
First we prove some standard properties (cf. proof of proposition \ref{subj-red-adm-prop}) and
a specific property on injective region substitutions.

\begin{description}

\item[Weakening]
If $\Gamma \rGives P:e$ then $\Gamma, x:A \rGives P:e$ with $x$ fresh.

\item[Variable substitution]
If $\Gamma,x:A \rGives P:e$ and $y:A\in \Gamma$ then 
$\Gamma \rGives [y/x]P:e$.

\item[Type substitution]
If $\Gamma \rGives P:e$ then $[B/t](\Gamma) \rGives P:e$.

\item[Injective region substitution]
If $\Gamma \rGives T:e$ and $\sigma$ is a (finite domain) region substitution which is injective on $\frv{T}\union e$ then $\sigma \Gamma \rGives \sigma T:\sigma e$.

\end{description}

We detail the proof of the last property which proceeds by induction on the typing
proof of $\Gamma \rGives T:e$. 

\begin{description}

\item[Application]
Suppose $\Gamma \rGives @(x,r^*,y^+):[r^*/r^*_1]e$ is derived from
$x:B,y^+:[r^*/r^*_1]A^+ \in \Gamma$, $B\equiv \forall r_1.A^+\act{e} R$,
$\frv{B}=\emptyset$, $r^*$ distinct variables.
Notice that $\frv{A^+}\union e\subseteq \set{r^*_1}$. It follows that
$\frv{@(x,r^*,y^+)}\union [r^*/r^*_1]e=\set{r^*}$.
So suppose $\sigma$ is an injective substitution on $r^*$ so that
$r'^*=(\sigma r)^*$. We remark:
\[
\begin{array}{ll}
\sigma B &\equiv B \\
\sigma [r^*/r^*_1]A^+ &\equiv [r'^*/r^*_1]A^+ \\
\sigma [r^*/r^*_1]e    &\equiv [r'^*/r^*_1]e  \\
\sigma @(x,r^*,y^+)    &\equiv @(x,r'^*,y^+)  \\
\sigma r^*             &\mbox{distinct}
\end{array}
\]
Then we can prove $\sigma \Gamma \rGives \sigma @(x,r^*,y^+): \sigma ([r^*/r^*_1]e)$
by the typing rule for application.

\item[Unit]
Suppose $\Gamma \rGives \lets{x}{()}{T}:e$ is derived from
$\Gamma,x:\times() \rGives T:e$ and $\sigma$ is injective
on $\frv{\lets{x}{()}{T}}\union e$. Then $\sigma$ is injective
on $\frv{T}\union e$, by inductive hypothesis 
$\sigma \Gamma,x:\times() \rGives \sigma T:\sigma e$, and we conclude
$\sigma \Gamma \rGives \sigma(\lets{x}{()}{T}):\sigma e$.

\item[Product]
Suppose 
\[
\Gamma \rGives \lets{x}{(y^+)\at{r}}{T}:e\union \set{r}
\]
is derived from
\[
\Gamma,x:\times(A^+)\at{r} \rGives T:e,
\]
$y^+:A^+\in \Gamma$ and $\sigma$ is injective on the set:
\[
\frv{\lets{x}{(y^+)\at{r}}{T}}\union e\union \set{r} = \frv{T}\union e \union \set{r}~.
\] 
Then $\sigma$ is injective on $\frv{T}\union e$. By inductive hypothesis:
\[
\sigma \Gamma,x:(\times(\sigma A^+))\at{\sigma r} \rGives \sigma T:\sigma e~.
\]
Moreover $y^+:(\sigma A)^+\in \sigma \Gamma$. So we conclude:
\[
\sigma\Gamma \rGives (\lets{x}{(y^+)\at{\sigma r}}{T}):\sigma e\union \set{\sigma r}~.
\]

\item[Existential] This case is similar to the previous one.
Suppose:
\[
\Gamma \rGives \lets{x}{(y)\at{r}}{T}:e\union \set{r}
\]
is derived from:
\[
\Gamma,x:(\exists t.A)\at{r} \rGives T:e~,
\]
$y:[B/t]A \in \Gamma$ and $\sigma$ is injective
on the set:
\[
\frv{\lets{x}{(y)\at{r}}{T}}\union e\union \set{r} = \frv{T}\union e \union \set{r}~. 
\]
Then $\sigma$ is injective on $\frv{T}\union e$. By inductive hypothesis:
\[
\sigma \Gamma,x:(\exists t.\sigma A)\at{\sigma r} \rGives \sigma T:\sigma e~.
\]
Moreover $y:\sigma [B/t]A \in \sigma \Gamma$.
We notice $\sigma [B/t]A \equiv [\sigma B/t]\sigma A$ and 
$\sigma (\exists t. A) \equiv \exists t.\sigma A$.
Then we conclude:
\[
\sigma\Gamma \rGives (\lets{x}{(y)\at{\sigma r}}{T}):\sigma e\union \set{\sigma r}~.
\]

\item[Projection]
Suppose:
\[
\Gamma \rGives \lets{x}{\prj{i}{y}}{T}:e\union \set{r}
\] 
is derived from
$y:\times(A_1,\ldots,A_n)\at{r}\in \Gamma$, $1\leq i \leq n$,
$\Gamma,x:A_i \rGives T:e$, and $\sigma$ is injective
on $\frv{\lets{x}{\prj{i}{y}}{T}}\union e\union \set{r}$.
Then $\sigma$ is injective on $\frv{T}\union e$ and by inductive hypothesis:
\[
\sigma \Gamma,x:\sigma (\times(A_1,\dots,A_n)\at{r}) \rGives \sigma T:\sigma e~.
\]
We conclude:
\[
\sigma \Gamma \rGives  \sigma (\lets{x}{\prj{i}{y}}{T}):\sigma e\union \set{\sigma r}~.
\]
The case where $y$  has an existential type is similar.

\item[Disposal]
Suppose $\Gamma \rGives \dis{r}{T}:e\union \set{r}$ is derived from
$\Gamma \rGives T:e$, $r\notin e$, and $\sigma$ is injective on 
$\frv{\dis{r}{T}}\union e \union \set{r}$. Then 
$\sigma$ is injective on $\frv{T}\union e$ and by inductive hypothesis
$\sigma \Gamma \rGives \sigma T: \sigma e$. Also $\sigma r \notin \sigma e$.
We conclude:
\[
\sigma \Gamma \rGives \sigma (\dis{r}{T}):\sigma e \union \set{\sigma r}~.
\]

\item[Allocation]
Suppose $\Gamma \rGives \letall{r}{T}:e$ is derived from
$\Gamma \rGives T:e\union \set{r}$, $r\notin e\union \frv{\Gamma}$, 
and $\sigma$ is injective on $\frv{\letall{r}{T}}\union e$. 
Up to renaming, we can choose $r$ so that it is not in the 
domain or image of $\sigma$.
Then  $\sigma$ is injective on $\frv{T}\union e\union \set{r}$ 
and by inductive hypothesis
$\sigma \Gamma \rGives \sigma T: \sigma e\union \set{\sigma r}$. 
Also, by the choice above, $\sigma r \notin \sigma e \union \sigma \Gamma$.
We conclude $\sigma \Gamma \rGives \sigma (\letall{r}{T}):\sigma e$.

\item[Labelling]
Suppose $\Gamma \rGives \lb{\ell}{T}:e$ is derived from 
$\Gamma \rGives T:e$ and $\sigma$ is injective on $\frv{\lb{\ell}{T}}\union \set{e}$. By inductive hypothesis $\sigma \Gamma \rGives \sigma T :\sigma e$ and 
we conclude $\sigma \Gamma \rGives \sigma (\lb{\ell}{T}): \sigma e$.

\item[Subeffect]
Suppose $\Gamma \rGives P:e$ is derived from $\Gamma \rGives P:e'$,
$e'\subseteq e$, and $\sigma$ is injective on $\frv{P}\union e$.
By inductive hypothesis $\sigma \Gamma \rGives \sigma P:\sigma e'$ and
we conclude $\sigma \Gamma \rGives \sigma P:\sigma e$.

\end{description}

If $\Gamma_P \rGives P:e_P$ then we know that $P\equiv F[H[\Delta]]$ 
and the reduced term has the shape $F[H[\Delta']]$.
For some $\Gamma$ we have $\Gamma\rGives \Delta:e'$.
We show that then $\Gamma\rGives \Delta':e'$ and 
$\frv{\Delta'}\subseteq \frv{\Delta}$.
Then we claim that the typing proof for the surrounding context
$F[H]$ can be ported to the program $F[H[\Delta']]$.

We proceed by case analysis on the reduction rule applied and
its typing. Notice that the typing is syntax directed except
for the subeffect rule. So for instance, if $\Gamma \rGives \Delta:e'$
and $\Delta$ is an application  then for some $e'' \subseteq e'$ 
we can derive $\Gamma \rGives \Delta:e''$
where the last rule being applied is the one for application.
A similar argument holds for the cases where $\Delta$ is a projection
or a labelling.

\begin{description}

\item[Application]
Suppose $\Gamma \rGives @(x,r^*,y^+):e''$ with $e''=[r^*/r^*_1]e$ is derived
from $x:B,y^+:[r^*/r^*_1]A^+ \in \Gamma$, $B\equiv \forall r_1.A^+\act{e} R$,
$\frv{B}=\emptyset$, $r^*$ distinct variables.
Since the program reduces, $x$ must be bound to a region closed function 
$\lambda r^*_1,z^+.T$ in the functional context $F$ and 
$\Gamma_1,z^+:A^+ \rGives T:e$ where 
$\Gamma_1$ is a prefix of $\Gamma$ and 
$\set{r^*_{1}}\inter \frv{\Gamma_1}=\emptyset$.
We notice that the substitution $\sigma = [r^*/r^*_{1}]$
is injective on $\frv{T}\union e \subseteq \set{r^*_{1}}$,
hence by the injective substitution property we derive:
\[
\Gamma_1,z^+:(\sigma A)^+ \rGives \sigma T : \sigma e~.
\]
Notice that we must have $y^+:(\sigma A)^+\in \Gamma_1$ hence
by the variable substitution property and weakening we derive:
\[
\Gamma \rGives \sigma [r^*/r^*_1,y^+/z^+]T: [r^*/r^*_1]e~.
\]
We conclude by noticing that:
\[
\frv{[r^*/r^*_1,y^+/z^+]T}\subseteq \frv{@(x,r^*,y^+)}=\set{r^*}~.
\]

\item[Projection]
Suppose $\Gamma \rGives \lets{x}{\prj{i}{y}}{T}:e\union \set{r}$ 
is derived from $y:\times(A_1,\ldots,A_n)\in \Gamma$, $1\leq i \leq n$ 
$\Gamma,x: A_i \rGives T:e$.
Since the program reduces, $y$ must be bound to a tuple 
$(z_1,\ldots,z_n)\at{r}$ in the heap context $H$ and
$z_1:A_1,\ldots,z_n:A_n\in \Gamma$.
Then by variable substitution we derive 
$\Gamma \rGives [z_i/x]T:e$. Also notice that 
$\frv{[z_i/x]T}=\frv{\lets{x}{\prj{i}{y}}{T}}$.

The case where $y$ has an existential type is similar except that it
relies on type substitution too (cf. proof of proposition
\ref{subj-red-adm-prop}).

\item[Labelling]
Suppose $\Gamma \rGives \lb{\ell}{T}:e$ is derived from 
$\Gamma \rGives T:e$. Since $\Delta \equiv \lb{\ell}{T} \act{\ell} T$, 
the conclusion is immediate.
\end{description}

\subsection*{Proof of theorem \ref{reg-sim-thm} [region simulation]}
First, we observe that the region erasure function commutes with variable
substitution: 
\[
[x/y]\rer{T} \equiv \rer{[x/y]T}~.
\]
By proposition \ref{decom-prop}, $P$ decomposes as $F[H[\Delta]]$ where
$\Delta$ is either an application, or a projection, or a labelling.
The region erasure function commutes with this decomposition too, so that we can write
$\rer{P}$ as $\rer{F}[\rer{H}[\rer{\Delta}]]$, where $\rer{F}[\rer{H}]$ is
an evaluation context. 
If $\rer{P} \act{\alpha} Q$ then we proceed by case analysis on the
reduction rule being applied. We detail the case where $\Delta$ is an application
$@(x,r^*,y^+)$.
Then we must have $F\equiv F_1[\lets{x}{\lambda r_1^*,z^+.T}{F_2}]$ and 
\[
\begin{array}{lll}
\rer{P} &\equiv &\rer{F_1}[\lets{x}{\lambda z^+.\rer{T}}{\rer{F_2}[\rer{H}[@(x,y^+)]]}  \\
        &\arrow &\rer{F_1}[\lets{x}{\lambda z^+.\rer{T}}{\rer{F_2}[\rer{H}[[y^+/z^+]\rer{T}]]} ~.
\end{array}
\]
Since $P$ is typable, the heap context is coherent and then $P$ 
can simulate the reduction  above as follows:
\[
P \arrow F[H[[r^*/r_1^*,y^+/z^+]T]]
\]
noticing that $\rer{[r^*/r_1^*,y^+/z^+]T}\equiv[y^+/z^+]\rer{T}$ (initial remark and invariance
of the region erasure function under region substitutions).

\end{document}